\documentclass[useAMS,usenatbib]{mn2e}

\usepackage{graphicx}
\usepackage{multirow}

\title[Background--Source Separation in astronomical
  images]{Background--Source Separation in astronomical images with Bayesian
  probability theory (I): the method}
\author[F. Guglielmetti et al.]{F.~Guglielmetti,$^{1}$\thanks{E-mail: 
Fabrizia.Guglielmetti@ipp.mpg.de} 
R.~Fischer,$^{1}$  and V.~Dose$^{1}$
\\
$^{1}$Max-Planck-Institut f\"ur Plasmaphysik, Boltzmannstrasse 2, 85748 Garching, Germany}
\begin{document}

\date{Accepted 2009 March 4. Received 2009 March 3; in original form 2008 October 14}

\pagerange{\pageref{firstpage}--\pageref{lastpage}} \pubyear{2002}

\maketitle

\label{firstpage}

\begin{abstract}
A probabilistic technique for the joint estimation of background and sources
with the aim of detecting faint and extended celestial objects is described. 
Bayesian probability theory is applied to gain insight into the coexistence 
of background and sources through a probabilistic two--component mixture model, 
which provides consistent uncertainties of background and sources.  
A multi--resolution analysis is used for revealing faint and extended objects 
in the frame of the Bayesian mixture model. 
All the revealed sources are parameterized automatically providing source
position, net counts, morphological parameters and their errors. \\
We demonstrate the capability of our method by applying it to three
simulated datasets characterized by different background and source
intensities. The results of employing two different prior knowledges on the source
signal distribution are shown. The probabilistic method allows
for the detection of bright and faint sources 
independently of their morphology and the kind of background.  
The results from our analysis of the three simulated datasets are compared
with other source detection methods. 
Additionally, the technique is applied to {\it ROSAT} all--sky survey data. 
\end{abstract}

\begin{keywords}
methods: data analysis -- methods: statistical -- techniques: image processing.
\end{keywords}

\section{Introduction}\label{introd}
Background estimation is an omnipresent problem for source detection methods 
in astrophysics, especially when the source signal is weak and difficult to  
discriminate against the background. 
An inaccurate estimation of the background may produce 
large errors in object photometry and the loss of faint and/or  
extended objects. 

Deep observations are commonly obtained by combining several individual
exposures to generate a final astronomical image. 
Often large exposure variations characterize these data and 
the background may vary significantly within the field. 
Hence, the background modelling has to incorporate the knowledge provided 
by the observatory's exposure time without compromising the statistical 
properties. 
In addition, instrumental structures, such as detector ribs or CCD gaps,
produce lack of data. 
The missing data must be handled consistently in the background
estimation to prevent undesired artificial effects. \\
Celestial objects exhibit a large variety of morphologies and apparent sizes. 
Thus, the search for sources should not be driven by any predefined morphology,
allowing proper estimate of their structural parameters. 
The instrumental point spread function (PSF) 
is often not known exactly for the whole field of view. 
So a source detection algorithm should be able to operate
effectively without the knowledge of the instrument characteristics
\footnote{If the instrumental PSF is known precisely, then that information
  can be taken into account in a further step 
of source characterization within the properties of a specific
observational setup.}. 

Conventional source detection methods employed in deep imaging surveys 
are: SExtractor (\citealt{bertin:1996}), sliding window technique 
(\citealt{harnden:1984}; \citealt{gioia:1990}) and maximum likelihood PSF
fitting (\citealt{hasinger:1994}; \citealt{boese:2001}), 
wavelet transformation (e.g.~\citealt{slezak:1990}; \citealt{rosati:1995}; \citealt{damiani:1997};
\citealt{starck:1998}; \citealt{freeman:2002}). 
A review of these techniques can be found in \citet{valtchanov:2001} and
\citet{becker:2007}. 
 
SExtractor is one of the most widely used source detection procedure in astronomy.  
It has a simple interface and very fast execution. It produces reliable aperture
photometry catalogues (\citealt{becker:2007}). 
It is applied in {\it X}--ray regime on filtered images 
(Valtchanov et al.~2001).

The sliding window technique is a fast and robust source detection method.  
This technique may fail while detecting extended sources,
sources near the detection limit and nearby sources (Valtchanov et al.~2001). 
This source detection method has been refined with more 
elaborated techniques, such as matched filters 
(e.g.~\citealt{vikhlinin:1995}; \citealt{stewart:2006}) 
and recently the Cash method (\citealt{stewart:2009}). 
The Cash method is a maximum likelihood technique. 
For source detection, the method employs a Cash 
likelihood--ratio statistic, that is an extended 
$\chi^{2}$ statistic for Poisson data (\citealt{cash:1979}). 
Both the matched filters and Cash methods are at
least by a factor of $1.2$  more sensitive than the sliding--window
technique (\citealt{stewart:2009}). Though, both methods are designed for the
detection of point sources.
The candidate sources are characterized in a further step using maximum
likelihood PSF fitting. 
The maximum likelihood PSF fitting procedure performs better than
other conventional techniques for flux measurements of point--like 
sources although accurate photometry is achieved when the PSF model used is close
to the image PSF (Valtchanov et al.~2001). 
In \citet{pierre:2004}, the maximum likelihood profile fit on photon images is
extended taking into account a spherically symmetric $\beta$--model (King
profile, see refs.~\citealt{king:1962}, \citealt{cavaliere-fusco:1978}) convolved 
with the instrumental PSF for improving the photometry of extended objects.

Wavelet transform (WT) techniques improve the detection of faint and extended
sources with respect to other conventional methods (see \citealt{starck:1998} 
for more details). 
In fact, WT techniques are able to discriminate structures
as a function of scale. Within larger scales, faint and extended sources are
revealed. WTs are therefore valuable tools for the
detection of both point--like and extended sources
(Valtchanov et al.~2001).   
Nonetheless, these techniques favor the detection of circularly
symmetric sources (Valtchanov et al.~2001). 
In addition, artefacts may appear around the detected structures in the
reconstructed image, and the flux is not preserved 
(\citealt{starck:1998}). In order to overcome these problems,
some refinements have been applied to the WT techniques. 
\cite{starck:1998}, for instance, employ a multi--resolution support filtering to preserve 
the flux and the adjoint WT operator 
to suppress artefacts which may appear around the objects.   
An advance on this method is presented in \cite{starck:2000}. A
space--variant PSF is incorporated in their WT technique. 
Object by object reconstruction is performed. For point sources
the flux measurements are close to that obtained by PSF fitting.

The detection of faint extended and point--like sources
is a non--trivial task for source detection methods. 
This task becomes more complicated when the detection rate and photometric reconstruction for the
sources are taken into account (Valtchanov et al.~2001).

A self--consistent statistical approach for background estimation and source
detection is given by Bayesian probability theory (BPT), which provides a 
general and consistent frame for logical inference. 
The achievement of Bayesian techniques on signal detections in astrophysics 
has already been shown, for example in the works of
\cite{gregory:1992}, \cite{loredo:1995} and 
\cite{scargle:1998}, for instance. In modern observational astrophysics, 
BPT techniques for image analysis have been extensively applied:  
e.g.~\citet{hobson:2003}, \citet{carvalho:2008}, \citet{savage:2007},  
\citet{strong:2003}.  
For the detection of discrete objects embedded in Gaussian noise (microwave regime), 
\cite{hobson:2003} utilizes a model--fitting methodology, where a parameterized 
form for the objects of interest is assumed.
Markov--chain Monte Carlo (MCMC) techniques are used to explore the parameter
space. 
An advance on this work is provided by Carvalho et al.~(2008). For speeding up the method of
\cite{hobson:2003}, Carvalho et al.~(2008) proposes to use Gaussian approximation
to the posterior probability density function (pdf) peak when performing a 
Bayesian model selection for source detection. 
The work of \cite{savage:2007} is developed within Gaussian statistics (infrared data). 
At each pixel position in an image, their method estimates the probability of 
the data being described by point source
or empty sky under the assumptions that the background is uniform and the
sources have circular shapes. The Bayesian information criterion is used for
the selection of the two models. Source parameters are estimated in a second
step employing Bayesian model selection. \cite{strong:2003} developed a 
technique for image analysis within Poisson statistics. The technique is 
instrument specific and is applied to $\gamma$--ray data. The first objective
of this technique is to reconstruct the intensity in each image pixel given a
set of data. The Maximum Entropy method is used for selecting from the data an
image between all the available ones from a multi--dimensional space. 
The dimension of the space is proportional to the number of image pixels. 

We propose a new source detection method based on BPT combined with the
  mixture--model technique. 
The algorithm allows one to estimate the background and its uncertainties 
and to detect celestial sources jointly. 
The new approach deals directly with the statistical nature
  of the data.  
Each pixel in an astronomical image is probabilistically assessed to contain 
background only or with additional source signal. 
The results are given by probability distributions quantifying our state of 
knowledge. The developed Background--Source separation (BSS) method 
encounters: background estimation, source detection and characterization.\\
The background estimation incorporates the knowledge of the exposure time map. 
The estimation of the background and its uncertainties is performed on 
the full astronomical image employing a two--dimensional spline. 
The spline models the background rate. The spline amplitudes and  
the position of the spline supporting points provides flexibility in the 
background model. 
This procedure can describe both smoothly and highly varying backgrounds.  
Hence, no cut out of regions or employment of meshes are needed for 
the background estimation. 
The BSS technique does not need a threshold level for separating the sources 
from the background as conventional methods do. 
The threshold level is replaced by a measure of probability. 
In conventional methods, the threshold level is described in terms of 
the noise standard deviation, then translated into a probability. 
The classification assigned to each pixel of an astronomical image 
with the BSS method allows one to detect sources without employing any predefined
morphologies. Only, for parametric characterization of the sources  
predefined source shapes are applied. 
The estimation of source parameters and their uncertainties 
includes the estimated background into a forward model.  
Only the statistics of the original data are taken into account. 
The BSS method provides simultaneously the advantages of a multi--resolution
analysis and a multi--color detection. 
Objects in astronomical images are organized in hierarchical structures 
\citep{starck:2006}. 
In order to quantify the multiscale structure in the data, a multi--resolution 
analysis is required (see \citealt{starck:2000}; \citealt{kolaczyk:2000}). 
In the BSS approach the multi--resolution analysis is incorporated in
combination with the source detection and background estimation technique 
  with the aim to analyse statistically source structures at multiple scales.  
When multiband images are available, the information contained in each image
  can be statistically combined in order to extend the detection limit of the
  data (see \citealt{szalay:1999}; \citealt{murtagh:2005}). \\
The capabilities of this method are best shown with the 
detections of faint sources independently of their shape and 
with the detections of sources embedded in a highly varying background. 
The technique for the joint estimation of background and sources in digital images
is applicable to digital images collected by a wide variety of sensors at any
wavelength. 
The {\it X}--ray environment is particularly suitable to our Bayesian approach
for different reasons. 
First, {\it X}--ray astronomy is characterized by low photon counts
even for relatively large exposures and the observational data are 
unique, i.e.~the experiment is rarely reproduced. 
Second, the {\it X}--ray astronomical images provided by new generation instruments 
are usually a combination of several individual CCD imaging
pointings.  
Last, there are few source detection algorithms developed 
so far for an automated search of faint and extended sources.

In this paper, our aim is to describe the developed BSS technique. 
The outline of this paper is as follows. In Section \ref{BI}, we 
briefly review the basic aspects of BPT. In Section \ref{technique}, we 
introduce the background estimation and source detection technique. 
In Section \ref{EOSP}, the BSS algorithm is extended in order to obtain 
an automated algorithm for source characterization. 
In Section \ref{rely}, the issue of false positives
in source detection is addressed. 
In Section \ref{simu_data}, the 
BSS method is applied to simulated data. We show results for two different
choices of prior pdf for the source signal.
In Section \ref{compa}, our results on the three simulated datasets are
compared with the outcome from {\sc wavdetect} algorithm (\citealt{freeman:2002}). 
In Section \ref{odata}, we test the BSS method on astronomical images 
coming from {\it ROSAT} all-sky survey (RASS) data. 
Our conclusions on the BSS technique are provided in Section \ref{conclu}. 
In Section \ref{acro}, a list of acronyms used throughout the paper is
reported. 

The flexibility of our Bayesian technique allows the investigation of 
data coming from past and modern instruments without changes in our algorithm. 
An extensive application of the BSS technique 
to real data is addressed in a forthcoming paper (Guglielmetti et al., in preparation). 
In the latter, the BSS algorithm is used for the analysis of a specific
scientific problem, i.e.~the search of galaxy clusters in the CDF--S data. 

\section{Bayesian probability theory}\label{BI}

In order to analyse the heterogeneous data present in astronomical images, we
employ BPT (see \citealt{jeffreys:1961};  
\citealt{bernardo:1994}; \citealt{gelman:1995}; \citealt{sivia:1996}; 
\citealt{jaynes:2003}; \citealt{dose:2003}; \citealt{ohagan:2004}; \citealt{gregory:2005}). 
BPT gives a calculus how to resolve an inference problem based on uncertain
information. The outcome of the BPT analysis is the pdf of the quantity of
interest, which encodes the knowledge to be drawn from the information
available ({\em a posteriori}). The posterior pdf comprises the complete
information which can be inferred from the data and the statistical model, and
supplementary information such as first--principle physics knowledge.  

This statistical approach is based on comparisons among alternative hypotheses
(or models) using the single observed data set. 
Each information entering the models that describe the data set is combined 
systematically. The combination includes all data sets coming from different
diagnostics, physical model parameters and measurement nuisance parameters.
Each data set and parameters entering the models are subject to uncertainties
which have to be estimated and encoded in probability distributions. Within BPT
the so--called statistical and systematic uncertainties are not 
distinguished. Both kinds of uncertainties are treated as a lack of knowledge.

Bayes' theorem states: 
\begin{equation}
P(H_{\rm i}|D,\sigma,I) = \frac{P(D|H_{\rm i},\sigma,I)P(H_{\rm i}|I)}{P(D|I)}, 
\label{BT}
\end{equation}
where the number of alternative hypotheses to be compared is larger than or equal
to $2$. 
Bayes' theorem is a consequence of the sum and product rules of probability
theory. 
The vertical bars in (\ref{BT}) denote conditionality property, based on
either empirical or theoretical information. 
Equation (\ref{BT}) relates the {\em posterior} pdf $P(H_{\rm i}|D,\sigma,I)$ to
known quantities, namely, the {\em likelihood} pdf $P(D|H_{\rm i},\sigma,I)$ and
the {\em prior} pdf $P(H_{\rm i}|I)$. $P(D|I)$ is the {\em evidence} of the data
which constitutes the normalization and will not affect the conclusions within
the context of a given model. The posterior pdf is the quantity to be
inferred. It depends on the full data set $D$, on the errors $\sigma$ entering the
experiment and on all relevant information concerning the nature of the physical situation and
knowledge of the experiment ($I$). 
The likelihood pdf represents the probability of finding the data $D$ for given
quantities of interest, uncertainties $\sigma$ and additional information
$I$. It reveals the error statistics of the experiment. 
The prior pdf is independent from the experimental errors. 
It represents physical constraints or additional information from other diagnostics. 
The terms `posterior' and `prior' have a logical rather than temporal 
meaning. The posterior and prior pdfs can be regarded as the knowledge `with'
and `without' the new data taken into account, respectively. 

In order to arrive at the pdf of any quantity $x$ in the model,
marginalization of the multi--dimensional pdf can be regarded as a projection
of the complete pdf on to that quantity. Marginalization is performed by
integration over the quantity $y$ one wants to get rid of:
\begin{equation}
P(x)=\int{P(x,y) dy}=\int{P(x|y)P(y)dy}.
\label{marru}
\end{equation}
Marginalization of a quantity $y$ thus takes into account the uncertainty of
$y$ which is quantified by the pdf P(y).
The uncertainty of $y$ propagates into the pdf P(x). 
Marginalization provides a way to eliminate variables 
which are necessary to formulate the likelihood but otherwise uninteresting. 

Another important property of BPT is the capability of modelling the data by
mixture distributions in the parameter space (\citealt{everitt:1981}; \citealt{neal:1992}). 
Mixture distributions are an appropriate tool for modelling processes whose
output is thought to be generated by several different underlying mechanisms,
or to come from several different populations. Our aim is to identify and
characterize these underlying `latent classes'. Therefore, we follow  
the standard Bayesian approach to this problem, which is to define a prior
distribution over the parameter space  of the mixture model and combine this
with the observed data to give a posterior distribution over the parameter
space. 
Essential for model comparison or object classification is the marginal
  likelihood ({\em evidence}, {\em prior predicted value}). Marginalization
  (integration) of the likelihood over parameter space provides a measure for
  the credibility of a model for given data. Ratios of marginal likelihoods
  ({\em Bayes factors}) are frequently used for comparing two models
  \citep{kass:1995}. 
Details of mixture modelling in the framework of BPT can be found in
von der Linden et al.~(1997, 1999) and Fischer et al.~(2000, 2001, 2002). 
In particular, \cite{fischer:2002} have demonstrated the capability of the
Bayesian mixture model technique even with
an unknown number of components for background separation from a measured
spectrum. The present approach follows these previous works.

\section{The Joint Estimation of Back\-ground and Sources with BPT} \label{technique}

The aim of the BSS method is the joint estimation of background and sources in
two--dimensional image data. We can identify two basic steps of our
algorithm: (A) background estimation and source detection, (B) 
calculation of source probability maps in a multi--resolution analysis. 

The input information of the developed algorithm is the experimental data, 
i.e.~the detected photon counts, and the observatory's exposure time.  

The background rate is assumed to be smooth, e.g.~spatially slowly varying
compared to source dimensions. To allow for smoothness the background rate is 
modelled with a two--dimensional thin--plate spline (TPS), 
(Section \ref{TPS_label}).
The number and the positions of the pivots, i.e.~the spline's supporting
points, decide what data structures are assigned to be
background. All structures which can not be described by the background model
will be assigned to be sources. 
The number of pivots required to model the background depends on the 
characteristics of the background itself. 
Though the minimum number of pivots is four (since the minimum expansion of
the selected spline has four terms), their number 
increases with increasing background variation.  

The coexistence of background and sources is described with a probabilistic
two--component mixture model (Section \ref{TCMM}) 
where one component describes background contribution only and the other component describes
background plus source contributions. 
Each pixel is characterized by the probability of belonging to one of the two 
mixture components. 
For the background estimation the photons contained in all pixels
are considered including those containing additional source
contributions. No data censoring by cutting out source areas is employed. 

For background estimation the source intensity is considered to be a nuisance 
parameter. According to the rules of BPT, the source signal distribution is 
described probabilistically in terms of a prior pdf. 
The prior pdf of the source signal is an approximation to the true
distribution of the source signal on the field. 
We studied two prior pdfs of the source signal: the exponential and
the inverse--Gamma function. 

The background and its uncertainties (Section \ref{sec:errbkg}) 
are estimated from its posterior pdf. Therefore, 
for each pixel of an astronomical image an estimate of its background and its 
uncertainties are provided. 

Moreover, the Bayesian approach introduces hyper--parameters, that are 
fundamental for the estimation of the posterior pdfs for the 
background and source intensities.
Specifically, in Section \ref{Hyperpar} 
we show that the hyper--parameters are estimated exclusively from the data. 

The source probability is evaluated with the mixture model technique for
pixels and pixel cells\footnote{We define pixels as the image finest resolution
  limited by instrumental design, while we define pixel cells as a group of
  correlated neighbouring pixels. \label{fn:domdef}} 
in order to enhance the detection of faint and extended sources in a
multi--resolution analysis. Pixels and pixel cells are treated identically
within the Bayesian formalism. For the correlation of neighbouring pixels, 
the following methods have been investigated: box filter with a square, box filter with 
a circle, Gaussian weighting filter (see Section \ref{TCMM} for more details). 

The BSS technique is morphology free, i.e.~there are no restrictions on the 
object size and shape for being detected. 
An analysed digital astronomical image is converted into the following: 
\begin{itemize}
\item[I)]{the background rate image, or {\em `TPS map'}, is
  an array specifying the estimated background rate at each image pixel for a
  given observation. The effects of exposure variations are consistently
  included in the spline model; }
\item[II)]{the background image, or {\em `background map'}, is an array
  specifying the estimated background amplitude at each image pixel for a
  given observation. It is provided by the TPS map multiplied with
  the telescope's exposure;
}
\item[III)]{the source probability images, or {\em `source probability maps'
  (SPMs)}, 
  display the probability that source counts are present in pixels and pixel
  cells for a given observation in a multi--resolution analysis.}
\end{itemize} 
Movies are produced with the SPMs obtained at different
resolutions. The moving images allow one to discern interactively the presence 
of faint extended sources in digital astronomical images. The size of faint 
extended sources is correlated with the scale of the resolution, used for 
their detection.
SPMs coming from other energy bands can be combined
statistically to produce conclusive SPMs at different
resolutions with the advantage to provide likelihoods for the detected sources 
from the combined energy bands (Section \ref{comb}).

\subsection{Two--component mixture model}\label{TCMM}

The general idea of the described Bayesian model is that a digital 
astronomical image consists of a smooth background with additive source
signal, which can be characterized by any shape, size and brightness. 
The background signal is the diffuse cosmic emission added to the instrumental
noise. The source signal is the response of the imaging system to a celestial object. 
A surface $b({\bf x})$ describes the background under the source signal, 
where ${\bf x}=(x,y)$ corresponds to the position on the grid in the image. \\
Therefore, given the observed data set $D=\{d_{\rm ij}\} \in \mathbf{N}_0$, where
$d_{\rm ij}$ is photon counts in pixel (or pixel cell) $\{ij\}$, 
two complementary hypotheses arise:
\begin{displaymath}
\left\{ \begin{array}{ll}
B_{\rm ij} : \quad d_{\rm ij}=b_{\rm ij}+\epsilon_{\rm ij} \\
\overline{B}_{\rm ij} : \quad d_{\rm ij}=b_{\rm ij}+s_{\rm ij}+\epsilon_{\rm ij}. 
\end{array} \right.
\end{displaymath}
Hypothesis $B_{\rm ij}$ specifies that the data $d_{\rm ij}$ consists only of
background counts $b_{\rm ij}$ spoiled with noise
$\epsilon_{\rm ij}$, i.e.~the (statistical) uncertainty associated with the
measurement process. 
Hypothesis $\overline{B}_{\rm ij}$ specifies the case where additional
source intensity $s_{\rm ij}$ contributes to the background.
Additional assumptions are that no negative values for source 
and background amplitudes are allowed and that
the background is smoother than the source signal. \\
The smoothness of the background is achieved by modelling the background count
rate with a bivariate TPS where the supporting points are chosen sparsely to 
ensure that sources can not be fitted. 
The spline fits the background component whereas count enhancements
classify pixels and pixel cells with source contributions.

In the following, pixel cells are subsumed by pixels. 
Pixel cells are collections of pixels where $d_{\rm ij}$ is the total photon count
in the cell $\{ij\}$. The photon counts of neighbouring pixels are added up
and the formed pixel cell is treated as a pixel. 
In principle, any cell shape can be chosen. In practice, 
two methods have been developed when pixels have weight of one within the cell   
(box filtering with cell shape of a square or of a circle) and one method when
pixels have different weights within the cell (Gaussian weighting). 
The box filter with cells of squared shape consists of taking
information of neighbouring pixels within a box. The cell size is
the box size. The box filter with cells of circular shape considers pixels 
with a weight of one if inside the cell size, otherwise zero. Pixels have a weight of one
when the cell size touches them at least at the centre. This method allows the
pixel cells to have varying shapes. The Gaussian weighting method provides 
Gaussian weights around a centre: weights are given at decreasing values according to the
distance from the centre.

As expressed in (\ref{BT}), we are interested in estimating the
probabilities of the hypotheses: $B_{\rm ij}$ and $\overline{B}_{\rm ij}$. \\ 
In this paper we address the problem when the photon counts are small and Poisson
statistics have to be used. 
The likelihood probabilities for the two hypotheses within Poisson statistics are:
\begin{eqnarray}
p(d_{\rm ij} \mid B_{\rm ij},b_{\rm ij}) = \frac{b_{\rm ij}^{d_{\rm ij}}}{d_{\rm ij}!} e^{-b_{\rm ij}}, 
\hspace{2.3cm} \label{like_bg} \\ 
p(d_{\rm ij} \mid \overline{B}_{\rm ij},b_{\rm ij},s_{\rm ij}) =  \frac{(b_{\rm ij}+s_{\rm ij})^{d_{\rm ij}}}{d_{\rm ij}!} e^{-(b_{\rm ij}+s_{\rm ij})}. 
\label{like_source}
\end{eqnarray} 
This technique is easily adaptable to other likelihoods, for instance
employing Gaussian statistics as given in \cite{fischer:2001a}. \\
The prior pdfs for the two complementary hypotheses are chosen to be 
$p(B_{\rm ij}) = \beta$ and $p(\overline{B}_{\rm ij}) = 1 - \beta$, independent of 
$i$ and $j$. Specifically, the parameter $\beta$ is the prior probability that a
pixel contains only background.

Since it is not known if a certain pixel (or pixel cell) contains purely background or
additional source signal, the likelihood for the mixture model is employed.
The likelihood for the mixture model effectively combines the probability 
distributions for the two hypotheses, $B_{\rm ij}$ and $\overline{B}_{\rm ij}$: 
\begin{eqnarray}
\nonumber
p(D|{\bf b}, {\bf s})= \hspace{3.5cm} \\
=\prod_{\rm ij}\{\beta\frac{b_{\rm ij}^{d_{\rm ij}}}{d_{\rm ij}!}e^{-b_{\rm ij}}+(1-\beta)\frac{(b_{\rm ij}+s_{\rm ij})^{d_{\rm ij}}}{d_{\rm ij}!}e^{-(b_{\rm ij}+s_{\rm ij})}\}, 
\label{mix_mod}
\end{eqnarray}
where ${\bf b}=\{b_{\rm ij}\}$, ${\bf s}=\{s_{\rm ij}\}$ and $\{ij\}$ corresponds to
the pixels of the complete field.

The probability of having source contribution in pixels and pixel cells is
according to Bayes' theorem (see \citealt{fischer:2001a}):
\begin{eqnarray}
\nonumber
p(\overline{B}_{\rm ij} \mid d_{\rm ij}, b_{\rm ij}, s_{\rm ij}) = \hspace{2.5cm}\\ 
= \frac{(1-\beta) {\bf \cdot} p(d_{\rm ij}
\mid \overline{B}_{\rm ij},b_{\rm ij},s_{\rm ij})}{\beta {\bf \cdot} p(d_{\rm ij} \mid
{B}_{\rm ij},b_{\rm ij})+(1-\beta) {\bf \cdot} p(d_{\rm ij}
\mid \overline{B}_{\rm ij},b_{\rm ij},s_{\rm ij})}.
\label{eqsp1}
\end{eqnarray}
This equation enables the data contained in an astronomical image to be 
classified in two groups: with and without source signal contribution. 

Equation (\ref{eqsp1}) is also used in the multi--resolution analysis. 
The SPM with the largest resolution is characterized by the probability of
uncorrelated pixels. 
At decreasing resolutions a correlation length is defined. 
Starting from a value of $0.5$, the correlation length increases in steps of 
$0.5$ pixel for decreasing resolution. 
The SPMs at decreasing resolutions are, therefore, characterized by the 
information provided by background and photon counts in pixel cells. 
Specifically, photon counts and background counts 
are given by a weighted integration over pixel cells. The integrated photon and background
counts enter the likelihood for the mixture model. Then, the source probability
is estimated for each image pixel in the multi--resolution analysis. The
multi--resolution algorithm preserves Poisson statistics.

\subsubsection{Source signal as a nuisance parameter}\label{nui}
\begin{figure}
\centering
\includegraphics[width=0.6\linewidth,angle=90]{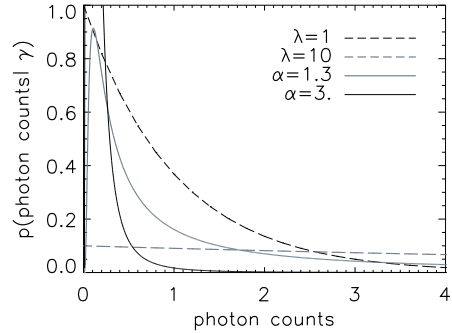} 
\caption{Representation of selected prior pdfs of the source 
  signal versus the photon counts. The {\em exponential prior pdf} is drawn at two  
  $\lambda$ values (dashed lines). The {\em inverse--Gamma function prior pdf} is plotted
  at two $\alpha$ values (continuous lines). 
On the ordinate, $\gamma$ indicates $\lambda$ or $\alpha$.}
\label{priors_fig}
\end{figure}

Following Fischer et al. (2000, 2001), 
the source signal in (\ref{like_source}) is a 
nuisance parameter, which is removed by integrating it out (marginalization
rule, eq.~(\ref{marru})):   
\begin{eqnarray}
\nonumber
p(d_{\rm ij}|\overline{B}_{\rm ij},b_{\rm ij})= \int_{0}^{\infty}
p(d_{\rm ij}|\overline{B}_{\rm ij},b_{\rm ij},s_{\rm ij}) 
{\bf \cdot} 
p(s_{\rm ij}|\overline{B}_{\rm ij},b_{\rm ij})ds_{\rm ij}.
\end{eqnarray}
A nuisance parameter is a parameter which is
important for the model (describing the data), but it is not of interest at
the moment. 
Following BPT, a prior pdf of the source signal has to be chosen. 
The final result depends crucially on the prior pdf set on the
source signal. In fact, in addition to the choice of the TPS pivots, the prior 
pdf of the source signal provides a description of
what is source. All that is not described as a source is 
identified as background and vice versa. 
Two approaches are presented: the first method accounts for the 
knowledge of the mean value of the source intensity over the complete field 
({\em exponential prior}), the second approach interprets the source signal 
distribution according to a power--law ({\em inverse--Gamma function prior}).  

\paragraph{Exponential prior} \label{EP}
Following the works of Fischer et al.~(2000, 2001, 2002), 
we choose a prior pdf of the source signal that is as 
weakly informative as possible. 
The idea follows a practical argument on the difficulty of providing sensible 
information. 
We describe the prior pdf on the source intensity by an exponential
distribution,  
\begin{equation}
p(s_{\rm ij} \mid \lambda) =
\frac{e^{-\frac{s_{\rm ij}}{\lambda}}}{\lambda}. 
\label{meprincip}
\end{equation}
This is the most uninformative Maximum Entropy distribution for known mean
value of the source intensity $\lambda$ over the complete field. \\
In Fig.~\ref{priors_fig}, equation (\ref{meprincip}) is drawn for two values of the mean source intensity:
$\lambda=1$ count and $\lambda=10$ counts. 
No bright sources are expected to appear in fields with $\lambda \sim 1$. 
In the case of values of $\lambda \gg 1$, bright and faint sources are 
present in these fields.\\
The \emph{marginal Poisson likelihood} for the hypothesis $\overline{B}_{\rm ij}$ has the form:
\begin{eqnarray} 
\nonumber
p(d_{\rm ij}|\overline{B}_{\rm ij},b_{\rm ij},\lambda)=\frac{e^{\frac{b_{\rm ij}}{\lambda}}}{\lambda(1+\frac{1}{\lambda})^{(d_{\rm ij}+1)}}
\hspace{2.cm}
\\
{\bf \times}
\frac{\Gamma[(d_{\rm ij}+1),b_{\rm ij}(1+\frac{1}{\lambda})]}{\Gamma(d_{\rm ij}+1)}, \hspace{.7cm}
\label{me_mpd}
\end{eqnarray}
where $\Gamma [a,x] = \int_{\rm x}^{\infty} e^{-t} t^{a-1} dt$ $(a > 0)$ is the 
incomplete--Gamma function and $\Gamma [a]=\Gamma[a,0]$ is the Gamma function 
(see refs.~Fischer et al.~2000, 2001). 

\begin{figure*}
\includegraphics[width=0.245\linewidth,angle=90]{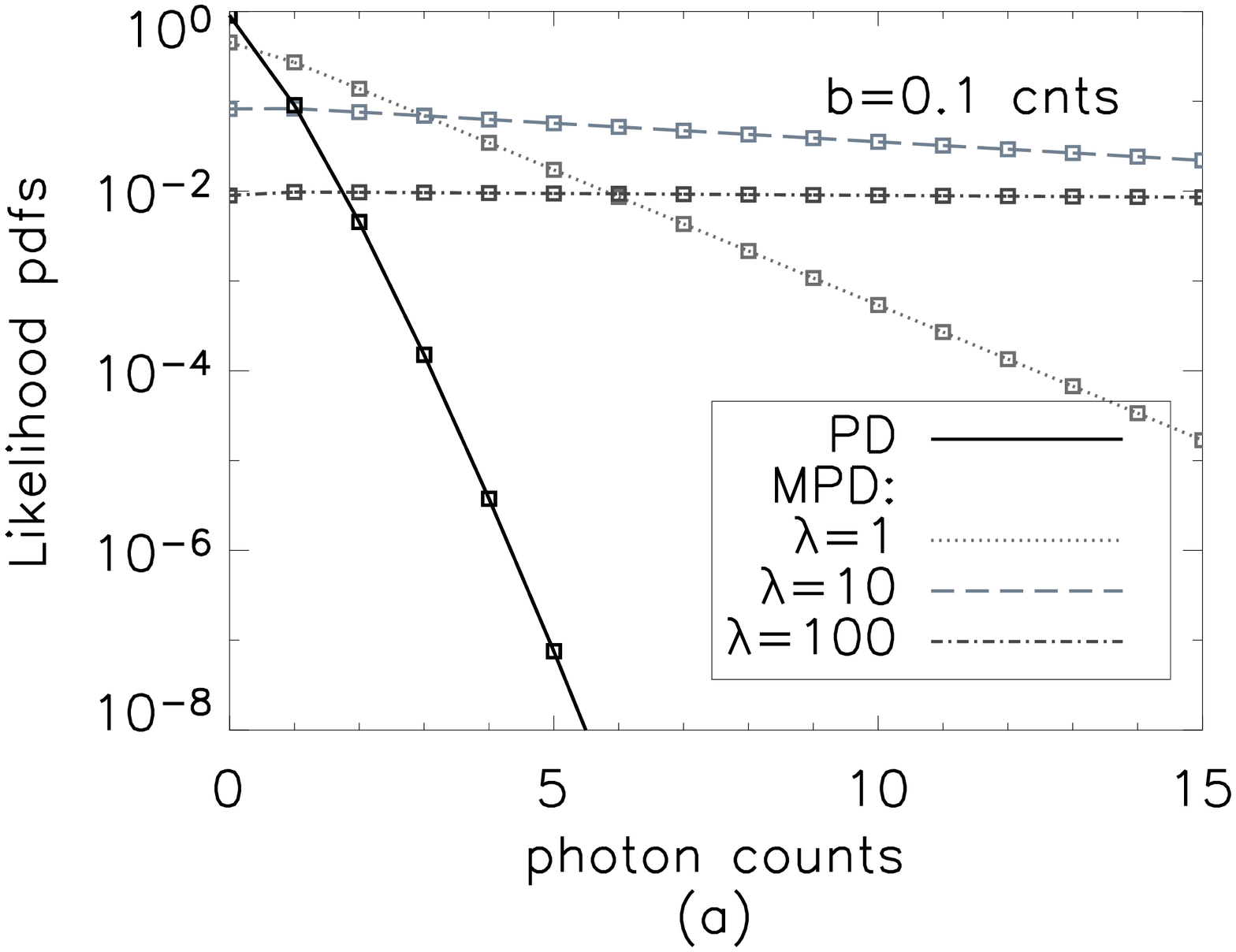}\includegraphics[width=0.245\linewidth,angle=90]{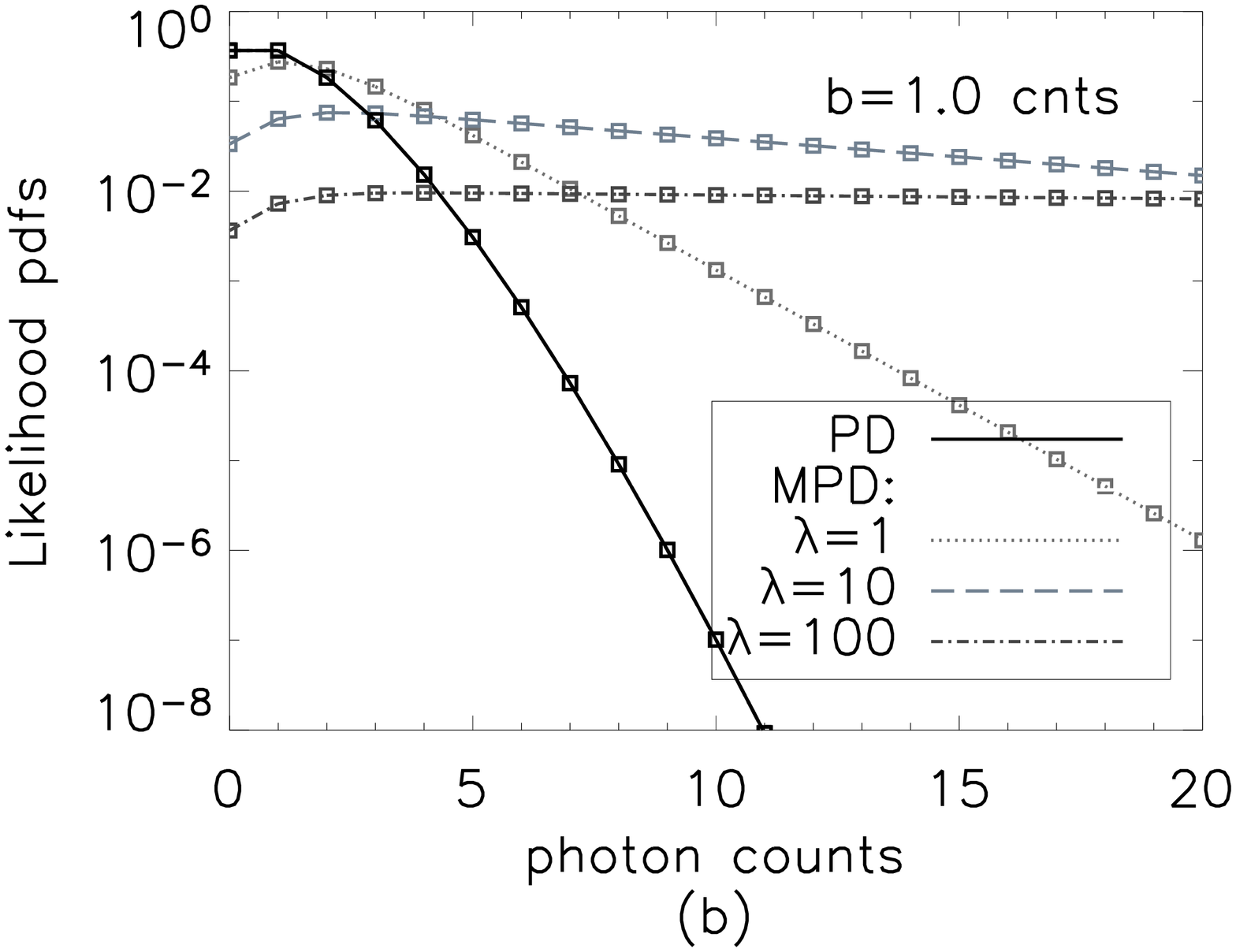}\includegraphics[width=0.245\linewidth,angle=90]{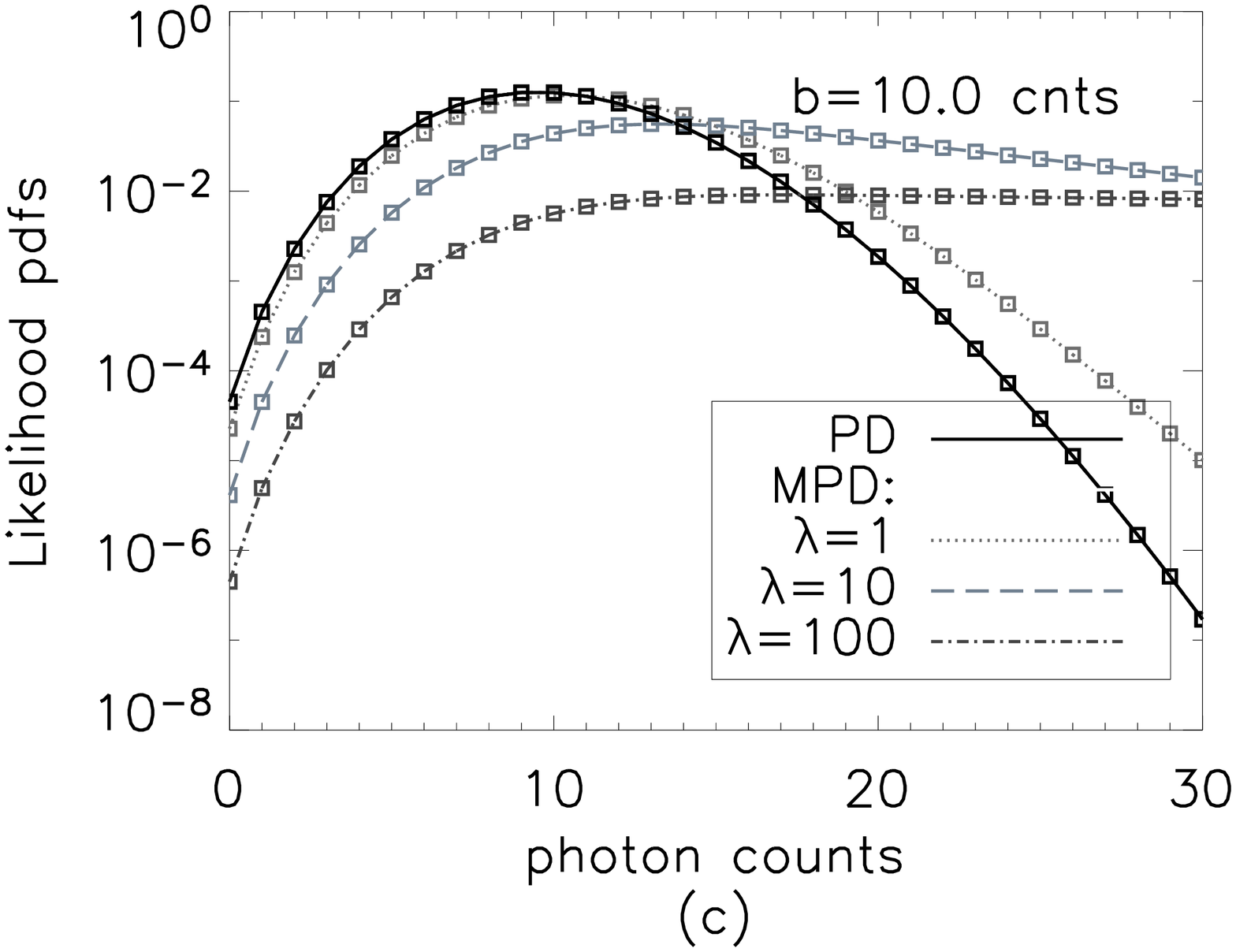}
\includegraphics[width=0.245\linewidth,angle=90]{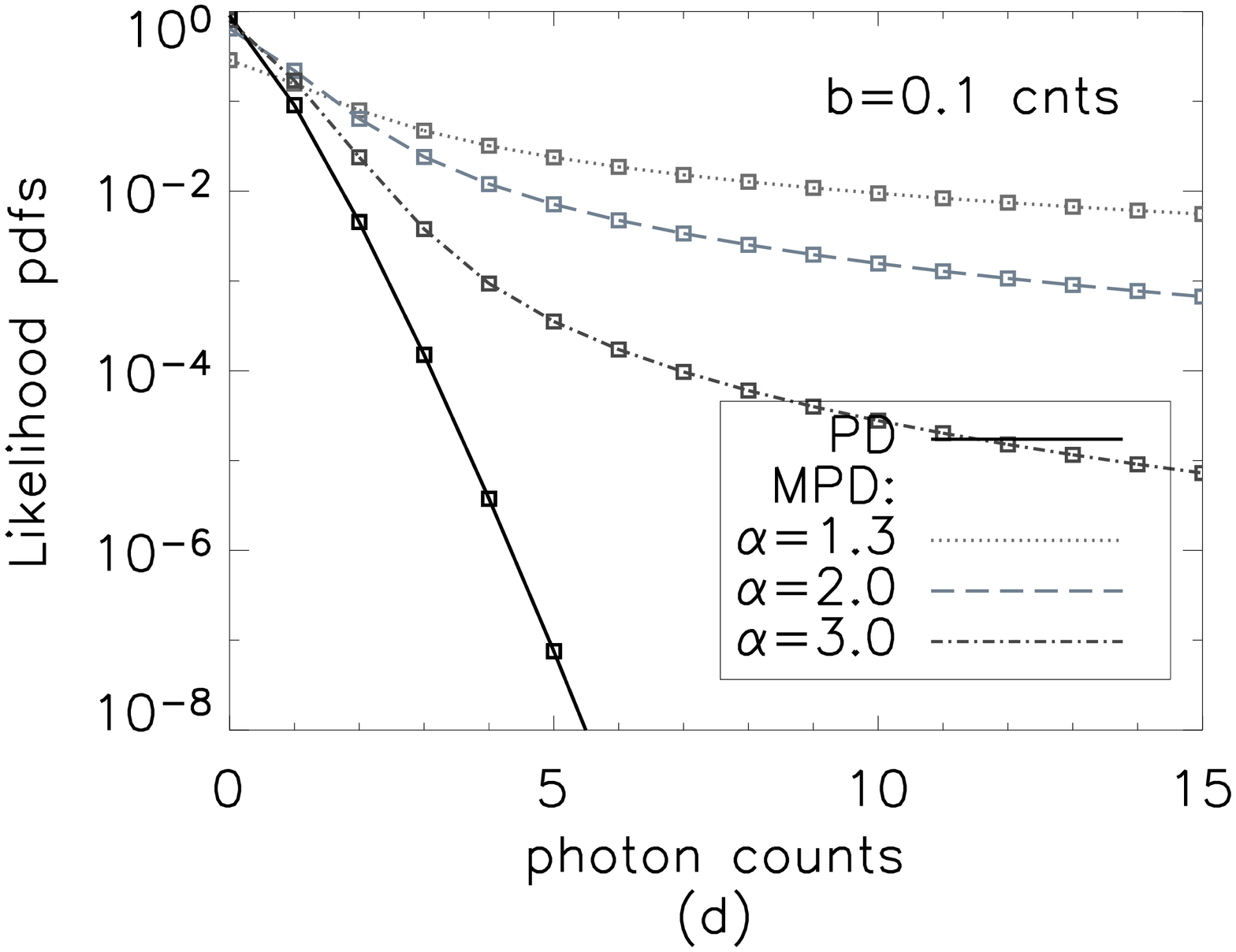}\includegraphics[width=0.245\linewidth,angle=90]{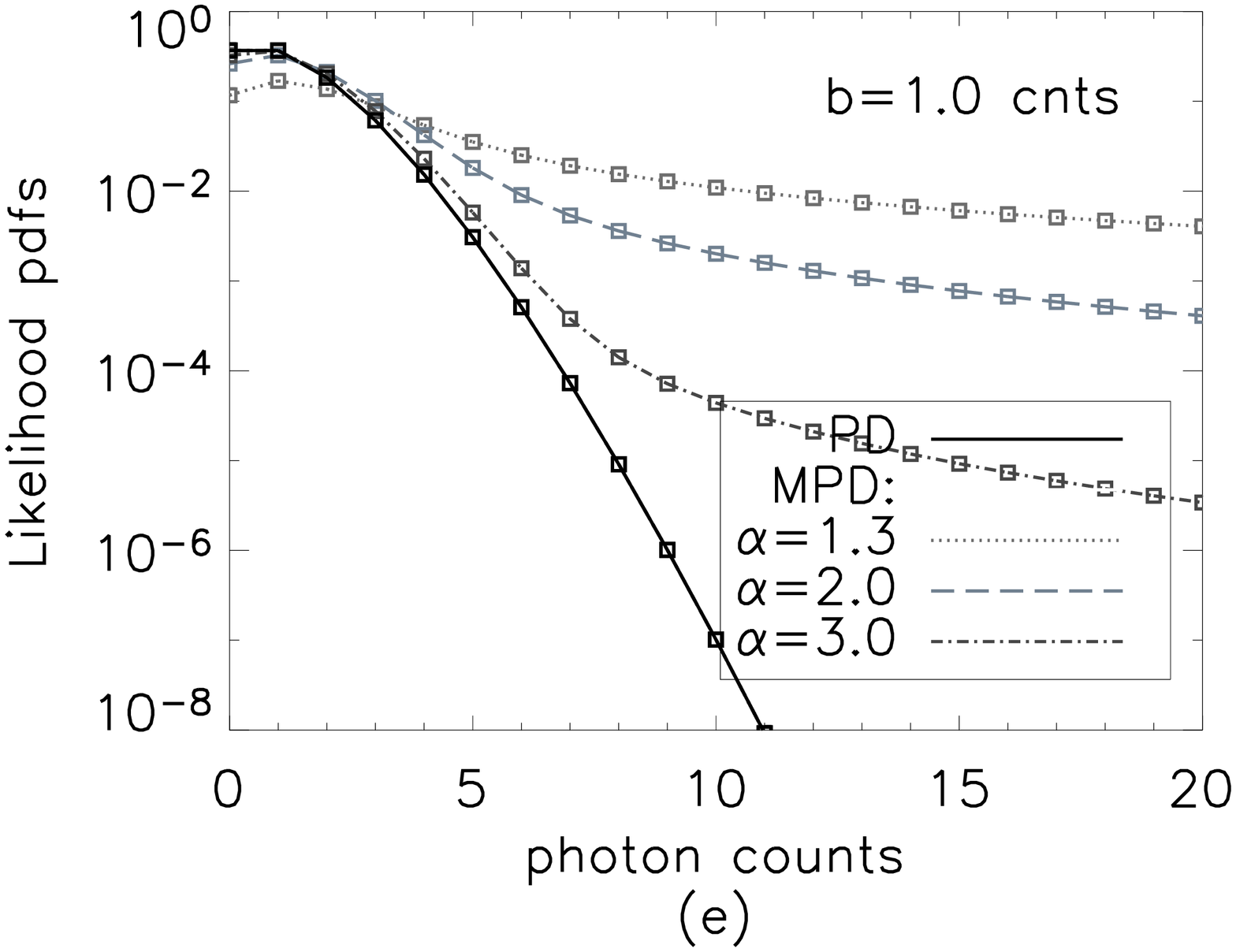}\includegraphics[width=0.245\linewidth,angle=90]{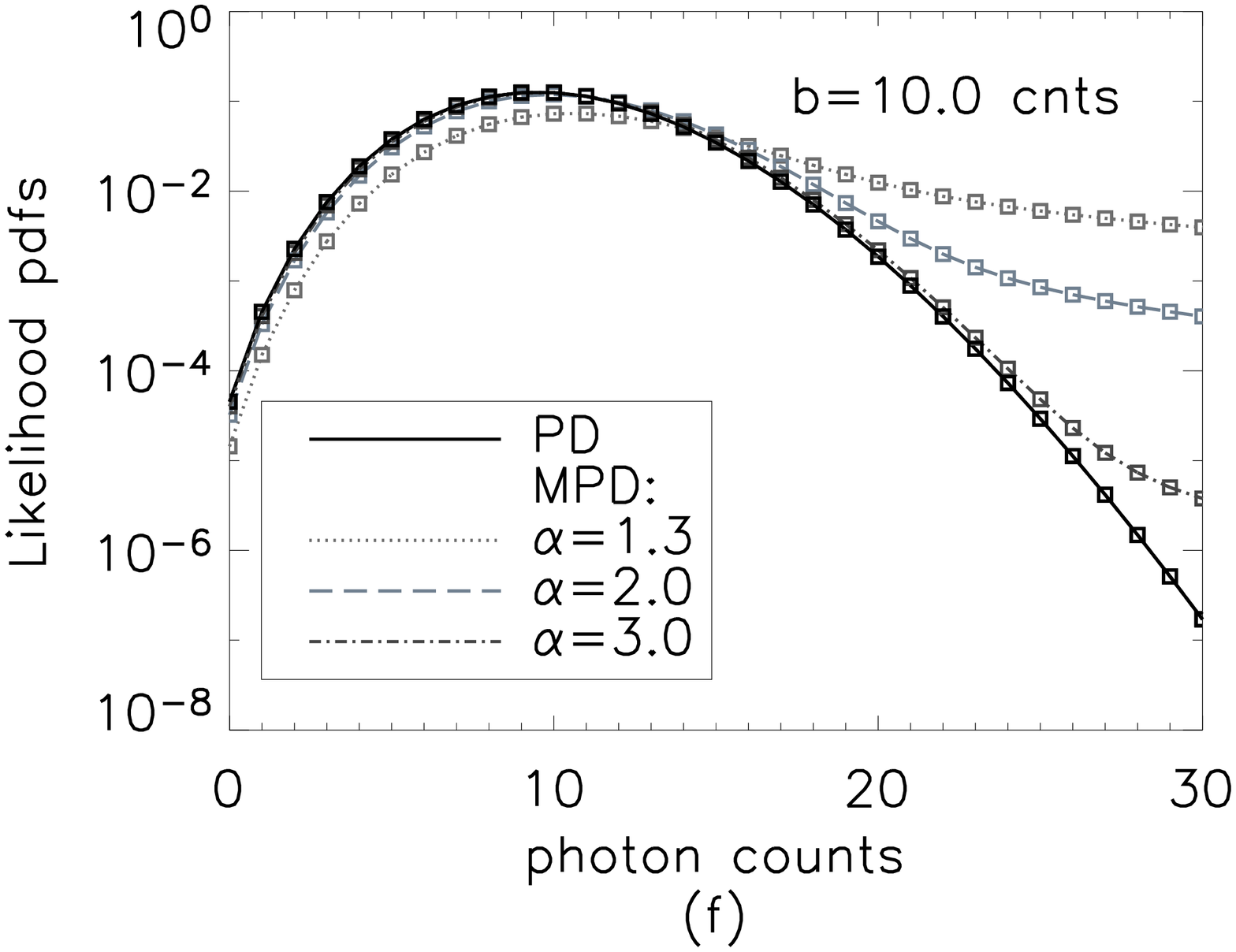}
\caption{Likelihood pdfs versus photon counts. 
  The Poisson distribution is indicated with PD. 
  The marginal Poisson distribution is indicated with MPD. 
Plots $(a)$, $(b)$ and $(c)$ are obtained using the {\em expo\-nential prior} pdf. 
Plots $(d)$, $(e)$ and $(f)$ are created using the {\em inverse--Gamma
  function prior} pdf.
}
\label{figpowp_problog}
\end{figure*}
The behaviour of the Poisson and the marginal Poisson
  distributions is depicted with a parameter study. For the parameter study three 
  background amplitudes $b$ are chosen: $0.1$, $1$ and $10$ counts. 
In Fig.~\ref{figpowp_problog} the Poisson distribution (\ref{like_bg}) 
and the marginal Poisson distribution (\ref{me_mpd}) are drawn on a 
logarithmic scale. These plots are indicated with $(a)$, $(b)$ and $(c)$, respectively. 
The parameter $\lambda$, which describes the mean intensity in a field,
has values of: $1$, $10$ and $100$ counts. 
The selected values for the background and for the parameter $\lambda$ are 
chosen such that the large variety of cases one encounters analysing
digital astronomical images are covered. For instance, $b=10$ counts and $\lambda=1$
count (plot $(c)$) corresponds to a field
when the source signal is much smaller than the background signal. 
On the other side, $b=0.1$ counts and $\lambda=100$ counts (plot $(a)$) 
corresponds to a field characterized by bright sources and small background amplitude.

The Poisson distribution is larger than the marginal Poisson distribution for
photon counts lower than or equal to the background intensity. 
Hence, hypothesis $B_{\rm ij}$ is more likely than the complementary 
hypothesis ${\overline B}_{\rm ij}$.  
This situation changes when the photon counts are larger than the
background amplitude.

The decay length of the marginal Poisson distribution is determined by the
expected source intensity $\lambda$. 
The probability to detect pixels satisfying hypothesis
${\overline B}_{\rm ij}$ is sensitive to the decay length of the marginal Poisson
distribution and to the background amplitude, that is recognizable in the 
distance between the Poisson and the marginal Poisson distributions.  
Hence, the BSS method allows probabilities to be sensitive to the parameters 
characterizing the analysed digital astronomical image.\\
Let's consider plot $(b)$ in Fig.~\ref{figpowp_problog} for photon counts
in the range ($0-10$). The background amplitude has a value of $1$ count. If 
the expected mean source intensity on the complete field has a
value of $1$ count, i.e.~$\lambda=1$ count, $3$ photon counts or more in a
pixel are classified as a source. The probability of detecting a source
increases with increasing counts in a pixel. This is due to the increasing
distance of the marginal Poisson likelihood from the Poisson likelihood. 
If an analyst allows for many bright sources distributed in the field, then the
relative number of faint sources is reduced. In fact, when a mean source
signal $100$ times larger than the background is expected, then $5$ photon 
counts or more in a pixel are needed to identify 
the event due to the presence of a source. When $\lambda=100$ counts, $5$ 
photon counts in a pixel reveal a source probability lower than the one
obtained when $\lambda=1$ count. This situation changes for $7$ or more
photon counts in a pixel.

\paragraph{Inverse--Gamma function prior} \label{IGFP}
Alternatively to the exponential prior, a power--law distribution can
be chosen to describe the prior knowledge on the source signal distribution. 

We are tempted to claim that any physical situation contains sensible 
information for a proper (normalizable) prior. 
We choose a prior pdf that is inspired by the cumulative signal number
counts distribution used often in astrophysics for describing the integral
flux distribution of cosmological sources, i.e.~a power--law function 
(see refs.~\citealt{rosati:2002ARA&A}; \citealt{brandt:2005araa} and references 
therein). The power law can not be employed as a prior pdf, because the power law is not 
normalized. We use instead a normalized inverse--Gamma function. It behaves at 
large counts as the power law, because it is described by a power law with an 
exponential cutoff. 

The prior pdf of the source signal, described by an inverse--Gamma
function, is:
\begin{equation}
p(s_{\rm ij} \mid \alpha, a) = e^{-\frac{a}{s_{\rm ij}}} {\bf \cdot} s_{\rm ij}^{-\alpha} {\bf \cdot}
\frac{a^{\alpha-1}}{\Gamma(\alpha-1)}; \hspace{.2cm} \alpha > 1; \hspace{.2cm}
a > 0,
\label{ps_igf}
\end{equation}
with slope $\alpha$ and cutoff parameter $a$. When $a$ has a small positive value, the 
inverse--Gamma function is dominated by a power--law behaviour. 

The parameter $a$ gives rise to a cutoff of faint sources. 
This parameter has important implications in the estimation of the
background. If $a$ is smaller than the
background amplitude, the BSS algorithm detects sources of the order of the
background. If $a$ is larger than the background amplitude, the BSS algorithm
assigns faint sources with intensities lower than $a$ to be background only.

In Fig.~\ref{priors_fig}, equation (\ref{ps_igf}) is drawn for 
two values of the parameter $\alpha$. For this example the cutoff parameter
$a$ has a value of $\sim 0.1$ count.
The distributions peak around $a$. 
The decay of each distribution depends on the value of $\alpha$. 
When $\alpha$ is large, i.e.~for values $\ge 2$, the distribution
drops quickly to zero. Instead the distribution drops slowly to zero, 
when $\alpha$ approaches $1$. Hence, small values of $\alpha$
indicate bright sources distributed on the field. 

The \emph{marginal Poisson likelihood} for the hypothesis $\overline{B}_{\rm ij}$
is now described by:
\begin{eqnarray} 
\nonumber
p(d_{\rm ij}|\overline{B}_{\rm ij},b_{\rm ij},\alpha,a)=\frac{2}{\Gamma(\alpha-1)} {\bf \cdot}
e^{-b_{\rm ij}} {\bf \cdot} \sum_{k=0}^{d_{\rm ij}} a^{\frac{k+\alpha-1}{2}} 
\\
\hspace{-5.5cm}
{\bf \times}
\frac{b_{\rm ij}^{d_{\rm ij}-k}}{\Gamma(k+1) \Gamma(d_{\rm ij}-k+1)}K_{k-\alpha+1}(2\sqrt{a}), 
\label{igf_mpd}
\end{eqnarray}
where $\alpha > 1$, $a > 0$ and $K_{k-\alpha+1}(2\sqrt{a})$ is the Modified
Bessel function of order $k-\alpha+1$. \\
In order to avoid numerical problems with the Bessel function, the following 
upward recurrence relation was derived:
\begin{eqnarray}
\nonumber
\widetilde{K}_{\nu}(z) \equiv
\frac{\widetilde{K}_{\nu-2}(z)}{(k-1)k}+\frac{2(\nu-1)}{z {\bf \cdot} k}\widetilde{K}_{\nu-1}(z)
\end{eqnarray}
where $\widetilde{K}_{\nu}(z):=K_{\nu}(z)/\Gamma(k+1)$ and
$\nu=k-\alpha+1$. 
$K_{\nu}(z)$ is the Bessel function of imaginary argument and it has the 
property: $K_{-\nu}(z)=K_{\nu}(z)$.

Fig.~\ref{figpowp_problog}, $(d)$--$(f)$, shows the corresponding
  parameter study for the inverse--Gamma function prior. 
The parameter $\alpha$ is chosen to be $1.3$, $2.0$ and $3.0$ and the 
cutoff parameter $a$ $\sim 0.1$ counts. 
The decay length of the marginal Poisson distributions (\ref{igf_mpd}) are now 
given by the value of $\alpha$. The decay length decreases with increasing $\alpha$ values.

\subsubsection{The likelihood for the mixture model} \label{LMM}
\begin{figure}
\includegraphics[width=0.6\linewidth,angle=90]{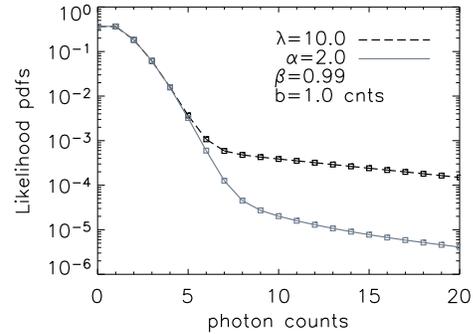}
\caption{Likelihood pdfs for the mixture model using the {\em exponential prior}
  (dashed line) and the {\em inverse--Gamma function prior} (continuous line). The
  Poisson and the marginal Poisson likelihood pdfs are weighted with
  $\beta$ and $(1-\beta)$, respectively.  
}
\label{LMM_exp}
\end{figure}

The marginal Poisson likelihood pdf will be indicated with
$p(d_{\rm ij}|\overline{B}_{\rm ij},b_{\rm ij},\gamma)$, where $\gamma$ indicates 
$\lambda$ or $\alpha$. In the case of the inverse--Gamma function prior pdf, the
cutoff parameter $a$ does not appear since the value of this parameter is
chosen such that the inverse--Gamma function is dominated by a power--law
behaviour.

The likelihood for the mixture model, as written in (\ref{mix_mod}), 
now reads:
\begin{eqnarray}
\nonumber
p(D \mid b,\gamma,\beta) = 
\prod_{\rm ij} [\beta {\bf \cdot} p(d_{\rm ij} \mid B_{\rm ij},b_{\rm ij}) + (1-\beta)\\
{\bf \times} p(d_{\rm ij} \mid \overline{B}_{\rm ij},b_{\rm ij},\gamma)]; \hspace{0.2cm}
D=\{d_{\rm ij}\}, \hspace{0.2cm} b=\{b_{\rm ij}\}.
\label{LMM_app}
\end{eqnarray} 
In Fig.~\ref{LMM_exp} we show the effect of the likelihood for the mixture
model on the data (semi--log plot). The likelihood pdf for the mixture model
is drawn for each prior pdf of the source signal employing the background 
value $b$ and the prior pdf $\beta$. The value chosen for the parameter 
$\beta$ indicates that $99$ per cent of the pixels distributed in the astronomical
image are containing only background. 
The likelihood pdfs are composed by a central peak plus a long tail. 
The central peak is primarily due to the presence of the Poisson
distribution. The long tail is caused by the marginal Poisson distribution. 
The presence of the long tail is essential in order to reduce the influence 
of the source signal for background estimation 
(see \citealt{fischer:2001} for more details). 

\subsection{Thin--plate spline} \label{TPS_label}

The TPS has been selected for modelling the structures 
arising in the background rate of a digital astronomical image. It is a type
of radial basis function. \\
The TPS is indicated with $t({\bf x})$, where ${\bf x}=(x,y)$ 
corresponds to the position on the grid in the detector field. The shape of 
the interpolating TPS surface fulfills a minimum curvature condition on
infinite support.

\begin{figure}
\centering
\includegraphics[width=0.9\linewidth]{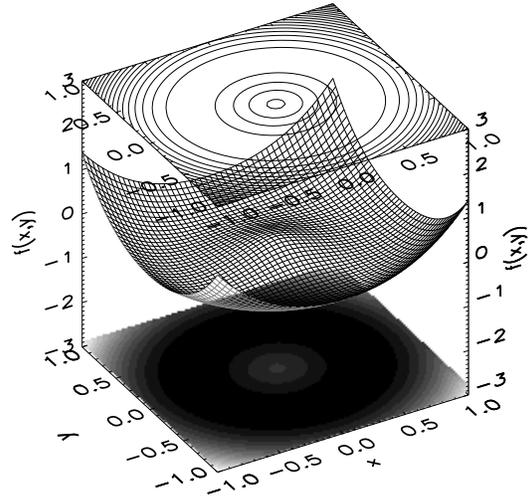} 
\caption{Example of thin--plate spline. 
$f({\bf x} - {\bf x} _{\rm l})= r^{2} {\rm ln}(r^{2})$ is drawn with one
  support point centered at the Cartesian origin. The projection of $f({\bf x}
  - {\bf x} _{\rm l})$ on the $(x,y)$ plane is placed below and
  its contour plot on top of the surface image.
}
\label{figRBF}
\end{figure}

More specifically, the TPS is a weighted sum of translations of radially 
symmetric basis functions augmented by a linear term 
(see \citealt{meinguet:1979} and ~\citealt{wahba:2006}), of the form
\begin{eqnarray}
\nonumber
t({\bf x}) =  E({\bf x}) + \sum_{l=1}^{N_{\rm r}} \lambda_{\rm l}
f({\bf x} - {\bf x}_{\rm l}), \hspace{0.2cm} 
 {\bf x} \in \mathnormal{R}^{2}. 
\end{eqnarray}
$E({\bf x})=c_{0}+c_{1}x+c_{2}y$ is the added plane. 
$N_{\rm r}$ is the number of support points (pivots).   
The weight is characterized by $\lambda_{\rm l}$. 
$f({\bf x} - {\bf x}_{\rm l})$ is a basis function, a function of 
real values depending on the distance between the grid points ${\bf x}$ 
and the support points ${\bf x}_{\rm l}$, such that $|{\bf x} - {\bf x}_{\rm l}| > 0$. \\
Given the pivots ${\bf x}_{\rm i}$ and the amplitude 
$z_{\rm i} = z({\bf x}_{\rm i})$, the TPS satisfies the interpolation conditions:
\begin{eqnarray}
\nonumber
t({\bf x}_{\rm i}) =  E({\bf x}_{\rm i}) + \sum_{l=1}^{N_{\rm r}} \lambda_{\rm l}
f({\bf x}_{\rm i} - {\bf x}_{\rm l}) = z_{\rm i}, \hspace{0.2cm} 
i=1,\dots,N
\end{eqnarray} 
and minimizes
\begin{eqnarray}
\nonumber
\Vert t \Vert^{2}=I[f(x,y)] = \int \int_{\mathrm{R}^{2}} (f_{\rm xx}^{2} + 2
f_{\rm xy}^{2} + f_{\rm yy}^{2}) dx dy.
\end{eqnarray} 
$\Vert t \Vert^{2}$ is a measure of energy in the second derivatives of
$t$. In other words, given a set of data points, a weighted combination 
of TPSs centered about each pivot gives the interpolation function that 
passes through the pivots exactly while minimizing the so--called 
`bending energy'. 
The TPS satisfies the Euler--Lagrange equation and its solution has the form:
$f({\bf x} - {\bf x}_{\rm l}) \simeq r^{2} {\rm ln}(r^{2})$, where $r^{2} = (x-x_{\rm l})^{2} + (y-y_{\rm l})^{2}$.   
This is a smooth function of two variables defined via Euclidean space 
distance. In Fig.~\ref{figRBF} an example of TPS with one pivot is pictured. \\
In order to fit the TPS to the data, it is necessary to solve for the weights 
and the plane's coefficients so that it is possible to interpolate the local TPS's
amplitude:
\begin{eqnarray}
\nonumber
t_{\rm ij}=t_{\rm ij}(N_{\rm r},\{{\bf x}_{\rm l},z_{\rm l},l=1,\ldots,N_{\rm r} \})
\end{eqnarray} 
which is the background rate. $b_{\rm ij}$ will
indicate the local background amplitude, i.e.~the multiplication of 
${t}_{\rm ij}$ and the local value of the telescope's exposure time ($\tau$): 
\begin{eqnarray}
\nonumber
b_{\rm ij}=t_{\rm ij} \times \mathrm {\tau}_{\rm ij}. 
\end{eqnarray} 
The TPS interpolant is defined by the coefficients, $c_{\rm i}$ of the plane 
$E({\bf x})$ and the weights $\lambda_{\rm l}$ of the basis functions.
The solution for the TPS has been evaluated on an infinite support, since no
solutions exist on a finite support, where the 
requirements for this function to be fulfilled are:
\begin{enumerate}
\item
$t({\bf x})$ is two times continuously differentiable,
\item
$t({\bf x})$ takes a particular value $z_{\rm i}$ at the point ${\bf x}_{\rm i}$, 
\item
$I[f(x,y)]$ is finite.
\end{enumerate}
Given the interpolation values ${\bf z}=(z_{1},\ldots,z_{N_{\rm r}})$, the weights 
$\lambda_{\rm l}$ and $c_{\rm i}$ are searched so that the TPS satisfies:
\begin{eqnarray}
\nonumber
t({\bf x}_{\rm l}) = z_{\rm l}, \hspace{0.2cm} l=1,\ldots,N_{\rm r} 
\end{eqnarray} 
and in order to have a converging integral, the following conditions need to 
be satisfied:
\begin{displaymath}
\left\{\begin{array}{ll}   
\sum_{l=1}^{N_{\rm r}} \lambda_{\rm l} &\hspace{-0.2cm} =0  \\
\sum_{l=1}^{N_{\rm r}} \lambda_{\rm l}x_{\rm l} &\hspace{-0.2cm} =0 \\
\sum_{l=1}^{N_{\rm r}} \lambda_{\rm l}y_{\rm l} &\hspace{-0.2cm} =0 .
\end{array} \right.
\end{displaymath} 
This means that we have $(N_{\rm r}-3)$ conditions on $t({\bf x}_{\rm l})=z_{\rm l}$. \\
The coefficients of the TPS, $\lambda_{\rm l}$, and the plane, $c_{\rm i}$, can be 
found by solving the linear system, that may be written in matrix form as:
\begin{eqnarray}
\nonumber
\left( \begin{array}{ll}
{\sf {\bf F}} & {\sf {\bf Q}} \\
{\sf {\bf Q}}^{T} & {\sf {\bf \cal O}}
\end{array} \right)
\left( \begin{array}{c}
{\it {\bf \lambda}} \\
{\it {\bf c}} \\
\end{array} \right) =
\left( \begin{array}{c}
{\it {\bf z}} \\
{\it {\bf 0}} \\
\end{array} \right)
\end{eqnarray}
where the matrix components are:\\
\hspace*{\fill}
\begin{minipage}{4cm}
 \begin{displaymath}
  \begin{array}{lll}
 {\it {\bf F_{\rm ij}}}&\hspace{-0.2cm} =&\hspace{-0.2cm} f({\bf x}_{\rm i}-{\bf x}_{j}) \\
\\
  {\it {\bf  z}}&\hspace{-0.2cm} =& \hspace{-0.2cm} (z_{1},\ldots,z_{N_{\rm r}})^{T} \\
\\
  {\it {\bf  0}}&\hspace{-0.2cm} =& \hspace{-0.2cm} (0,0,0)^{T} \\
\\
  {\it {\bf \lambda}} &\hspace{-0.2cm} =& \hspace{-0.2cm} (\lambda_{1},\ldots,\lambda_{N_{\rm r}})^{T} \\
\\
  {\it {\bf  c}} &\hspace{-0.2cm} =& \hspace{-0.2cm} (c_{0},c_{1},c_{2})^{T}. \\
  \end{array}
 \end{displaymath}
\end{minipage}
\vspace*{-3.cm}\\
\begin{minipage}{4cm}
 \begin{displaymath}
  {\sf {\bf Q}}=
  \left( \begin{array}{ccc}
  1 & x_{1} & y_{1} \\
  1 & x_{2} & y_{2} \\
  \vdots & \vdots & \vdots \\
  1 & x_{N_{\rm r}} & y_{N_{\rm r}}
  \end{array} \right)
 \end{displaymath}
\end{minipage}
\vspace{1.cm}
\\
\\
\\
After having solved $({\it {\bf \lambda,c}})^{T}$, the TPS can be evaluated at any point. 

The pivots can be equally spaced or can be located in
structures arising in the astronomical image. Following the works of
\cite{fischer:2001} and \cite{udo:2007}, the present work can be extended 
employing adaptive splines, i.e.~allowing the
number of pivots and their positions to vary in accordance with the
requirements of the data.

\subsection{Estimation of the background and its uncertainties} \label{sec:errbkg}

The posterior pdf of the background is, according to Bayes' theorem, proportional to the
product of the mixture likelihood pdf, eq.~(\ref{LMM_app}), and the prior
pdf $p(b)$, that is chosen
constant for positive values of $b$ and null elsewhere.
Its maximum with respect to $b$, $b^{\ast}$, gives an estimate of the
background map which consists of the TPS combined with the observatory's 
exposure map. The estimation of the background considers all pixels,
i.e. on the complete field, because we tackle the source signal implicitly as outlier. 

The posterior pdf for $b$ is given by:
\begin{eqnarray}
\nonumber
p(b \mid D) = \int p(b \mid D,z) p(z \mid D) d^{N_{\rm r}}z= \\
\nonumber
= \int \delta(b-b(z)) \frac{p(D \mid z) p(z)}{p(D)}d^{N_{\rm r}}z . 
\end{eqnarray}
This integral is complicated due to the presence of the delta function. This
is, however, of minor importance since our interest focuses on the expectation
values of some functionals of $b$, say $g(b)$. Therefore:
\begin{eqnarray}
\nonumber
E(g(b) \mid D) = \int g(b) p(b \mid D) db =  \hspace{1.5cm} \\
\nonumber
= \int g(b(z)) \frac{p(D \mid z) p(z)}{p(D)} d^{N_{\rm r}}z =  \\
= \frac{\int g(b(z)) p(D \mid z) p(z) d^{N_{\rm r}}z}{\int p(D \mid z) p(z) d^{N_{\rm r}}z}.
\label{expect}
\end{eqnarray}
Assuming the maximum of $p(D \mid z) p(z)$ is well determined we can apply the
Laplace approximation:
\begin{equation}
p(D \mid z) p(z) \simeq p(D \mid z^{\ast}) p(z^{\ast}) {\rm exp} \{ -\frac{1}{2}
\Delta{\bf z}^{T} H \Delta{\bf z} \}, 
\label{lapap}
\end{equation}
 that means we approximate the integral function by a Gaussian at the
 maximum $z^{\ast}$ and we compute the volume under
 that Gaussian. The covariance of the fitted Gaussian is determined by the
 Hessian matrix, as given by (\ref{lapap}), 
where $\Delta{\bf z} = {\bf z} - {\bf z}^{\ast}$ and 
$H_{\rm ij} := -\frac{\partial^{2}{\rm ln}[p(D \mid z)p(z)]}{\partial z_{\rm i}
  \partial z_{\rm j}}$ is element $\{ij\}$ of the Hessian matrix.
This approximation is the $2^{nd}$ order
Taylor expansion of the posterior pdf around the optimized pivots amplitude's
values. For more details see \cite{ohagan:2004}.\\ 
Then equation (\ref{expect}) becomes:
\begin{eqnarray}
\nonumber
E(g(b) \mid D) = \frac{\int g(b(z)) {\rm exp}\{ -\frac{1}{2} \Delta{\bf z}^{T} H
  \Delta{\bf z} \} d^{N_{\rm r}}z}{(2\pi)^{\frac{N_{\rm r}}{2}} ({\rm det} H)^{-\frac{1}{2}}}.
\end{eqnarray} 
Therefore, the posterior mean of $b$ is:
\begin{eqnarray}
\nonumber
E(b_{\rm ij} \mid D) = \frac{\sqrt{{\rm det} H}}{(2\pi)^{\frac{N_{\rm r}}{2}}} \int b_{\rm ij}(z)
{\rm exp}\{ -\frac{1}{2} \Delta{\bf z}^{T} H \Delta{\bf z} \} d^{N_{\rm r}}z = \hspace{-3.95cm}\\
= {\bf T}_{\rm ij}^{T} {\bf z}^{\ast} = <b_{\rm ij}>, 
\nonumber
\end{eqnarray}
where $b_{\rm ij}(z)={\bf T}_{\rm ij}^{T}\cdot{\bf z}$, and the variance is:
\begin{eqnarray}
\nonumber
E(\Delta b_{\rm ij} \Delta b_{\rm lk} \mid D) =  
\frac{\sqrt{{\rm det} H}}{(2\pi)^{\frac{N_{\rm r}}{2}}} \int
(b_{\rm ij}(z)-<b_{\rm ij}>) \\
\nonumber
\times (b_{\rm lk}(z)-<b_{\rm lk}>){\rm exp}\{ -\frac{1}{2} \Delta{\bf z}^{T} H
\Delta{\bf z} \} d^{N_{\rm r}}z=\\
={\bf T}_{\rm ij}^{T} H^{-1} {\bf T}_{\rm lk}. 
\label{error}
\end{eqnarray}
The $1\sigma$ error on the estimated background function is therefore
calculated by the square root of equation (\ref{error}).

\subsection{Determining the hyper--parameters} \label{Hyperpar}

The two hyper--parameters $\gamma$ and $\beta$ have so far been assumed to
be fixed. However, these parameters can be appropriately estimated from the
data. 

Within the framework of BPT the hyper--parameters (nuisance parameters) $\gamma$ and $\beta$ 
have to be marginalized.\\
Alternatively, and not quite rigorous in the Bayesian sense, the 
hyper--parameters can be estimated from the marginal posterior pdf, 
where the background and source parameters are integrated out,
\begin{eqnarray}
\max_{\beta,\gamma} p(\beta,\gamma \mid D) \to \beta^{\ast},\gamma^{\ast} .
\nonumber
\end{eqnarray}
Hence, the estimate of the hyper--parameters is the maximum of their joint posterior.

The basic idea is to use BPT to determine the hyper--parameters explicitly,
i.e.~from the data. This requires 
the posterior pdf of $\gamma$ and $\beta$. Bayes' theorem gives:
\begin{equation}
p(\gamma,\beta \mid D) = \frac{p(D \mid \gamma,\beta)p(\gamma)p(\beta)}{p(D)}.
\label{btl}
\end{equation}
The prior pdfs of the hyper--parameters are independent because the
hyper--parameters are logically independent. 
These prior pdfs are chosen uninformative, because of our lack of knowledge on these
parameters. 
The prior pdf for $\beta$ is chosen to be constant in $[0,1]$. Since $\gamma$
is a scale parameter, the appropriate prior distribution is Jeffrey's prior: $p(\gamma) \sim 1/\ \gamma$.
Equation (\ref{btl}) can be written as follows: 
\begin{eqnarray}
\nonumber
p(\gamma,\beta \mid D) \propto p(D \mid \gamma,\beta)p(\gamma) = 
\hspace{2.cm} \\
= p(\gamma) \int p(D \mid z,\gamma,\beta)p(z)dz. 
\label{btl1}
\end{eqnarray}
Assuming the maximum of $p(D \mid z,\gamma, \beta)p(z)$ is well determined, we can apply the Laplace 
approximation
\begin{eqnarray}
\nonumber
p(D \mid z,\gamma,\beta)p(z) \simeq p(D \mid z^{\ast},\gamma,\beta) 
p(z^{\ast}) {\rm exp}\{-\frac{1}{2} \Delta {\bf z}^{T} H \Delta {\bf z} \} 
\end{eqnarray}
where $\Delta {\bf z} = {\bf z}-{\bf z}^{\ast}$, ${\bf z}^{\ast}$ corresponds
to the maximum value of the integrand in (\ref{btl1}), and 
$H_{\rm ij} := -\frac{\partial^{2}{\rm ln}[p(D \mid z,\gamma, \beta)p(z)]}{\partial z_{\rm i}
  \partial z_{\rm j}}$  is element $\{ij\}$ of the Hessian matrix. Considering
$\mathrm{dim}({\bf z})=N_{\rm r}$, where $N_{\rm r}$ is the number of support points,
equation (\ref{btl1}) can be written as follows:
\begin{eqnarray}
p(\gamma, \beta \mid D) 
= p(\gamma) p(D \mid z^{\ast},\gamma, \beta) p(z^{\ast})
\frac{(2\pi)^{\frac{N_{\rm r}}{2}}}{(\mathrm{{\rm det}} H)^{\frac{1}{2}}}. 
\nonumber
\end{eqnarray}
$p(z^{\ast})$ is chosen to be constant. The last term corresponds 
to the volume of the posterior pdf of $\gamma$ and $\beta$ for each $\gamma$,
$\beta$ estimates.

\subsection{Probability of hypothesis $\overline{B}$} \label{probbar}
\begin{figure*}
\includegraphics[width=0.245\linewidth,angle=90]{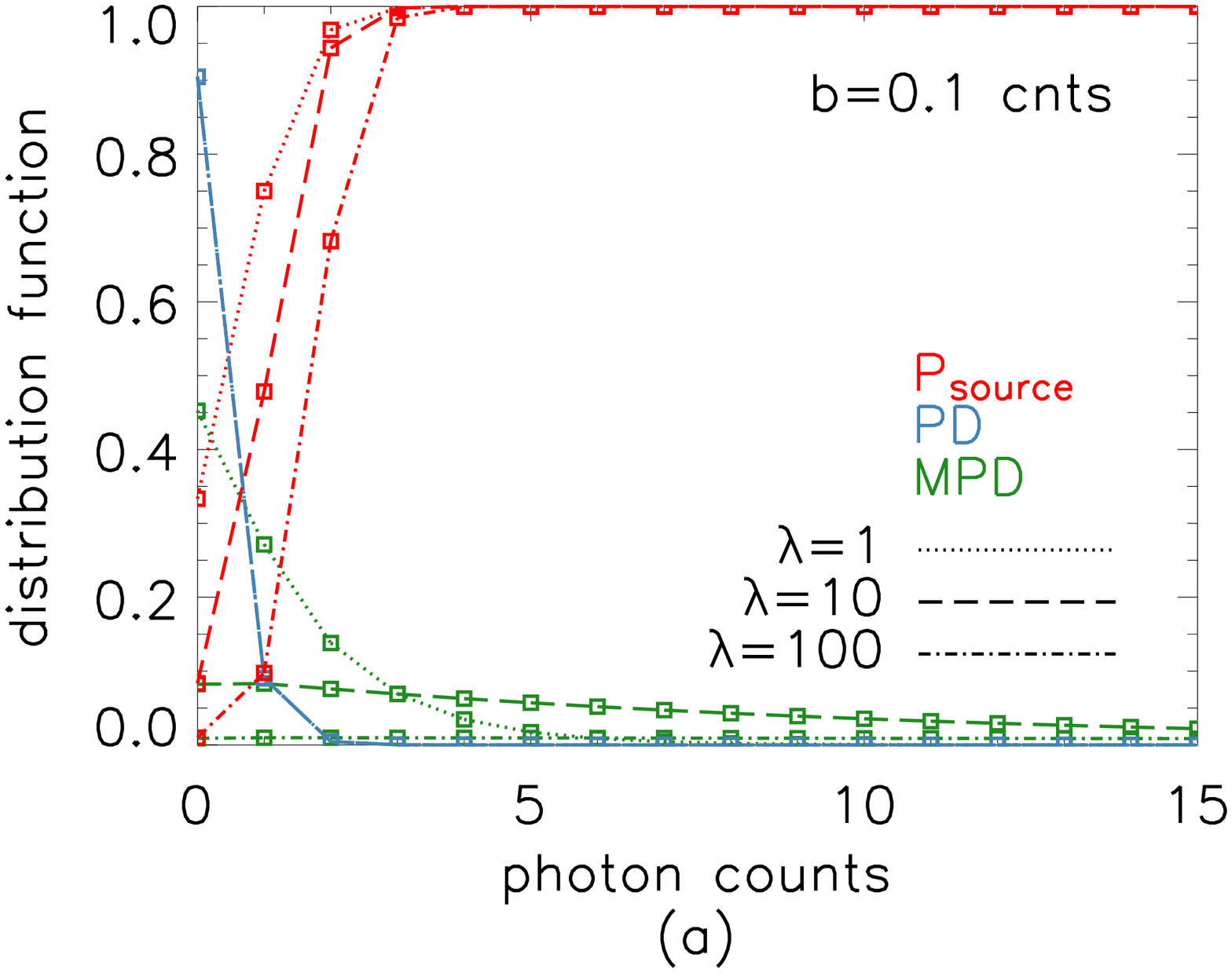}\includegraphics[width=0.245\linewidth,angle=90]{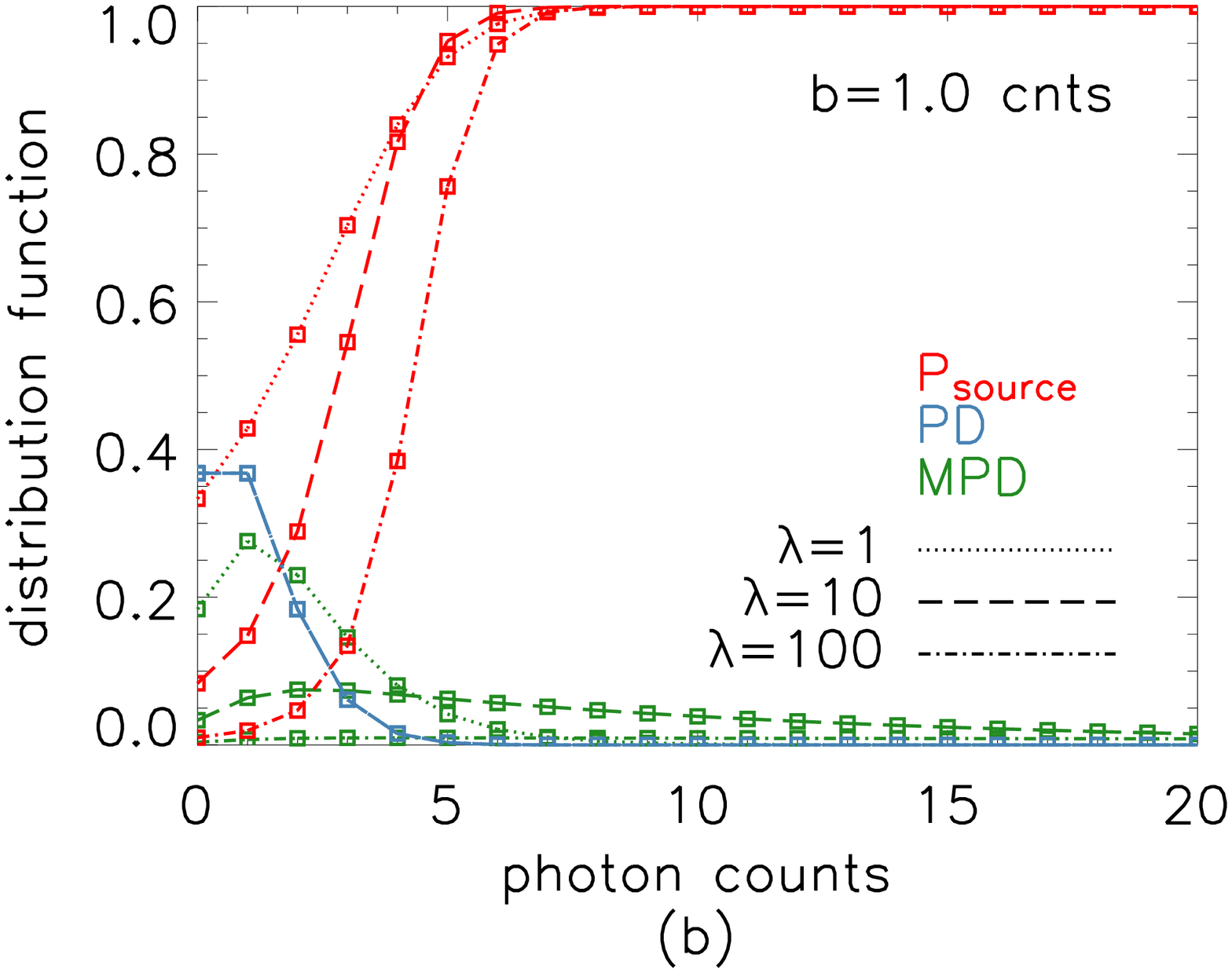}\includegraphics[width=0.245\linewidth,angle=90]{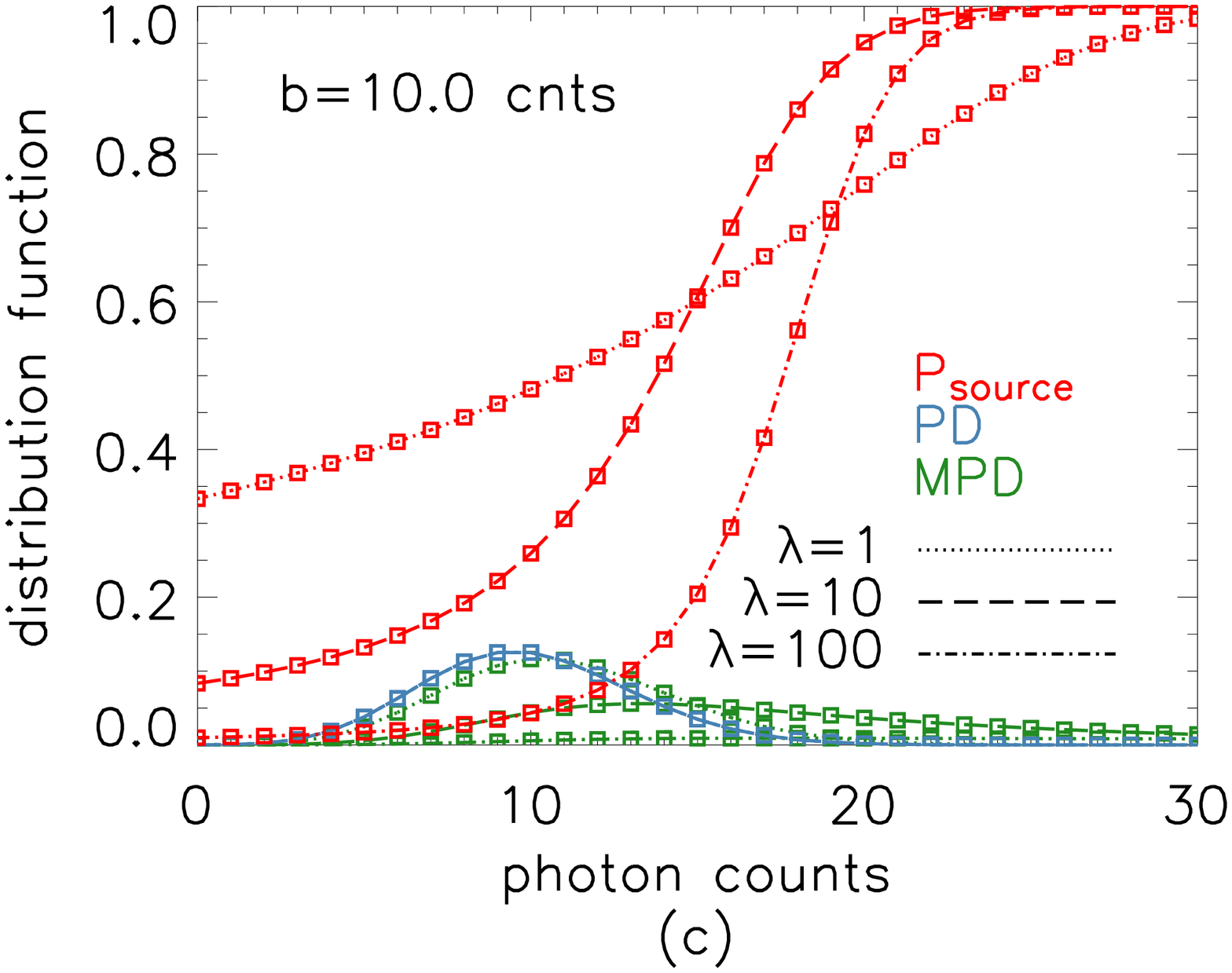}
\includegraphics[width=0.245\linewidth,angle=90]{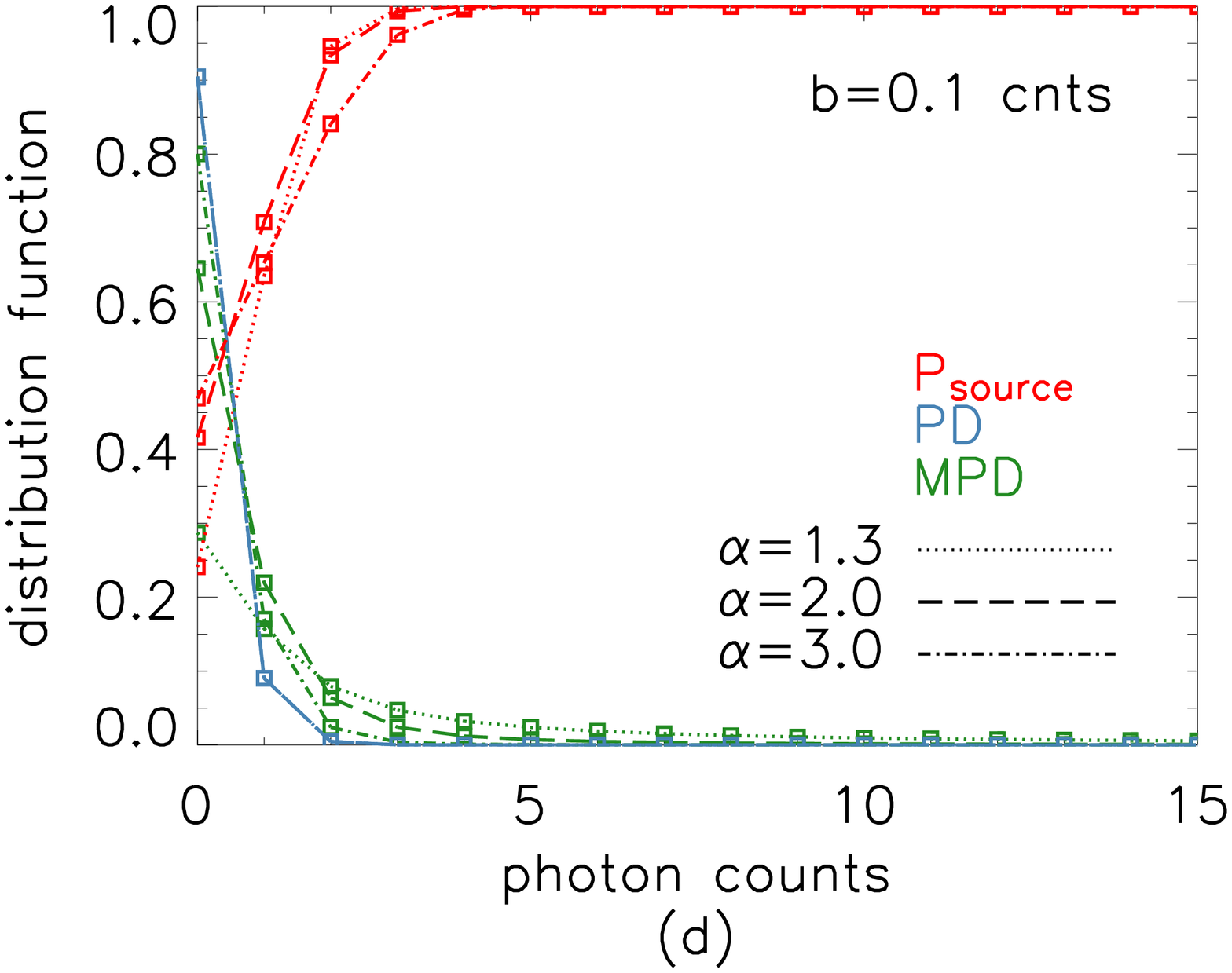}\includegraphics[width=0.245\linewidth,angle=90]{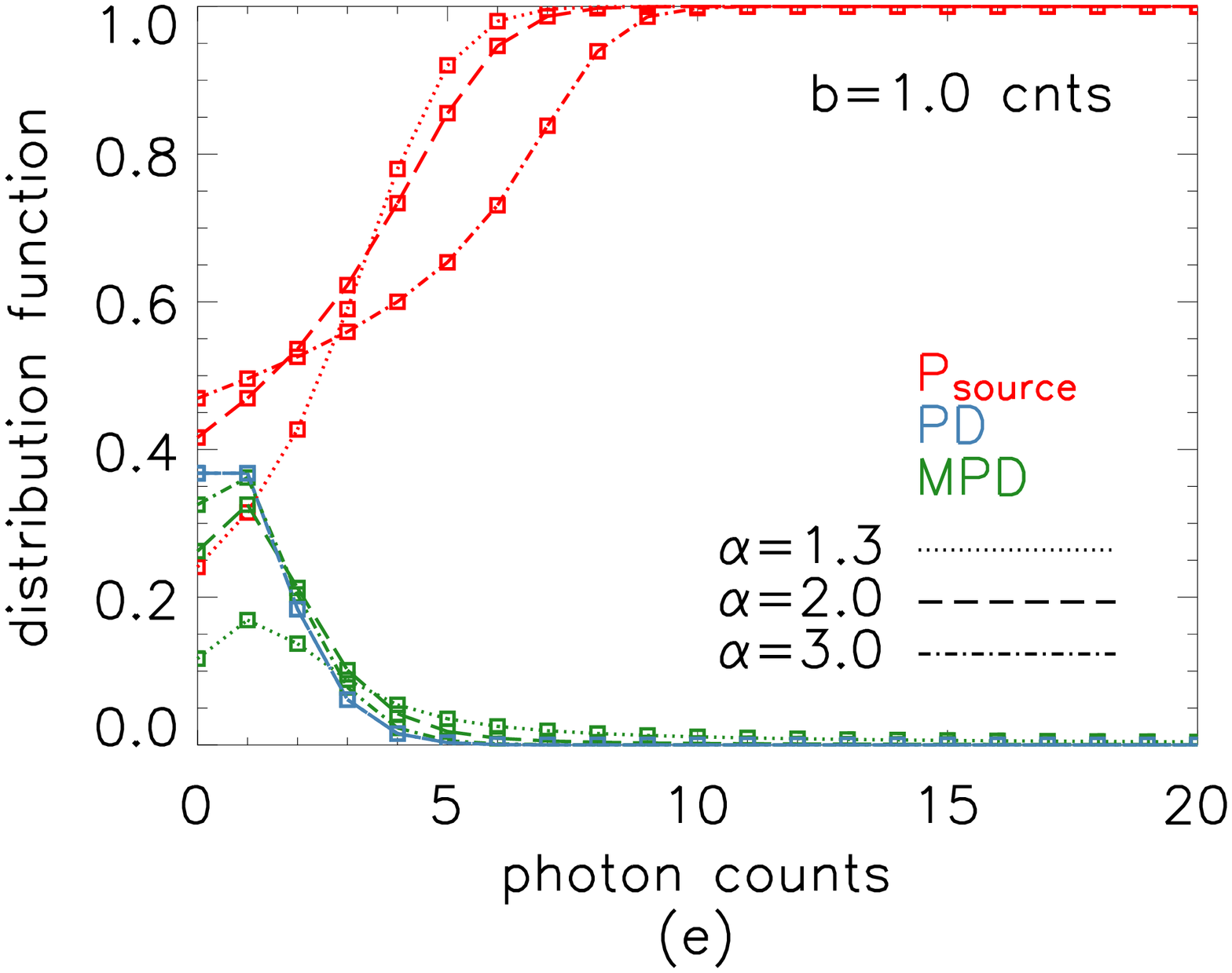}\includegraphics[width=0.245\linewidth,angle=90]{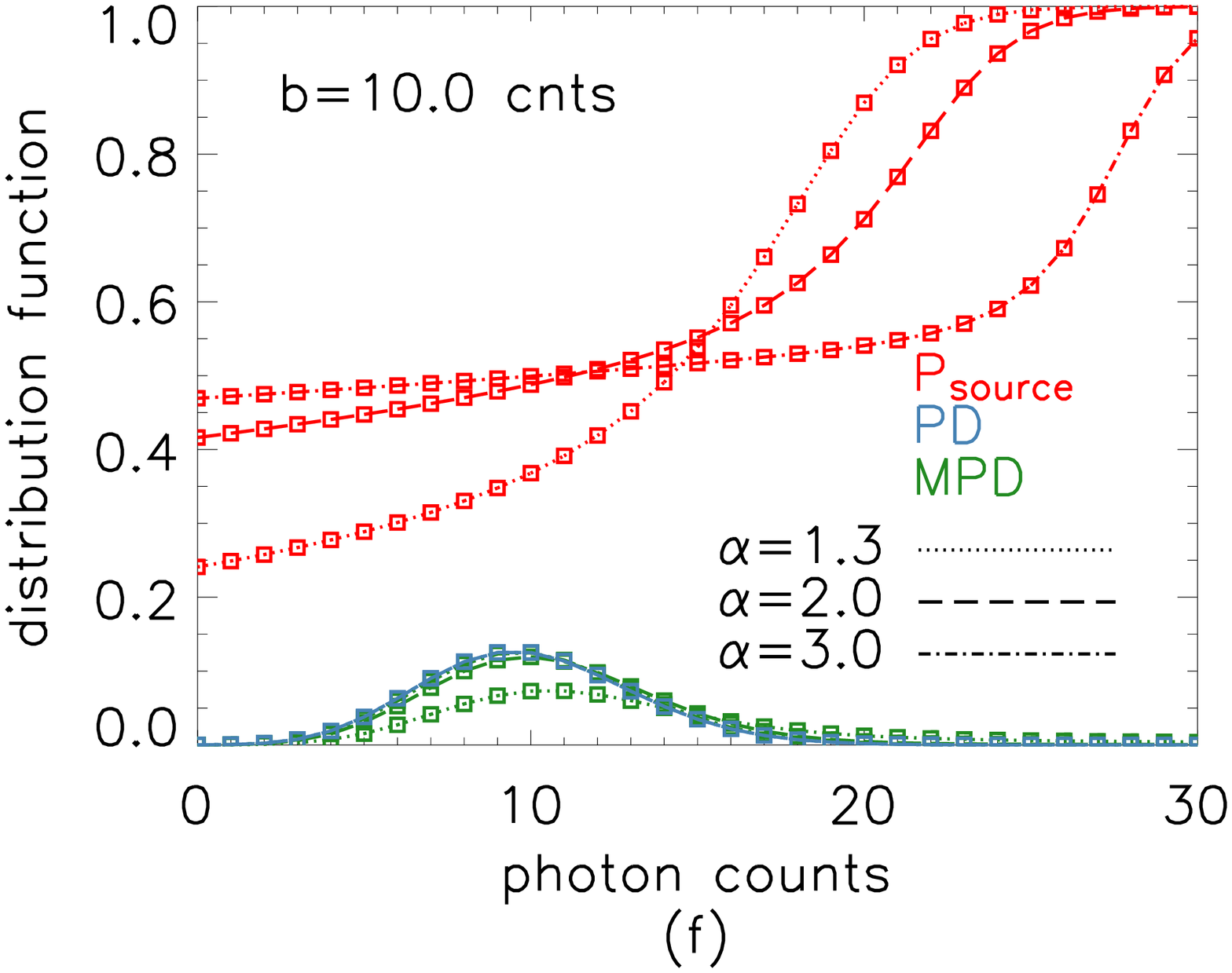}
\caption{Source probability ($P_{\rm source}$), Poisson distribution (PD) and 
Marginal Poisson distribution (MPD) versus photon counts as given by: 
eqs.~(\ref{eqsp}), (\ref{like_bg}) and (\ref{me_mpd}) for the {\em exponential
  prior} pdf, panels (a)--(c); eqs.~(\ref{eqsp}), (\ref{like_bg}) and
(\ref{igf_mpd}) for the {\em inverse--Gamma function prior} pdf, panels (d)--(f).}
\label{figexppow_prob}
\end{figure*}

In principle, the probability of detecting source signal in addition to the 
background should be derived marginalizing over the background 
coefficients and the hyper--parameters. Following the works of 
\cite{vonderl:1999} and \cite{fischer:2001}, the 
multidimensional integral, arising from the marginalization, can be 
approximated at the optimal values found of the background coefficients 
and the hyper--parameters. 

Equation (\ref{eqsp1}) is approximated with:
\begin{eqnarray}
p(\overline{B}_{\rm ij} \mid d_{\rm ij}) \approx 
\frac{1}{1 + \frac{\beta}{1-\beta}\cdot\frac{p(d_{\rm ij} \mid
{B}_{\rm ij},b_{\rm ij}^{\ast})}{ p(d_{\rm ij} \mid
    \overline{B}_{\rm ij},b_{\rm ij}^{\ast},\gamma)}}=P_{\rm source} ,
\label{eqsp}
\end{eqnarray}
where $b^{\ast}=\{b_{\rm ij}^{\ast}\}$ is the estimated background amplitude, as
  explained in Section \ref{sec:errbkg}. SPMs are estimated employing this
  formula.

The BSS method does not incorporate the shape of the PSF. 
When the PSF FWHM is smaller than the image pixel size, then
one pixel contains all the photons coming from a point--like source. 
Otherwise point--like sources are detected on pixel cells as large as the
PSF FWHM. 
Extended objects are detected in pixel cells large enough that the source size is
completely included. The pixel cell must be larger than the PSF FWHM and it
can exhibit any shape. 

Equation (\ref{eqsp}) shows that the source probability strictly depends on the ratio
between the Poisson likelihood, $p(d_{\rm ij} \mid {B}_{\rm ij},b_{\rm ij})$, and the
marginal Poisson likelihood, $p(d_{\rm ij} \mid \overline{B}_{\rm ij},b_{\rm ij},\gamma)$  
(Bayes factor). Bayes factors offer a way of evaluating evidence in favor of 
  competing models.

Fig.~\ref{figexppow_prob} shows the effect of the mixture model
  technique on the probability of having source contribution in pixels and
  pixel cells for the exponential and the inverse--Gamma function prior pdfs. 
  For the parametric studies, the parameter $\beta$ is chosen to be $0.5$. 
This noncommittal value of $\beta$ arises if each pixel (or pixel cell) is equally 
likely to contain source signal contribution or background only. 
$P_{\rm source}$ depends on the likelihood for the mixture model, which is the
  weighted sum of the Poisson distribution and the marginal Poisson distribution. 
For photon counts of about the mean background intensity, $P_{\rm source}$ is
  small. 
  $P_{\rm source}$ increases with increasing photon counts due to 
the presence of the long tail in the marginal likelihood. This allows efficient separation 
of the source signal from the background.\\
In panels $(a)$--$(c)$, the distribution function of 
  $P_{\rm source}$, the Poisson pdf ($PD$) and the marginal Poisson pdf ($MPD$)
  are drawn using the exponential prior (see also Figs.~\ref{priors_fig} and \ref{figpowp_problog}). 
In the case of fields with bright sources ($\lambda > 10$ times the background intensity), 
$P_{\rm source}$ is nearly zero for photon counts less than or 
equal to the mean background intensity. $P_{\rm source}$ increases
rapidly with increasing photon counts. 
In the case of fields where the mean source intensity is similar to the mean
background intensity, pixels containing photon counts close to the mean background
intensity have probabilities about $50$ per cent. 
In these cases, $P_{\rm source}$ increases slowly with increasing photon
counts, because the two Poisson distributions are similar. 
In the case of fields dominated by large background signal ($\lambda <$ mean
background amplitude), $P_{\rm source}$ increases very slowly with
increasing photon counts. In this case, the decay of the marginal Poisson
distribution follows closely the decay of the
Poisson distribution (e.g.~for $b=10$ counts and $\lambda=1$ count).\\
In Fig.~\ref{figexppow_prob} panels $(d)$--$(f)$, the distribution functions 
are shown using the inverse--Gamma function prior (see also Figs.~\ref{priors_fig} and
\ref{figpowp_problog}). 
The steepness of the slope depends on the parameter $\alpha$ 
(Fig.~\ref{priors_fig}). The source probability curve increases
faster at decreasing $\alpha$ values, because small $\alpha$ values indicate
bright sources distributed in the field. 
In panels $(e)$ and $(f)$, $P_{\rm source}$ provides values close to
$50$ per cent at low numbers of photon counts. This effect addresses the cutoff
parameter $a$. In fact, in these plots the cutoff parameter $a$ is smaller
than the background amplitude. 
The situation is different in the simulations with small background amplitude
(panel $(d)$),
where the source probabilities decrease below $50$ per cent at low numbers of photon
counts. In these simulations the cutoff parameter $a$ is chosen larger than the
mean background amplitude. Faint sources with intensities lower than
$a$ are described to be background. 
\begin{table}
\caption{Interpretation of source probability values.}
\label{interp}
\begin{center}
\begin{tabular}{@{}lccccc}
\hline
 & \textless 50\% & 50\% & 90\% & 99\% & 99.9\% \\
\hline \hline
$P_{\rm source}$ & backg. & indif. & weak & strong & very\\
 & only & & & & strong\\
\hline
\end{tabular}
\end{center}
\end{table}

The interpretations of the probability of having source contributions in
pixels and pixel cells are shown in Table \ref{interp}. 
Source probabilities $\textless 50$ per cent indicates the detection of background
only. $P_{\rm source}$ is indifferent at values of $50$ per cent. In
both cases, $P_{\rm source}$ might contain sources but they can not be
distinguished from the background due to statistical fluctuations. 
Source probability values $\gg 50$ per cent indicate source detections. 
False sources due to statistical fluctuations may occur especially for 
values $\textless 99$ per cent. For more details about the interpretations
  provided in Table \ref{interp} see \cite{jeffreys:1961} and \cite{kass:1995}. 

\subsubsection{Statistical combination of SPMs at different energy bands}\label{comb}
An astronomical image is usually given in various energy bands. 
SPMs, obtained with (\ref{eqsp}), acquired at different energy bands can be combined
statistically. 
The probability of detecting source signal in addition to the background for
the combined energy bands $\{k\}$ is:
\begin{eqnarray}
p(\overline{B}_{\rm ij} \mid d_{\rm ij})_{\rm comb} =  
1 - \prod_{{\rm k}=1}^{{\rm n}} (1-p(\overline{B}_{\rm ij} \mid d_{\rm ij})_{\rm k}),
\label{combeq}
\end{eqnarray}
where $n$ corresponds to the effective energy band. 
 
Equation (\ref{combeq}) allows one to provide conclusive posterior pdfs for the detected sources 
from combined energy bands. It results, as expected, that if an object is detected in
  multiple bands, the resulting $p(\overline{B}_{\rm ij} \mid d_{\rm ij})_{\rm comb}$ is larger than the
  source probability obtained analysing a single band.
An application of this technique is shown in Section \ref{twocolor_search}. This
analysis can be extended to study crowded fields or blends by investigating
source colours.

\section{Source characterization} \label{EOSP}

Following the estimation of the background and the identification of
  sources, the sources can be parameterized.\\
SPMs at different resolutions are investigated first. 
Sources are catalogued with the largest source probability reached in one of the
SPMs. Local regions are chosen around the detected sources. The
region size is determined by the correlation length where the maximum of the source
probability is reached. 

Although any suitable source shape can be used, 
we model the sources by a two--dimensional Gaussian. 
The data belonging to a source detection area '$k$' are given by:
\begin{equation}
D_{\rm ij} = b_{\rm ij} + G_{\rm ij} \qquad \forall \{ij\} \in \{k\}.  
\label{moddata}
\end{equation}  
$D_{\rm ij}$ are the modelled photon counts in a pixel $\{ij\}$ spoiled with the
background counts $b_{\rm ij}$. $G_{\rm ij}$ is the function which describes the photon counts 
distribution of the detected source:
\begin{eqnarray}
\nonumber
G_{\rm ij}=\frac{I}{2\pi\sigma_{\rm x}\sigma_{\rm y}\sqrt{1-\rho^{2}}} \cdot
{\rm exp}\{-\frac{1}{2(1-\rho^{2})} [(\frac{x_{\rm ij}-x_{\rm s}}{\sigma_{\rm x}})^{2}+\\
\nonumber
+(\frac{y_{\rm ij}-y_{\rm s}}{\sigma_{\rm y}})^{2} -2\rho
(\frac{x_{\rm ij}-x_{\rm s}}{\sigma_{\rm x}})(\frac{y_{\rm ij}-y_{\rm s}}{\sigma_{\rm y}})
]\}. \hspace{1.4cm}
\end{eqnarray}
$I$ is the intensity of the source, i.e.~the net source counts. 
$\sigma_{\rm x},\sigma_{\rm y}$ and $\rho$ provide the source shape. 
$x_{\rm s}$, $y_{\rm s}$ is the source pixel position on the image. 

Position, intensity and source shape are provided max\-i\-miz\-ing the likelihood
function assuming constant priors for all parameters:
\begin{eqnarray}
\nonumber
p(x_{\rm s}, y_{\rm s}, I, \sigma_{\rm x}, \sigma_{\rm y}, \rho| b,d) \propto 
\prod_{\rm ij} \frac{D_{\rm ij}^{d_{\rm ij}}}{d_{\rm ij}!} {\rm exp}\{-D_{\rm ij}\} \\\qquad \forall \{ij\} \in \{k\},
\label{likedata}
\end{eqnarray}
where $d_{\rm ij}$ are the photon counts detected on the image. 

According to (\ref{moddata}) and (\ref{likedata}) the source fitting
is executed on the sources for given background. 
This is reasonable since the uncertainty of the estimated background is small. 
No explicit background subtraction from the photon counts is needed for
estimating the source parameters. 

Source position and extensions are converted from detector space to sky
space. Source fluxes are provided straightforwardly.

The rms uncertainties of the source parameters are   
estimated from the Hessian matrix, where $H_{\rm ij} :=
-\frac{\partial^{2}{\rm ln}[p(D \mid
    \xi)p(\xi)]}{\partial \xi_{\rm i} \partial \xi_{j}}$ is element $\{ij\}$ of
    the Hessian matrix and $\xi$ indicates the source parameters. 
The square root of the diagonal elements of the inverse of the Hessian
    matrix provides the $1\sigma$ errors on the estimated parameters.

The output catalogue includes source positions, source counts, source count
rates, background counts, source shapes and their errors. The
source probability and the source detection's resolution are incorporated. 

Source characterization can be extended with model selection. With the
  Bayesian model selection technique, the most 
suitable model describing the photon count profile of the detected sources can
  be found.
The models to be employed are, for instance, Gaussian profile, King profile
\citep{cavaliere-fusco:1978}, de Vaucouleurs model, Hubble model. 
Such an extention to the actual method would allow an improvement in the
estimation of the shape parameters of faint and extended sources.

\section{Reliability of the detections} \label{rely}

There are several methods for reducing the number of false positives, one is
to use $p$--values from statistical significance tests (\citealt{linnemann:2003}). 
$P$--values are used to a great extent in many fields of science. 
Unfortunately, they are commonly misinterpreted. 
Researches on $p$--values (e.g.~\citealt{berger-sellke:1987}) showed that
$p$--values can be a highly misleading measure of evidence. 

In Section \ref{sigtest} we provide the definition of $p$--values. 
In Section \ref{bayeview}, we express our general view on how this problem can 
be tackled with BPT. The commonly used measure of statistical significance 
with Poisson $p$--values is introduced in Section \ref{sst_linn}. 
In Section \ref{spurSS} simulations are used for comparing the Poisson $p$--values with
the Bayesian probabilities. 

\subsection{$P$--values} \label{sigtest}

In hypothesis testing one is interested in making inferences about the truth
of some hypothesis $H_{0}$, given a set of random variables ${\bf X}$: 
$\bf{X} \sim f({\bf x})$, where $f({\bf x})$ is a continuous density and {\bf
  x} is the actual observed values. A statistic $T({\bf X})$ is 
chosen to investigate compatibility of the model with the observed data ${\bf
  x}$, with large values of T indicating less compatibility 
(\citealt{sellke:2001}). 
The $p$--value is then defined as:
\begin{eqnarray}
\nonumber
{\it p} = Pr(T({\bf X}) \ge T({\bf x})|H_{0}).   
\end{eqnarray}

The significance level of a test is the maximum allowed probability, 
assuming $H_{0}$, that the statistic would be observed. 
The $p$--value is compared to the significance level. If the
$p$--value is smaller than or equal to the significance level then $H_{0}$
is rejected. The significance level is an arbitrary number between
$0$ and $1$, depending on the scientific field one is working in. However,
classically a significance level of $0.05$ is accepted. Unfortunately, the
significance level of $0.05$ does not indicate a particular strong
evidence against $H_{0}$, since it just claims an even chance. 

An extensive literature dealing with misinterpretations about 
$p$--values exists, see e.g.~\cite{berger-sellke:1987}, \cite{fds30852},
\cite{berger-berry:1988}, \cite{delampady-berger:1990}, Loredo (1990, 1992), 
\cite{kass:1995}, Sellke et al.~(2001), \cite{gregory:2005} and references therein.  

\subsection{The Bayesian viewpoint}\label{bayeview}

Since the state of knowledge is always incomplete, a hypothesis can never be
proved false or true. One can only compute the 
probability of two or more competing hypotheses (or models) on the basis of the 
only data available, see e.g.~\cite{gregory:2005}. 

The Bayesian approach to hypothesis testing is conceptually
straightforward. Prior probabilities are assigned to all unknown hypotheses. 
Probability theory is then used to compute the posterior 
probabilities of the hypotheses given the observed data (\citealt{fds30916}). 
This is in contrast to standard significance testing which does not provide 
such interpretation. In fact, in the classic approach the truth of a 
hypothesis can be only inferred indirectly.   

Finally, it is important to underline that the observed data and parameters
describing the hypotheses are subject to uncertainties which have to be
estimated and encoded in probability distributions. With BPT 
there is no need to distinguish between statistical (or random) and 
systematic uncertainties. Both kinds of uncertainties are treated as lack of
knowledge. For more on the subject see \cite{fischer:2003}.

\subsection{Significance testing with $p$--values} \label{sst_linn}
Several measures of statistical significance with $p$--values have been
developed in astrophysics.
A critical comparison and discussion about methods for measuring 
the significance of a particular observation can be found 
in \cite{linnemann:2003}. 
Following Linnemann, our attention is focused on the Poisson 
probability $p$--value:
\begin{equation}
{\it p}_{P}=P(\ge d | b) = \sum_{j=d}^{\infty}\frac{e^{-b}b^{j}}{j!}.
\label{poisspval}
\end{equation}
${\it p}_{P}$ is the probability of finding $d$ or more (random) events under
a Poisson distribution with an expected number of events given by
$b$. 
Linnemann remarks that Poisson probability $p$--value estimates lead to
overestimates of the significance since the uncertainties on the background are ignored.

\subsection{Comparing threshold settings for source reliability} \label{spurSS}
\begin{figure}
\centering
\includegraphics[scale=0.28,angle=90]{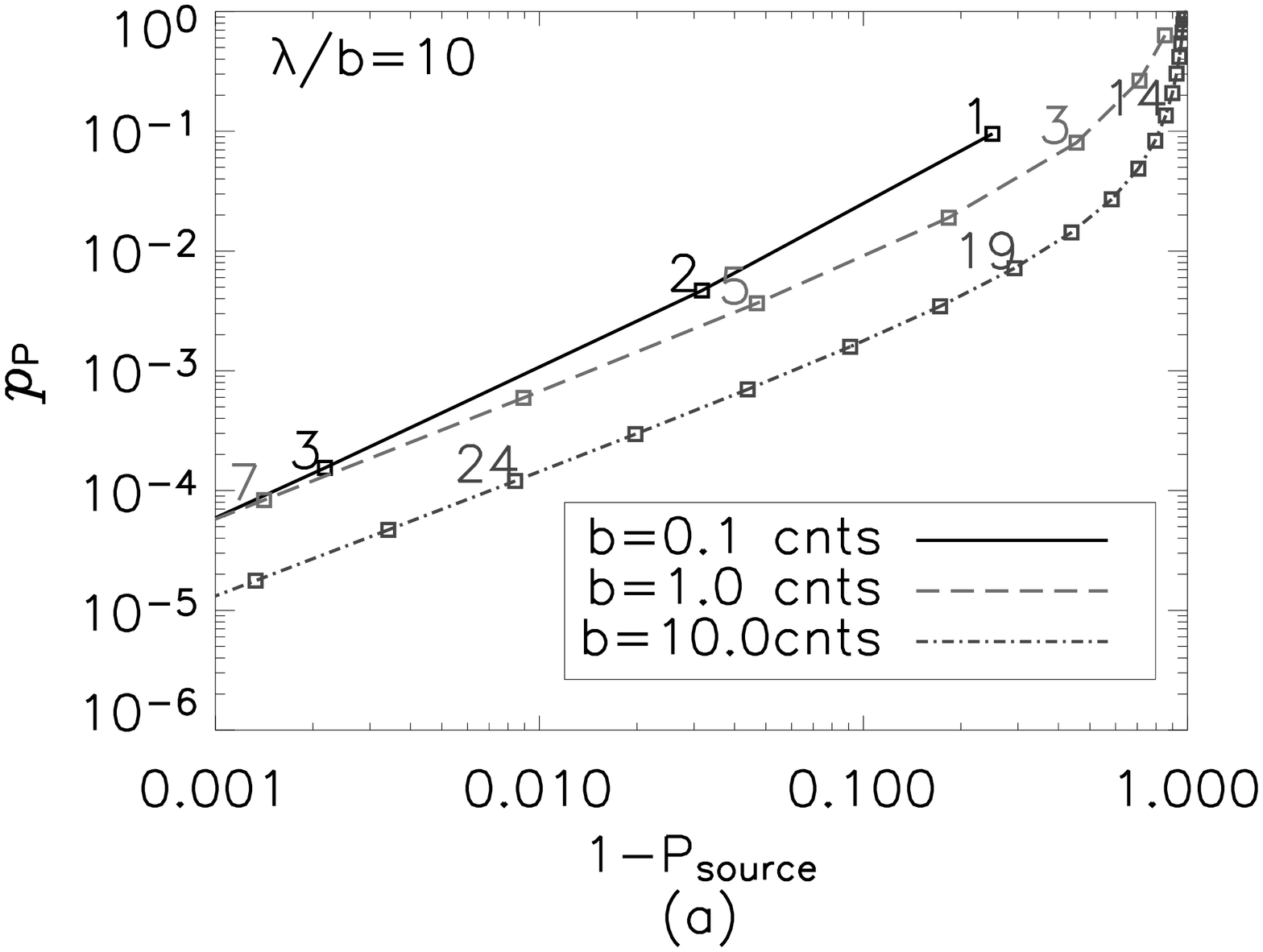}\\ 
\includegraphics[scale=0.28,angle=90]{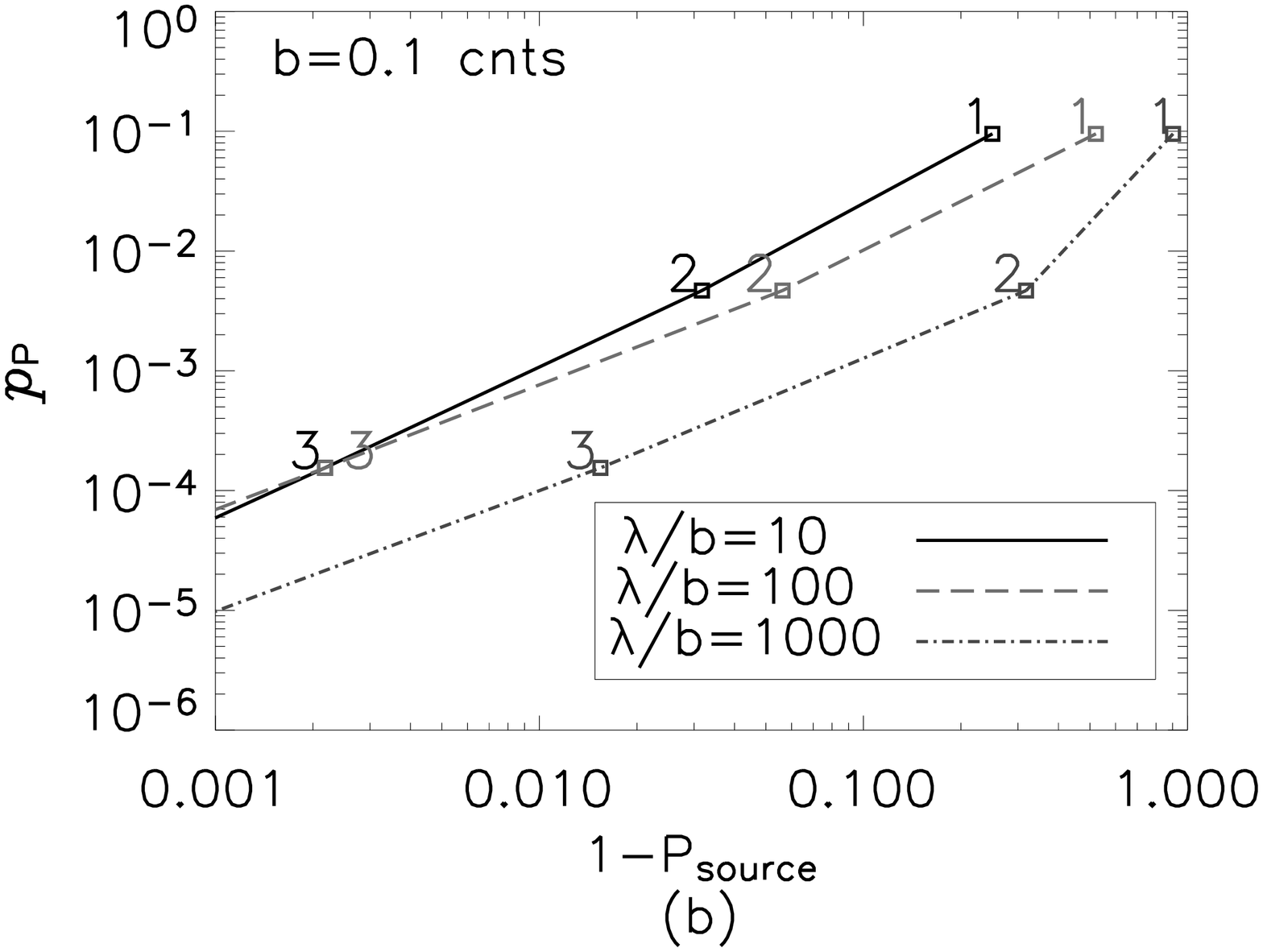}
 \caption{
 Classic hypothesis testing versus Bayesian approach.
}
\label{spur_exp}
\end{figure}
In order to restrain the rates of false source detections per field, a 
threshold on probabilities is commonly set according to the goal of a specific research. 
For instance, \cite{freeman:2002} have chosen an upper limit of $1$ spurious
detection per analysed {\it CHANDRA} field. 
This value corresponds to a false--positive probability threshold of 
$10^{-7}$. The method developed by Freeman et al.~is a wavelet--based
source detection algorithm ({\sc wavdetect}). 
For the XMM--$Newton$ serendipitous source catalogue, \cite{watson:2003} have
chosen a detection likelihood $L=10$, corresponding to $\approx 4 \sigma$.
$L$ is given by $-{\rm ln}(1-P)$, where $P$ is the probability of source detection 
obtained by a maximum likelihood method (\citealt{cruddace:1988}). 
The selected likelihood threshold corresponds to the detection of $< 1$
spurious source per field.   
In any systematic investigation to source detection, the threshold level is a
tradeoff between the detection power and false detections rate. 

\begin{table}
\caption{Bayesian source probability ($P_{\rm source}$) vs Poisson $p$--value ($p_{P}$).} 
\label{exa1}
\begin{tabular}{@{}lcccccc}  \hline 
$b$ & $\lambda$ & $d$ & $P_{\rm src}$ & $1-p_{P}$ &
  $1-P_{\rm src}$ & $p_{P}$\\ 
(cnts) & (cnts) & (cnts) &  & & & \\ 
\hline  \hline
 0.1 & 1.0 & 2 & 0.97 & 0.995 & 0.03 & 0.005 \\ \hline
 1.0 & 10.0 & 2 & 0.29 & 0.74 & 0.71 & 0.26 \\ \hline
 10.0 & 100.0 & 2 & 0.01 & 0.0005 & 0.99 & 0.9995 \\ \hline
\end{tabular}
\medskip 
 This example shows the variation of $P_{\rm source}$, here indicated with $P_{\rm src}$, 
and $p_{P}$ for detecting $2$ photon
  counts in a pixel at three background levels. 
$P_{\rm source}$ represents the source probability of detecting $2$
  photon counts ($d$) in a pixel according to the Bayesian technique. 
(1-$p_{P}$) is the cdf. It provides the probability of detecting $1$ photon count 
or less in a pixel according to Poisson statistics.
(1-$P_{\rm source}$) is the background probability estimated with the Bayesian
method. $p_{P}$ provides the probability of detecting $2$ photon counts 
or more in a pixel according to Poisson statistics.
\end{table}
Following this idea, the Poisson probability $p$--value (\ref{poisspval}) 
is compared to the Bayesian source probability (\ref{eqsp}).
 Fig.~\ref{spur_exp} compares the two statistics. 
Panel $(a)$ shows the relation between $p_{P}$
  and ($1-P_{\rm source}$) for varying background amplitudes and source intensities. 
Panel $(b)$ displays the same data but with fixed background value and varying source
intensities. These results are obtained employing the exponential prior. 
A more detailed study, including the
  inverse--Gamma function prior, can be found in \cite{guglielmetti:2009}.
The squares on each plot indicate the photon counts $d$ chosen for
calculating (\ref{poisspval}).
On the abscissa, the background probability is calculated as the 
complementary source probability provided by the Bayesian method.  
The value close to unity corresponds to a source probability, $P_{\rm source}$,
which goes to zero, instead a value of $0.1$ corresponds to $90$ per cent source probability and
$0.01$ to $99$ per cent source probability.\\
Each plot shows a general tendency. 
For a given count number $d$, $P_{\rm source}$ and (1-$p_{P}$) increase with decreasing 
background intensity. However, $P_{\rm source}$ is more conservative.
This is due to the dependency of $P_{\rm source}$ not only on the mean background
intensity but also on the source intensity distribution. This
dependency is expressed by the likelihood for the mixture model
(\ref{mix_mod}), that plays a central role for the estimation of the source 
probability (\ref{eqsp}). 
An example of the different interpretations of source detection
employing the two statistics is provided in Table \ref{exa1}.

\begin{figure}
  \centering
\includegraphics[scale=0.28,angle=90]{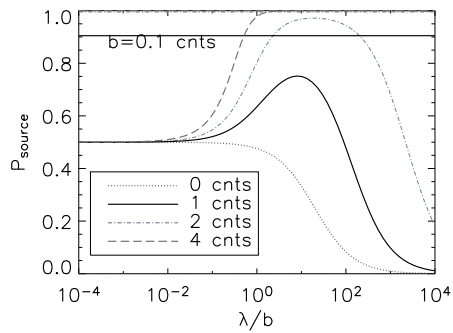} 
  \caption{
Dependency of the Bayesian source probability on the source intensities 
distributed on a field.
}
\label{varsp}
\end{figure}
In Fig.~\ref{varsp}, $P_{\rm source}$ for a
  given number of photon counts in a pixel versus $\lambda / b$, the ratio between 
the mean source intensity and the background intensity, is drawn for a fixed background value.  
The abscissa is drawn on a logarithmic scale. 
For a given number of photon counts, the value of $P_{\rm source}$ varies with 
the source intensities expected in the astronomical image and with the
background amplitude. 
For $2$ photon counts, $P_{\rm source}$ reaches a maximum where
the mean source intensity in the field has values in the range $(1-5)$
counts. In this part of the curve, 
$2$ photon counts in a pixel are discriminated best from the background. 
Away from this range, the source probability decreases. 
For small $\lambda / b$ values, $P_{\rm source}$ approaches $0.5$ because source
and background can not be distinguished. For large $\lambda / b$ values,
$P_{\rm source}$ decreases since more sources with large intensities are expected
relative to small intensities. Therefore, a signal with $2$ counts is assigned
to be background photons only. \\
($1-p_{P}$) is calculated for the same values of background and photon counts
as for $P_{\rm source}$. 
($1-p_{P}$) is constant, since it does not depend on the source
intensities expected in the field. 
Its value for $2$ photon counts, as seen in Table \ref{exa1}, is larger than
the maximum of $P_{\rm source}$. 
In general, $(1-p_{P})$ is larger than $P_{\rm source}$ for values of photon
counts larger than the mean background intensity. 
If the values of the photon counts are lower than or equal to the mean background intensity, 
$(1-p_{P})$ is lower than the maximum of $P_{\rm source}$.

The comparison shows that it is not possible to calibrate $p_{P}$ with
$P_{\rm source}$ because of the intrinsic difference in the nature 
of the two statistics. 
In fact, Poisson $p$--values do not provide an interpretation about background 
or sources and do not include uncertainties on the background. 
The Bayesian method, instead, gives information about background and sources and their
uncertainties.

The comparison between the two statistics reveals that
slightly different answers are arising for the two priors of the source signal. When
the exponential prior is employed, fields with large intensities 
are less penalized by false positives caused by random Poisson noise 
than fields with source signal very close to the background amplitude. 
When the inverse--Gamma function prior is used, 
false positives detections depend on the cutoff parameter $a$. This is because 
the cutoff parameter has an effect on faint sources. 
The same behaviour is expected on false positives in source detection. 
The exponential prior, instead, does not exclude small source intensities. 
Note that the choice of the source signal prior pdf is crucial for
source detection. For a reliable analysis the source signal prior chosen has
to be as close as possible to the true one. 

\section{Simulated data} \label{simu_data}

Artificial data are used for performance assessment of the BSS technique. 
Three simulations are analysed utilizing the exponential and the inverse--Gamma
function prior pdfs of the source signal.
The datasets, described in Section \ref{setupsim}, are meant to test the capabilities of
the BSS method at varying background values. The idea is to cover different cases 
one encounters while surveying different sky regions or employing
instruments of new and old generations.  
In Section \ref{resu} we review the outcome of our analysis on the three
simulated datasets. Comments are given for each feature of the developed 
technique. In Section \ref{summar} we provide a summary on the outcome of our analysis.

\subsection{Simulations setup}\label{setupsim}

\begin{figure*}
  \centering
\includegraphics[scale=0.375,angle=90]{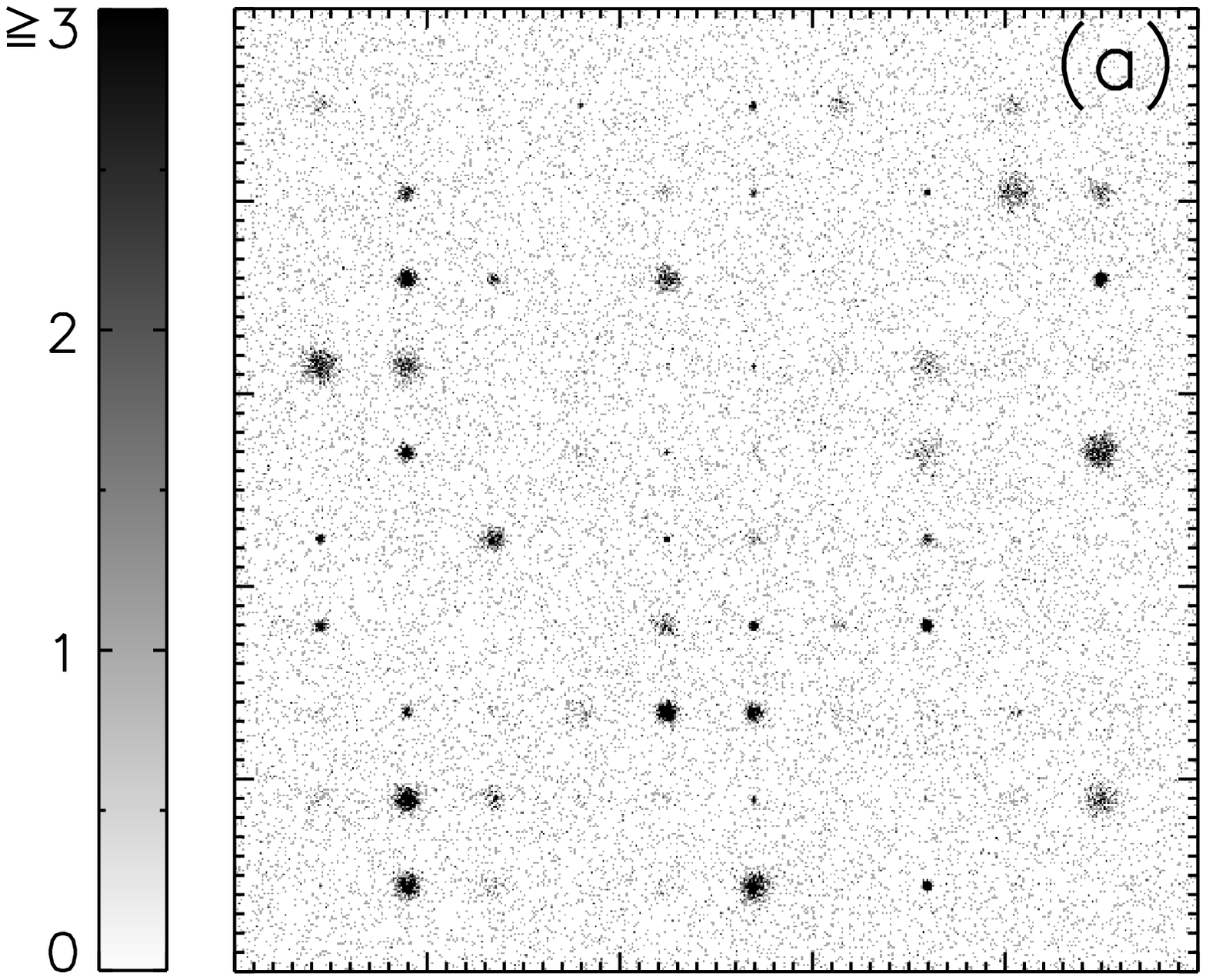}\includegraphics[scale=0.375,angle=90]{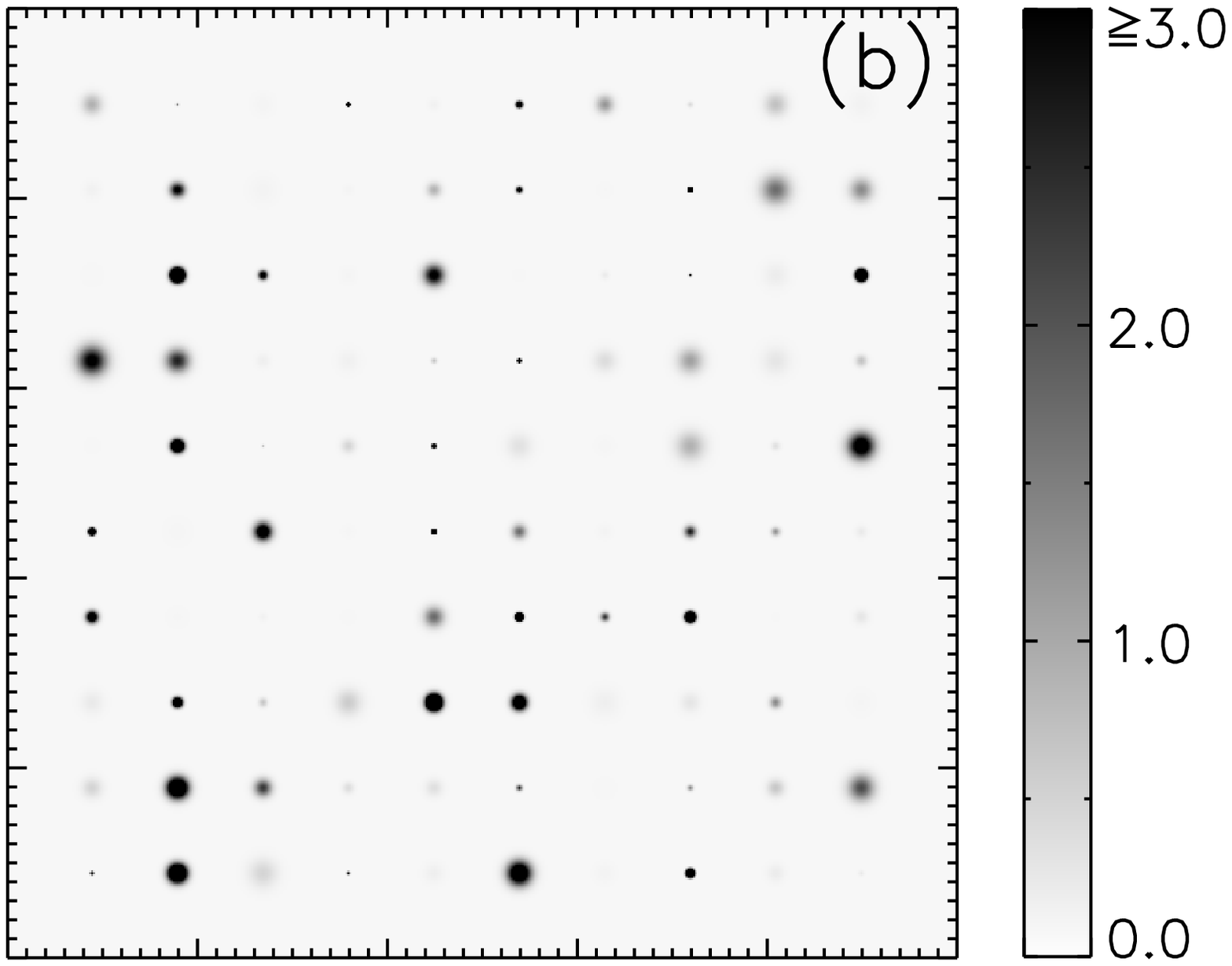} 
\includegraphics[scale=0.375,angle=90]{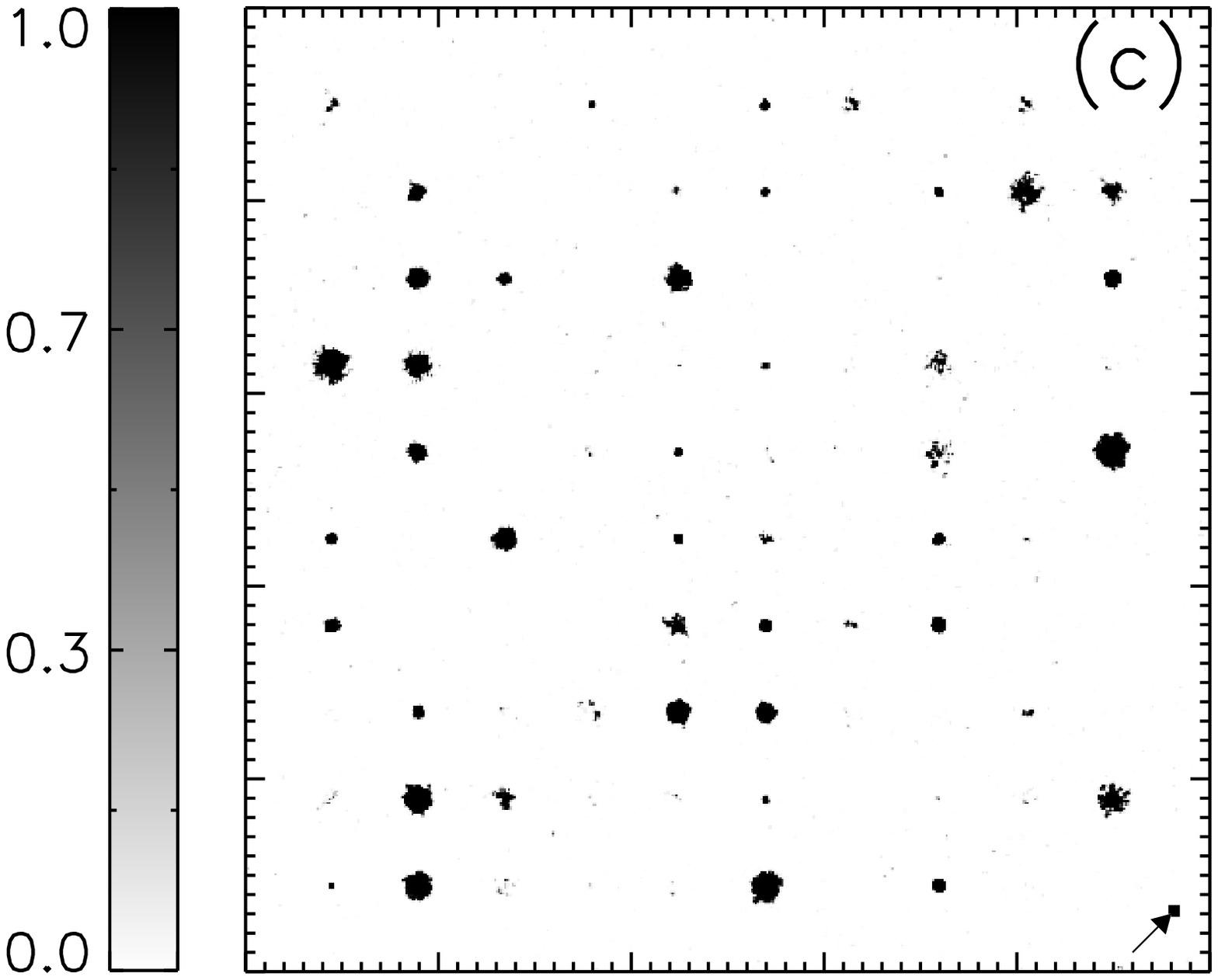}\includegraphics[scale=0.375,angle=90]{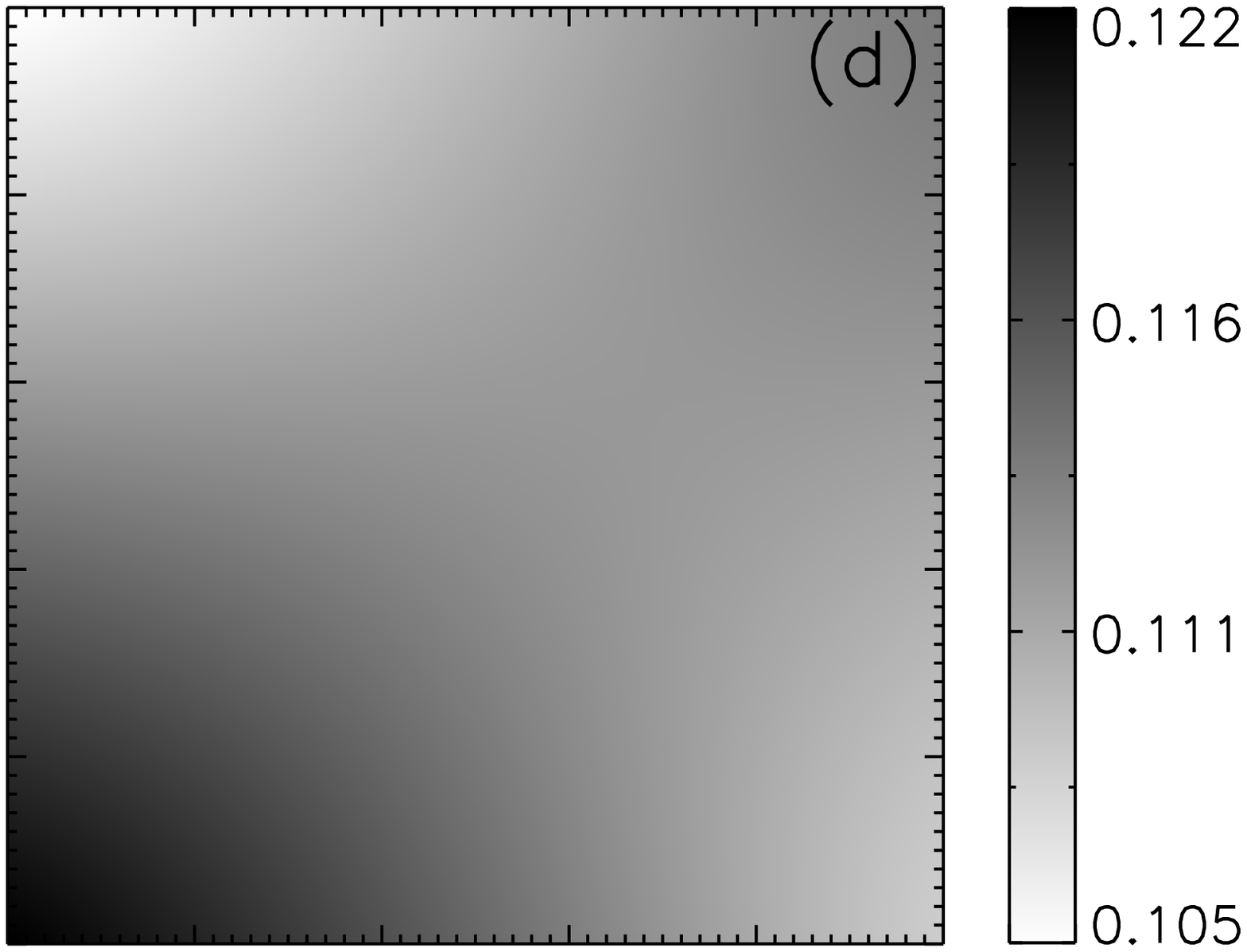}
\includegraphics[scale=0.375,angle=90]{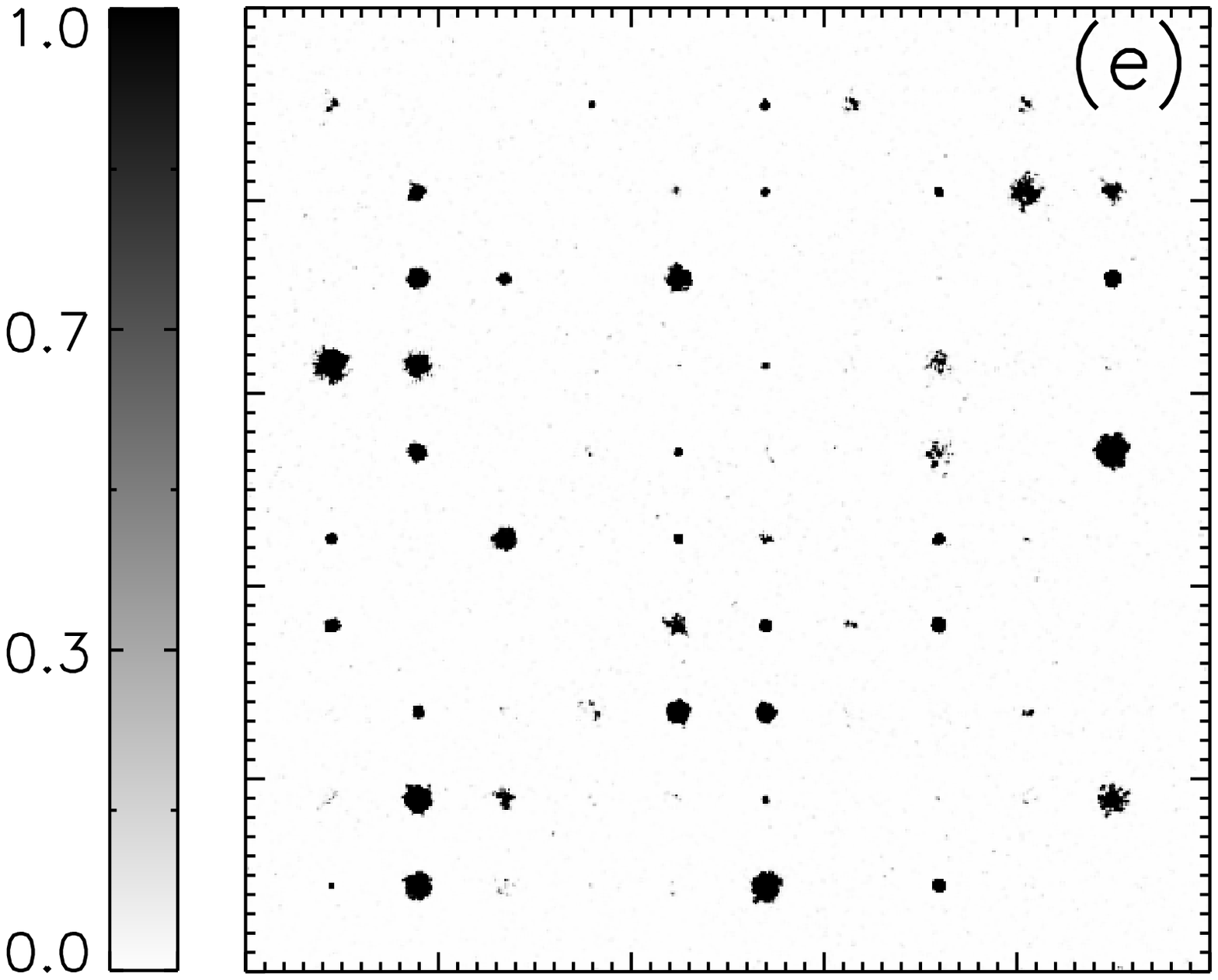}\includegraphics[scale=0.375,angle=90]{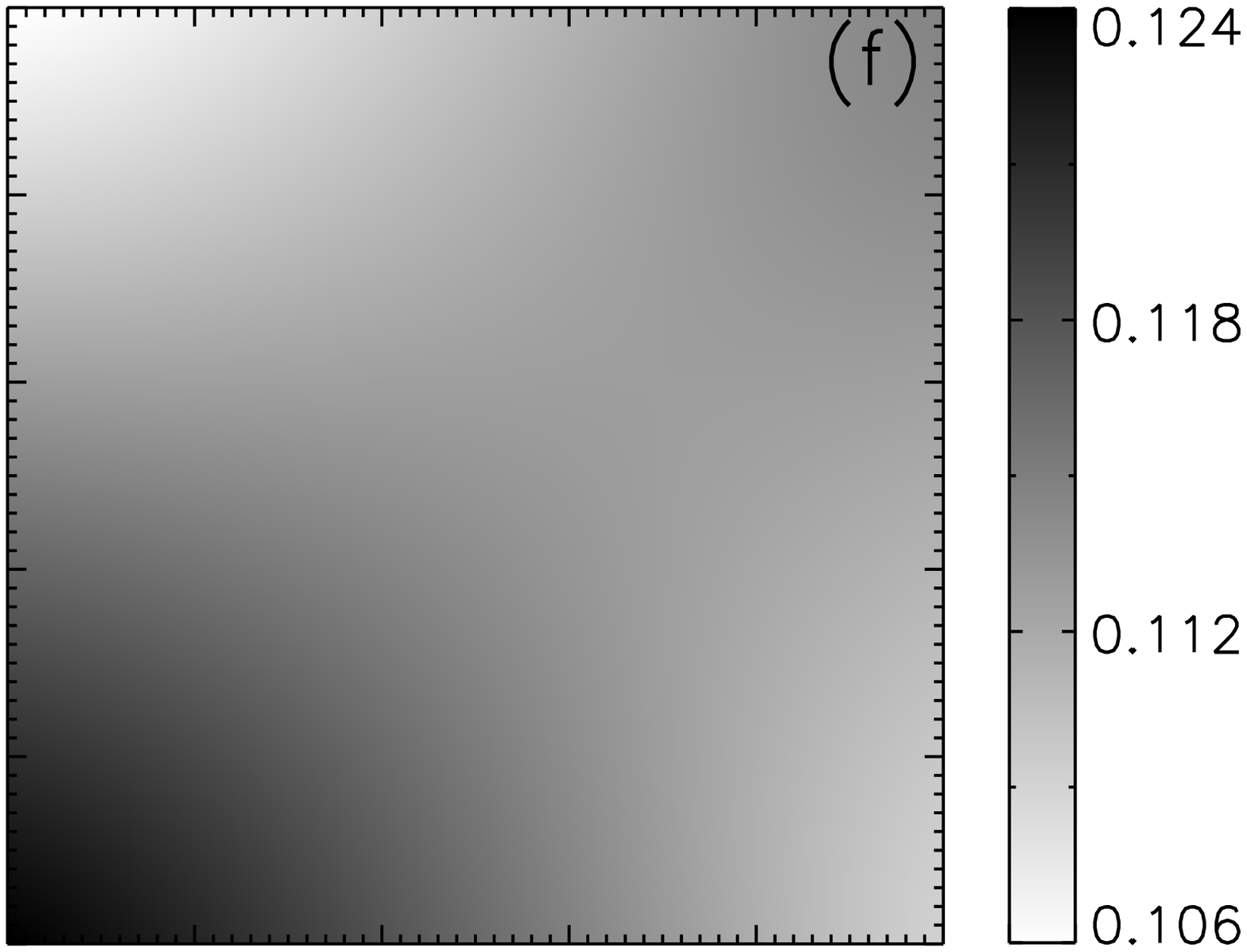} 
\caption{
Panels $(a)$  and $(b)$, simulated data with small background: image with
Poisson noise, image without Poisson noise, respectively. Panels $(c)$ and $(d)$, results 
with {\em exponential prior} pdf: SPM with $3$ pixels resolution, background map,
respectively. 
Panels $(e)$ and $(f)$, results with {\em inverse--Gamma function prior} pdf. 
}
\label{simb01_datares}
\end{figure*}
Three sets of simulated fields composed of $100$ sources modelled on a 
constant background with added Poisson noise are generated. 
Groups of ten sources are characterized by the same number of photon counts
but with different sizes. 
A logarithmic increment in photon counts per group is chosen ranging from $1$ to $512$. 
The shape of each source is characterized by a two--dimensional circular
Gaussian.
The source extensions, given by the Gaussian standard deviation, increase from 
$0.5$ to $5.0$ pixels in steps of $0.5$. 
Sources are located equidistantly on a grid of $500 \times 500$ pixels. 
Initially, the simulated sources are located on the grid such that the source intensities 
  increase on the abscissa, while the source extensions increase on the
  ordinate. Subsequently, the $100$ sources are randomly permuted on the
  field.
A background is added on each simulated field with values of $0.1$, $1$ and 
$10$ counts respectively. 
We assume a constant exposure.

In Fig.~\ref{simb01_datares}, we show one out of three datasets: the simulated
data with small background. Image $(a)$ represents the simulated data with
added Poisson noise. The image indicated with $(b)$ is the simulated data
without Poisson noise. It is placed for comparison. 
The simulated datasets $(a)$ and $(b)$ are scaled in the range ($0-3$) photon
counts$\cdot$pixel$^{-1}$ in order to enhance faint sources. 
The original scale of the simulated data with Poisson noise is ($0-317$)
photon counts$\cdot$pixel$^{-1}$. 
The simulated data without Poisson noise show counts$\cdot$pixel$^{-1}$ in the range
($0.1-326.0$). \\ 
The images representing the other two datasets for $b=1$ and $10$ counts are 
similar to the one shown in Fig.~\ref{simb01_datares}. In these datasets, 
the number of sources to be separated from the background decreases 
with increasing background intensity.

The cutoff parameter $a$ is chosen to be $0.14$ counts in the three simulated
datasets. This is to show the effect of $a$ when the background is smaller or
larger than $a$.

\subsection{Results}\label{resu}
\subsubsection{Background estimation}

For the background modelling, only four pivots located at the field's corners 
are used. This choice is driven by the presence of a constant background. 
An optimization routine is used for maximizing the likelihood for the mixture
model (\ref{mix_mod}). The solution of the optimization routine is the
pivots amplitude's estimates from which the background is calculated. 

The three setups are designed such that half of the $100$ simulated sources
are characterized by $\le 16$ photon counts. Some of these simulated sources are
too faint for being detected. These sources may contribute to the background model.

In Fig.~\ref{simb01_datares}, the estimated background maps are displayed when
employing the exponential prior pdf (image $(d)$) and the inverse--Gamma 
function prior pdf (image $(f)$) for the simulated data with small background. \\
The two images show that the background intensity decreases slightly toward the upper left
and lower right corners of about $5$ per cent. The same trend is seen also in the
estimated backgrounds with intermediate and large values. 
Evidently, this effect is not introduced by the prior over the signal. Also, it is not 
introduced by the selected pivots positions. If that were the case, then the
same magnitude is expected at each image corner.   
Instead, this is an overall effect induced by the simulated sources. All
simulated sources are randomly permuted. In the upper left and lower right
corners are located numerous faint sources. In the lower left corner many
bright sources are clustered. 
The increment in the background intensity is due to the statistical
distribution of the sources. This explains why the same trend in background intensities is
seen in all background models. 

When employing the exponential prior pdf, the estimated background intensities are
in agreement with the simulated background amplitudes. In the case of the
inverse--Gamma function prior pdf, the estimated backgrounds are sensitive to the cutoff 
parameter $a$. 
When $a$ is set larger than the mean background (i.e.~simulated data with
  small background), the background is overestimated. The overestimated
  background is due to the presence of source signal below the cutoff
  parameter. 
Hence, no source intensities below $0.14$ counts are allowed. 
It results that the estimated background is $40$ per cent larger than the simulated
one. 
For simulated data with intermediate background, the cutoff parameter
$a$ is fixed to a value lower than the simulated background. 
The background is underestimated by only $\sim 1$ per cent with respect to the simulated one.   
For simulated data with large background, the cutoff parameter $a$ is much 
lower than the simulated mean background value. 
The estimated background is in agreement with the simulated one.
  
The background uncertainties are quite small compared with the background
itself, on the order of few a percent. 
This effect holds because the background is estimated on the full field. 
However, the errors increase where the estimates deviate from the simulated
background. In addition, when applying the inverse--Gamma function prior
pdf, the errors are larger than those found utilizing the exponential prior
pdf. 

\subsubsection{Hyper--parameter estimation}
\begin{figure}
\centering
\includegraphics[scale=0.3,angle=90]{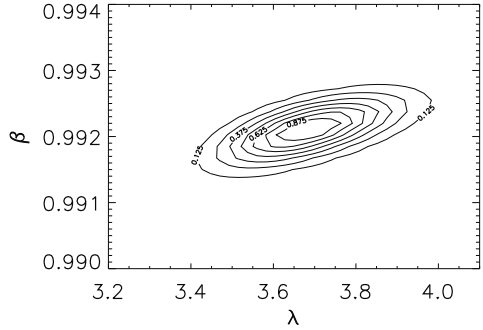}
\caption{Contour plot of the posterior 
pdf for the hyper--parameters, $p(\lambda, \beta | D)$, estimated from the simulated
  field with small background using the {\em exponential prior} pdf. }
\label{simb01_2}
\end{figure}

In Fig.~\ref{simb01_2} the contour plot in ($\lambda$, $\beta$) parameter space for
the joint probability distribution is shown for the hyper--parameters evaluated from the
simulated data with small background. The contour levels indicate the
credible regions. 
The values of the estimated hyper--parameters are: $\beta=(99.2 \pm 0.03)$ per
cent, $\lambda=(3.68 \pm 0.1)$ counts. 
The estimated $\beta$ value provides the information that only $0.8$ per cent of
the pixels in the field contains sources. A similar answer is found with the 
other simulated data: the $\beta$ value increases slightly at increasing
background amplitudes. 
$\lambda$, instead, provides the mean source intensity in the field. 
The estimated value of $\lambda$ increases with increasing
background amplitudes because small intensities are assigned to be background. 

When employing the inverse--Gamma function prior pdf, the hyper--parameter $\alpha$
is found with the smaller value in the simulated data with small background. 
The largest value of $\alpha$ is found in the simulated data with intermediate
background. Large values of $\alpha$ indicates that more faint sources and
less bright sources are expected in the field (Fig.~\ref{priors_fig}). 
These results do not contradict our expectations on the hyper--parameter
estimates, since the cutoff parameter  
selects the source signal distribution at the faint end.  

\subsubsection{The components of the mixture model}

\begin{figure}
\centering
\includegraphics[scale=0.28,angle=90]{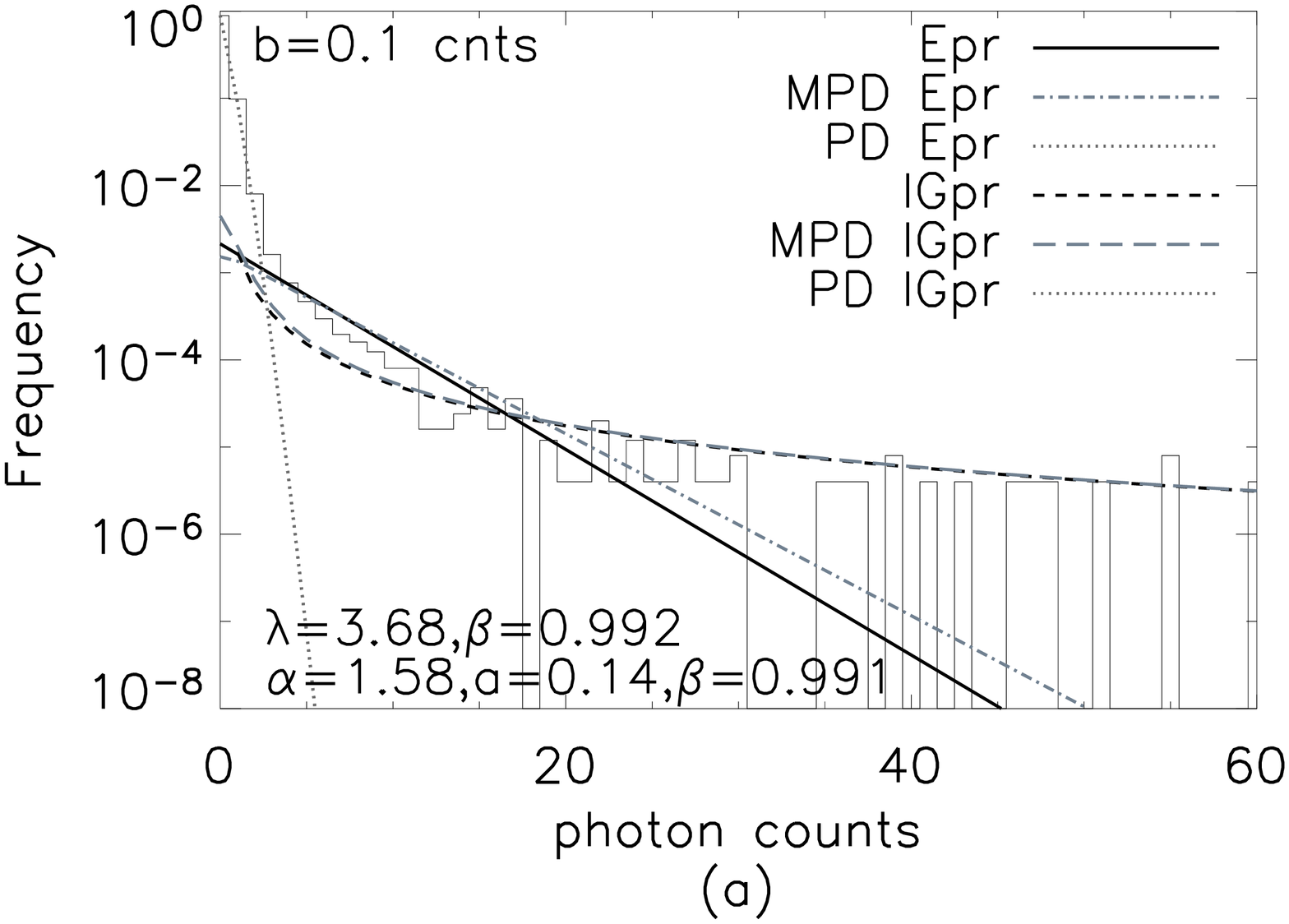} 
\includegraphics[scale=0.28,angle=90]{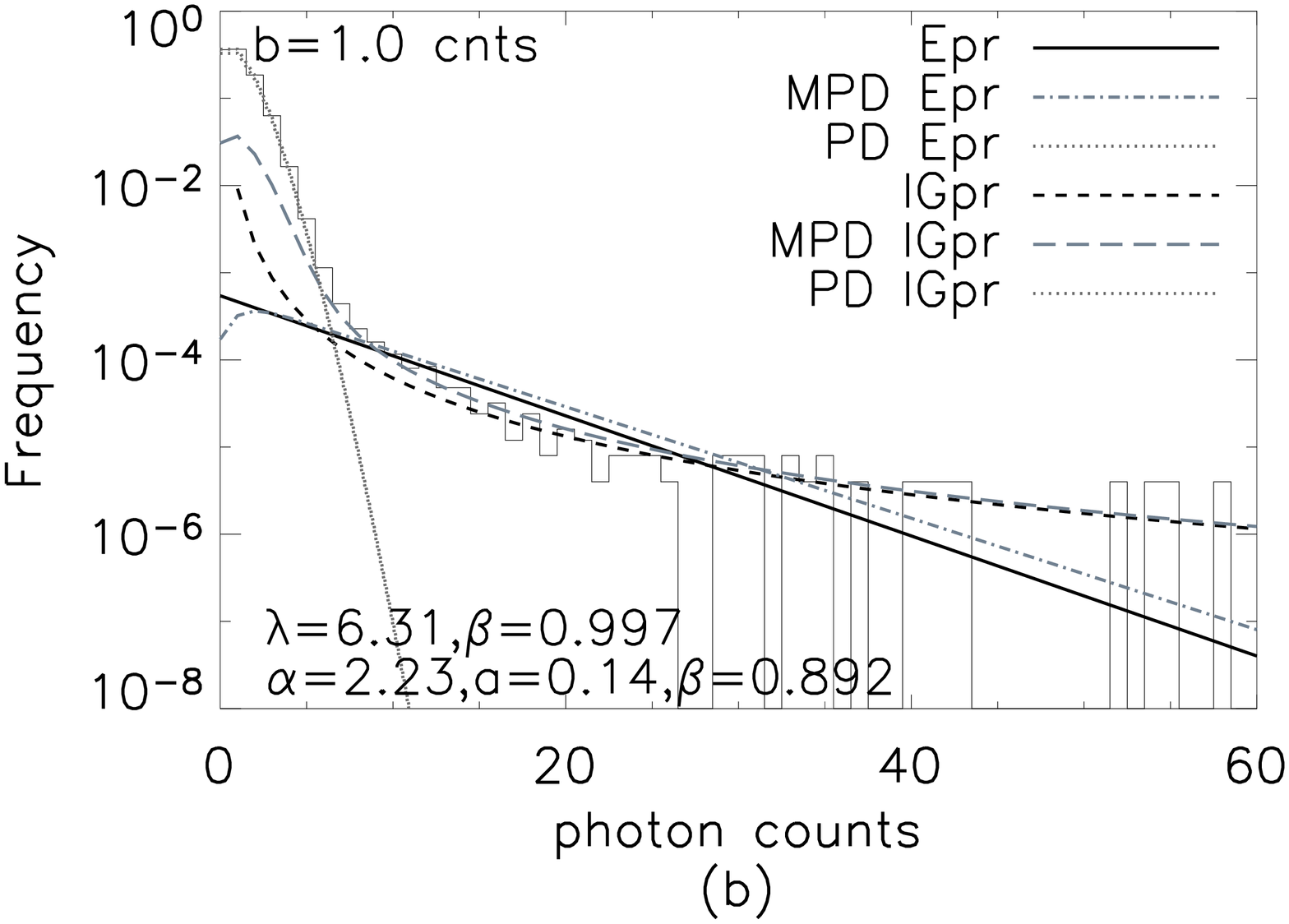}
\includegraphics[scale=0.28,angle=90]{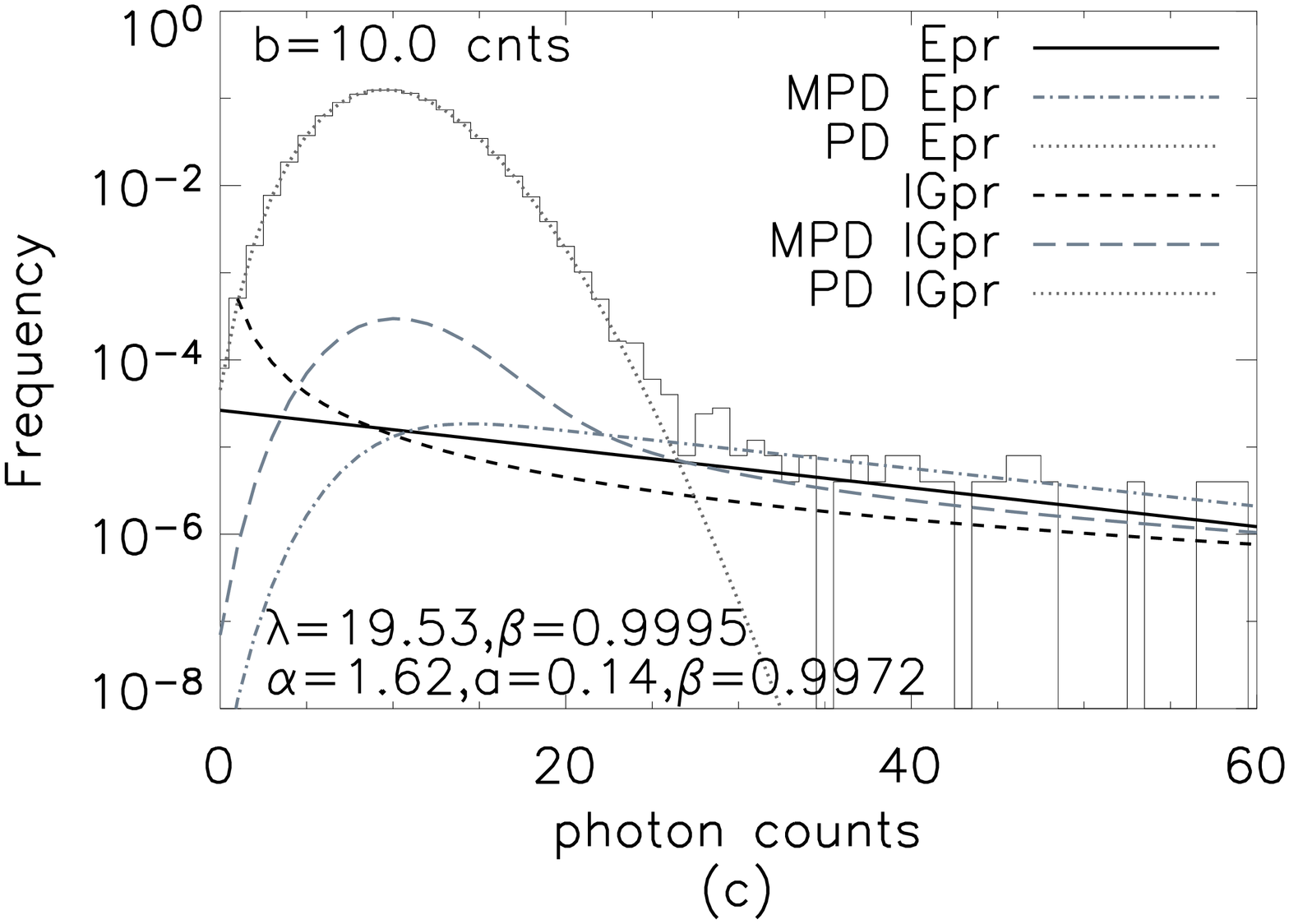}
\caption{Normalized histograms of the
  simulated data with small (panel $(a)$), intermediate (panel $(b)$) and
  large (panel $(c$)) backgrounds are displayed. The exponential prior
  ($Epr$), inverse--Gamma function prior ($IGpr$) and related 
marginal Poisson ($MPD$) and Poisson ($PD$) distributions are
  plotted over the data. The ordinates are in logarithmic scale.}
\label{histogr}
\end{figure}
\begin{figure*}
\centering
\includegraphics[scale=0.55]{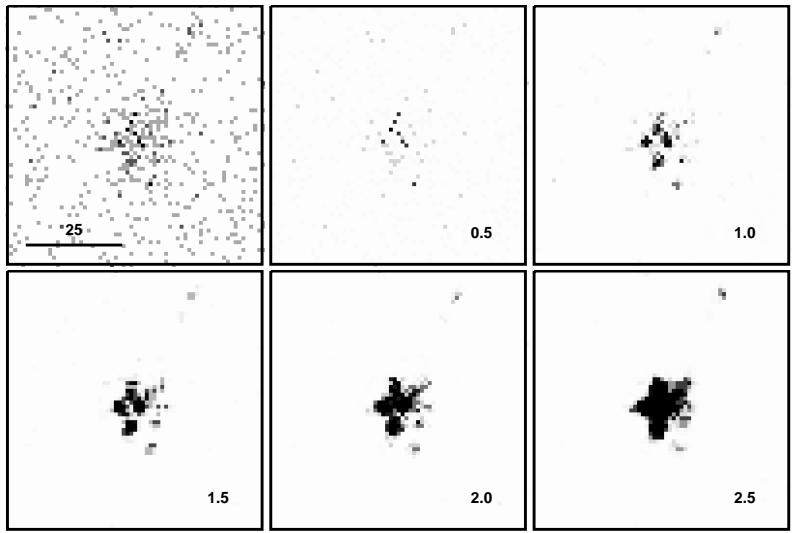} 
\caption{The upper left image is a zoom in of the photon count image 
(panel $(a)$, Fig.~\ref{simb01_datares}) on a simulated source located at
  (x,y)=(360,270). 
The width of the photon count image is $65$ pixels.
  The following images are SPMs at decreasing resolutions. 
  The correlation length of each SPM is written on the lower right
  corner of each image in pixel units.
}
\label{simb01_seq53} 
\end{figure*}
In Fig.~\ref{histogr}, the Poisson and the marginal Poisson distributions 
multiplied with their prior knowledge on the model are plotted 
over the normalized histogram of each simulated dataset. 
The likelihoods are drawn for the exponential and inverse--Gamma function prior
pdfs. The values of the estimated hyper--parameters are shown in each
image. 

The exponential prior pdf is plotted over the histogram with a continuous line. 
The inverse--Gamma function prior pdf, instead, is plotted with a dashed line. The
simulated data are neither distributed exponentially nor as an inverse--Gamma
function. Hence, the prior pdfs of the source signal are not expected to fit the data
exactly. 

The marginal Poisson distribution weighted with ($1-\beta$) is drawn with a
dash--dot line when employing the exponential prior and with long dashes line 
in the case of the inverse--Gamma function prior. The Poisson
distribution (dotted line) is weighted with $\beta$ for the
exponential and the inverse--Gamma function prior pdfs. 

The intersection between the Poisson pdf and the marginal Poisson pdf 
indicates the source detection sensitivity. 
When employing the exponential prior in the simulated data with small 
background (panel $(a)$), the exponential prior enables the detection of fainter
sources than the inverse--Gamma function prior. This is expected since 
the cutoff parameter occurs at a value larger than the simulated mean background. 
Considering the simulated data with intermediate background (panel $(b)$), the
detection is more sensitive to faint sources when employing the inverse--Gamma
function prior compared to the exponential prior. In fact, the cutoff
parameter allows to describe as source signal part of the simulated background
amplitude. 
For the simulated data with large background (panel $(c)$), the same sensitivity in source
detection is expected when employing the two priors over the signal
distribution.     

\subsubsection{Source probability maps}
The box filter method with cell shape of a circle is used in the three simulations for the
multi--resolution analysis. 
Examples of SPMs are shown in Fig.~\ref{simb01_datares} for the simulated data
with small background. 
Images $(c)$ and $(e)$ are obtained employing the exponential and 
the inverse--Gamma function prior pdfs, respectively. 
These images represent the probability of having source contributions in pixel
cells with a resolution of $1.5$ pixels. 
At this resolution a pixel cell is composed by $9$ pixels. 
A pixel cell with a correlation radius of $1.5$ pixels is drawn in the lower
right corner of image $(c)$ (Fig.~\ref{simb01_datares}). It is indicated with an arrow.

The multi--resolution technique provides an analysis of source probabilities
variation and of source features of the detected sources.
In Fig.~\ref{simb01_seq53}, we display the photon count image and the SPMs 
zoomed in a bright extended source. 
This source is detected with the largest source probability ($\sim 1$) at
$2.5$ pixels resolution. At this resolution the source is detected as one
unique object, as given by the simulation. 
At larger resolutions, instead, the source is dissociated in small parts as
seen in the photon count image. 
This indicates that the source counts are not distributed uniformly. 
In an astronomical observation, more information is required in order to
understand the nature of such sources. Secondly, the
maximum in source probability is reached at a correlation length that is 
smaller than the source size. This is due to the source brightness relatively
to the small background value. Within the range of resolutions studied, the 
source probability is constant at correlation lengths larger than $2.5$ pixels. \\
This example shows that the multi--resolution technique combined with the BSS
method is particularly appropriate for the search of extended and non--symmetrical
sources. 

\subsubsection{Comparison between estimated and simulated source parameters}
\begin{figure*}
\centering
\includegraphics[scale=0.35]{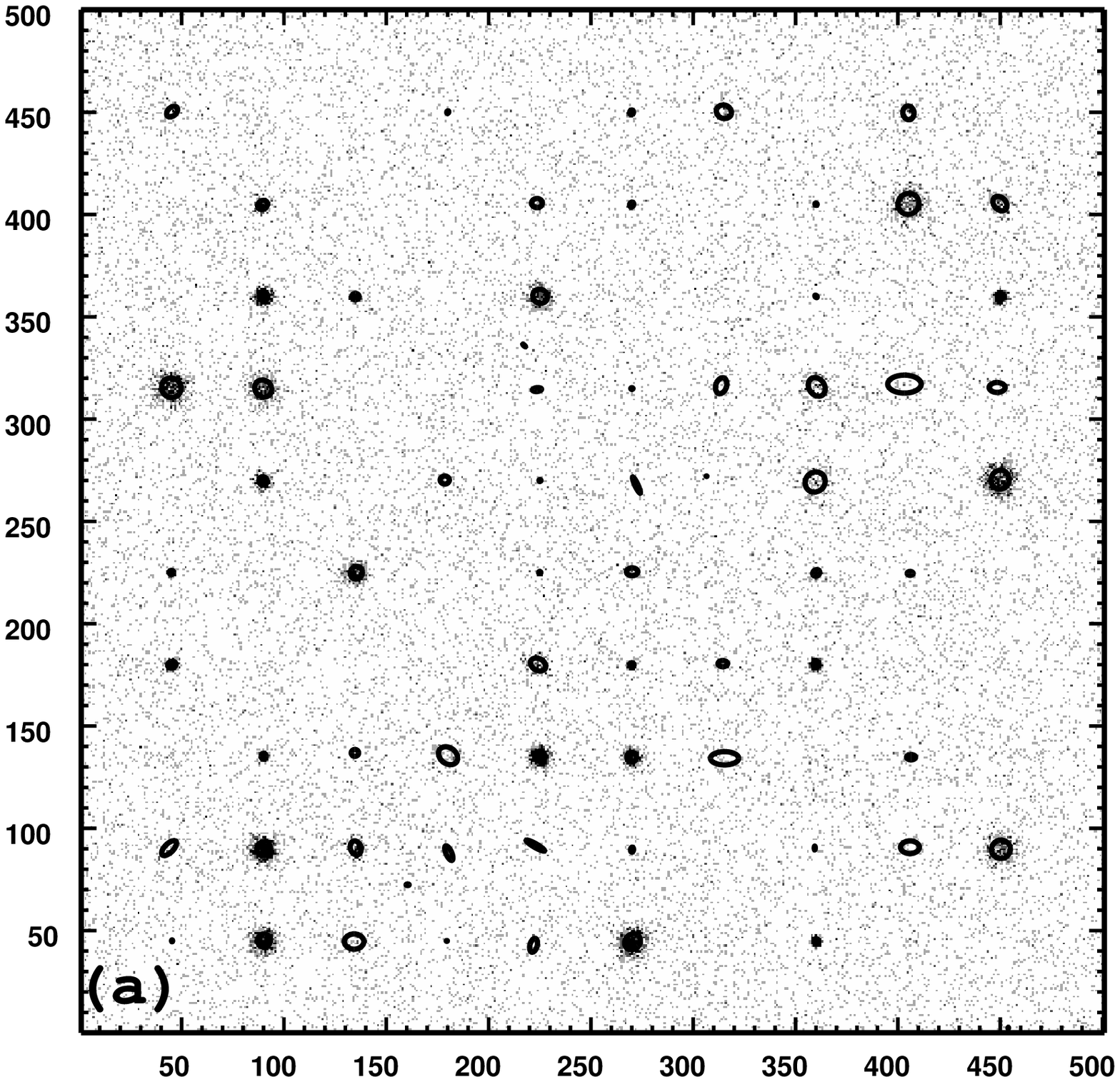} 
\includegraphics[scale=0.35]{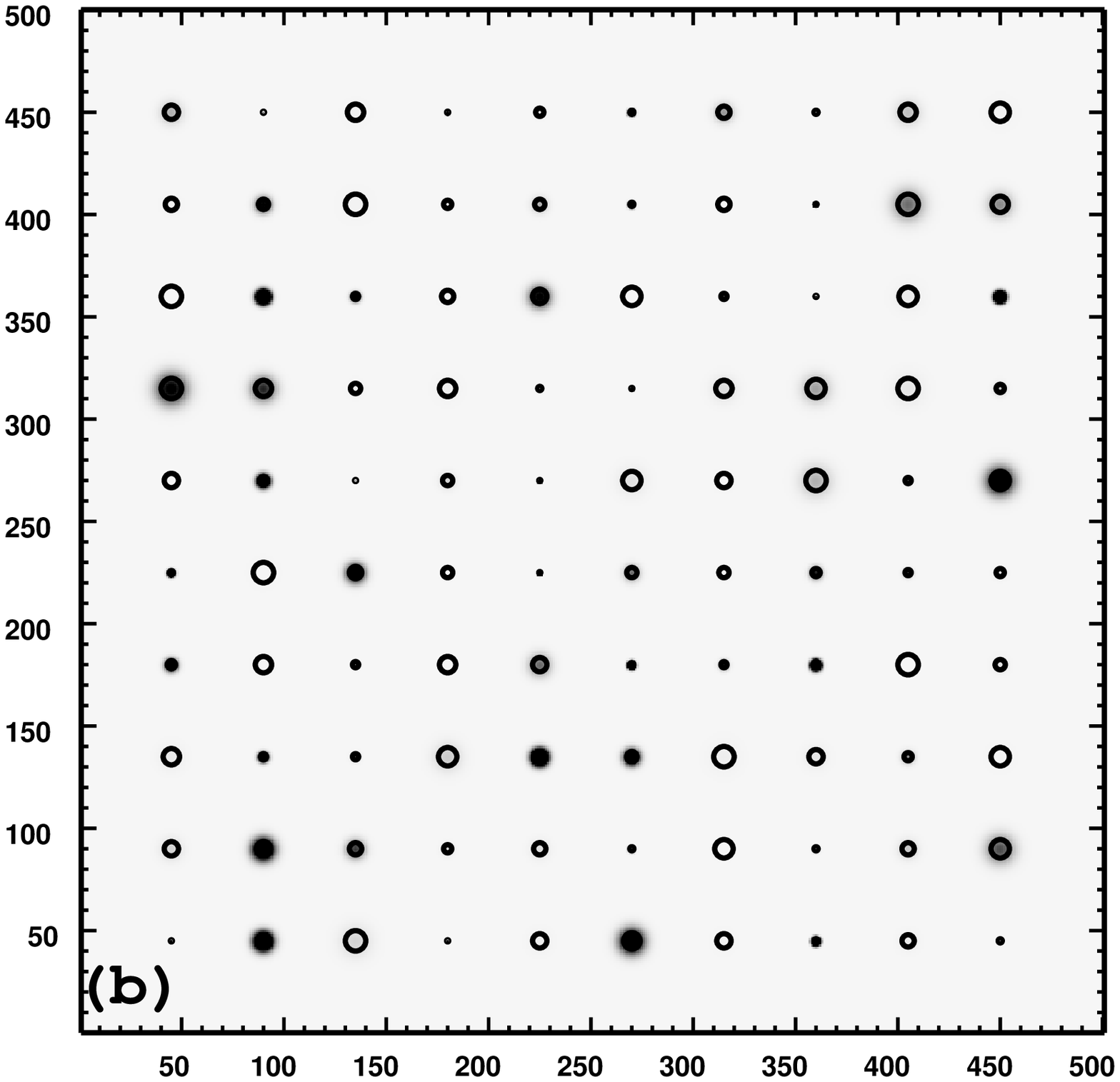}
\caption{
Panels $(a)-(b)$: simulated data with small background with and without
Poisson noise, respectively, and scaled in the range ($0-3$) counts. 
The catalogue obtained employing the {\em exponential prior} pdf {\em (a)} and 
the shapes of the region of the simulated sources {\em (b)} are superposed. 
}
\label{simb01_cata} 
\end{figure*}
Source parameters and their uncertainties are derived as described in Section
\ref{EOSP}. Sources are catalogued when a probability larger than $50$ per
cent is reached at least in one of the SPMs. 
This threshold is chosen within these simulated datasets in order to clarify
the interpretations provided in Table \ref{interp}.

The parameters of bright sources are precisely estimated. 
In Fig.~\ref{simb01_seq53} we give an example of detection of a bright
extended source employing SPMs in the multi--resolution technique. 
The true parameters of this source are: $128$ photon counts,
($\sigma_{\rm x}$,$\sigma_{\rm y}$)=($5$,$5$), ($x$,$y$)=($360$,$270$). 
The estimated parameters of this source are: ($129.79 \pm 23.70$) net source 
counts, ($\sigma_{\rm x}$,$\sigma_{\rm y}$)=($4.92 \pm 0.67$,$4.96 \pm 0.71$) pixels,
($x$,$y$)=($359.57 \pm 0.92$,$269.29 \pm 0.98$) pixels. 
Instead, the effect of background fluctuations on faint source estimates 
can be quite pronounced.

In Fig.~\ref{simb01_cata}, 
we provide an example of the estimated source positions and extensions on 
the simulated data with small background, utilizing the exponential prior pdf. 
The errors on the estimated parameters are not considered in this plot. 
Some of the detected faint sources look uncentered and distorted. 
Four false positives in source detection are found.
The simulated data without Poisson noise with the simulated
source shapes superimposed are shown for comparison. 

Table \ref{summres} reports the number of detected sources for each
simulation. Different columns are used for accounting true detections and 
false positives separately. The number of detected sources employing the
inverse--Gamma function prior is larger with respect to the exponential prior
case only when the cutoff parameter is set lower than the mean background
amplitude. 
\begin{table}
\caption{Source detections on simulated data employing different prior pdfs
  of the source signal.}
\label{summres}
\begin{center}
\begin{tabular}{@{}lccccc}
\hline
 simulated & prior pdf & true & \multicolumn{3}{c}{false positives} \\
  data & & detect & $\ge50\%$ & $\ge90\%$ & $\ge99\%$\\
\hline \hline
\multirow{2}{*}{0.1} & Epr & 64 & 4 & 3 & 1\\ 
 & IGpr & 57 & 7 & 1 & 0\\
\hline
\multirow{2}{*}{1.0} & Epr & 41 & 6 & 3 & 0\\ 
 & IGpr & 42 & 10 & 2 & 0\\
\hline
\multirow{2}{*}{10.0} & Epr & 25 & 0 & 0 & 0 \\ 
 & IGpr & 26 & 2 & 2& 2 \\
\hline
\end{tabular}
\end{center}
\medskip
The background value of the simulated data are reported in counts$\cdot$pixel$^{-1}$. 
$Epr$ and $IGpr$ have same meaning as given in Fig.~\ref{histogr}. 
The term $true$ $detect$ provides the number of detected sources matched with the
simulations. They are catalogued if their probabilities are larger than $50$ per cent.
The number of false positives in source detection are listed
considering $50$, $90$ and $99$ per cent probability thresholds.   
\end{table}

\begin{figure*}
\centering
\includegraphics[scale=0.28,angle=90]{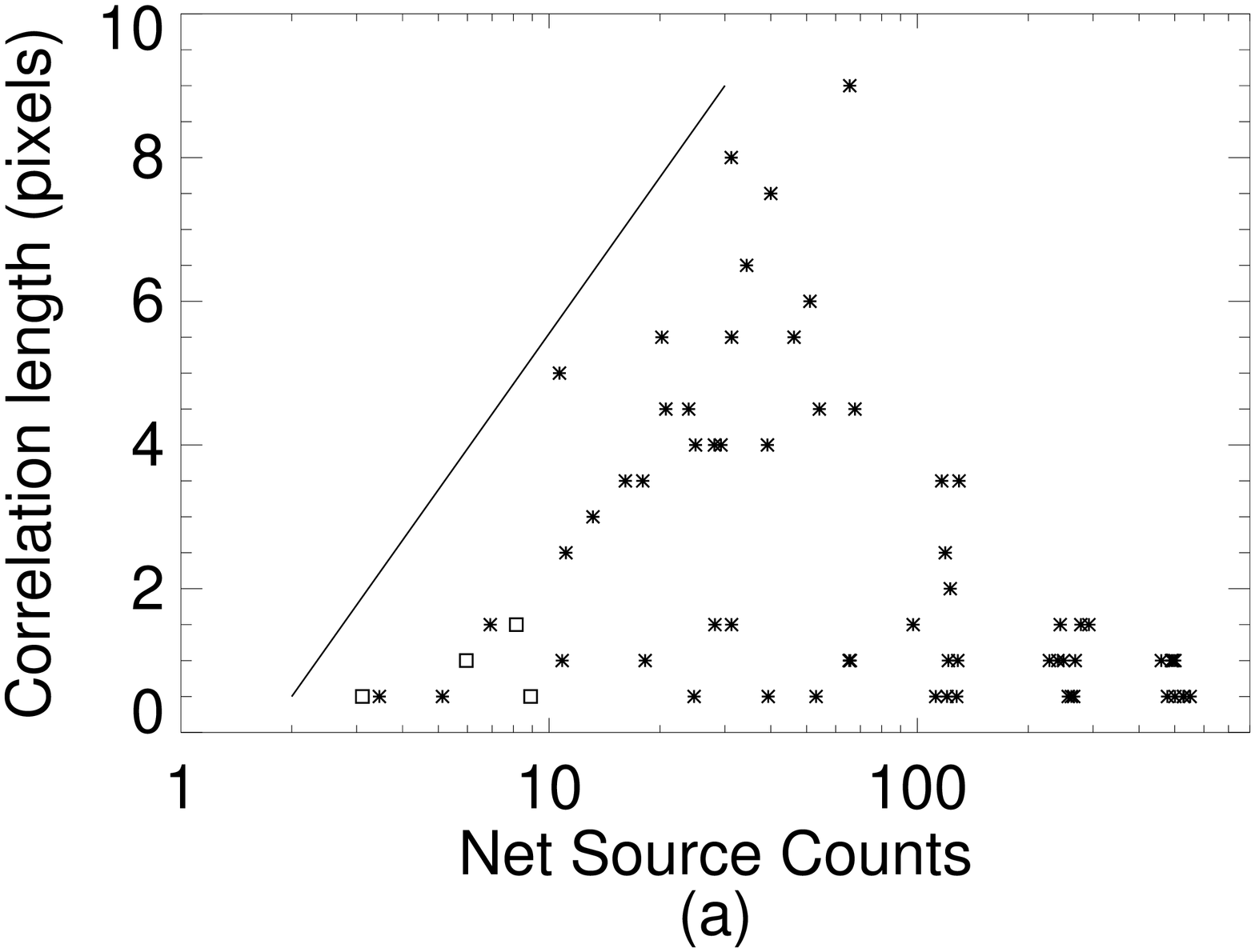}\includegraphics[scale=0.28,angle=90]{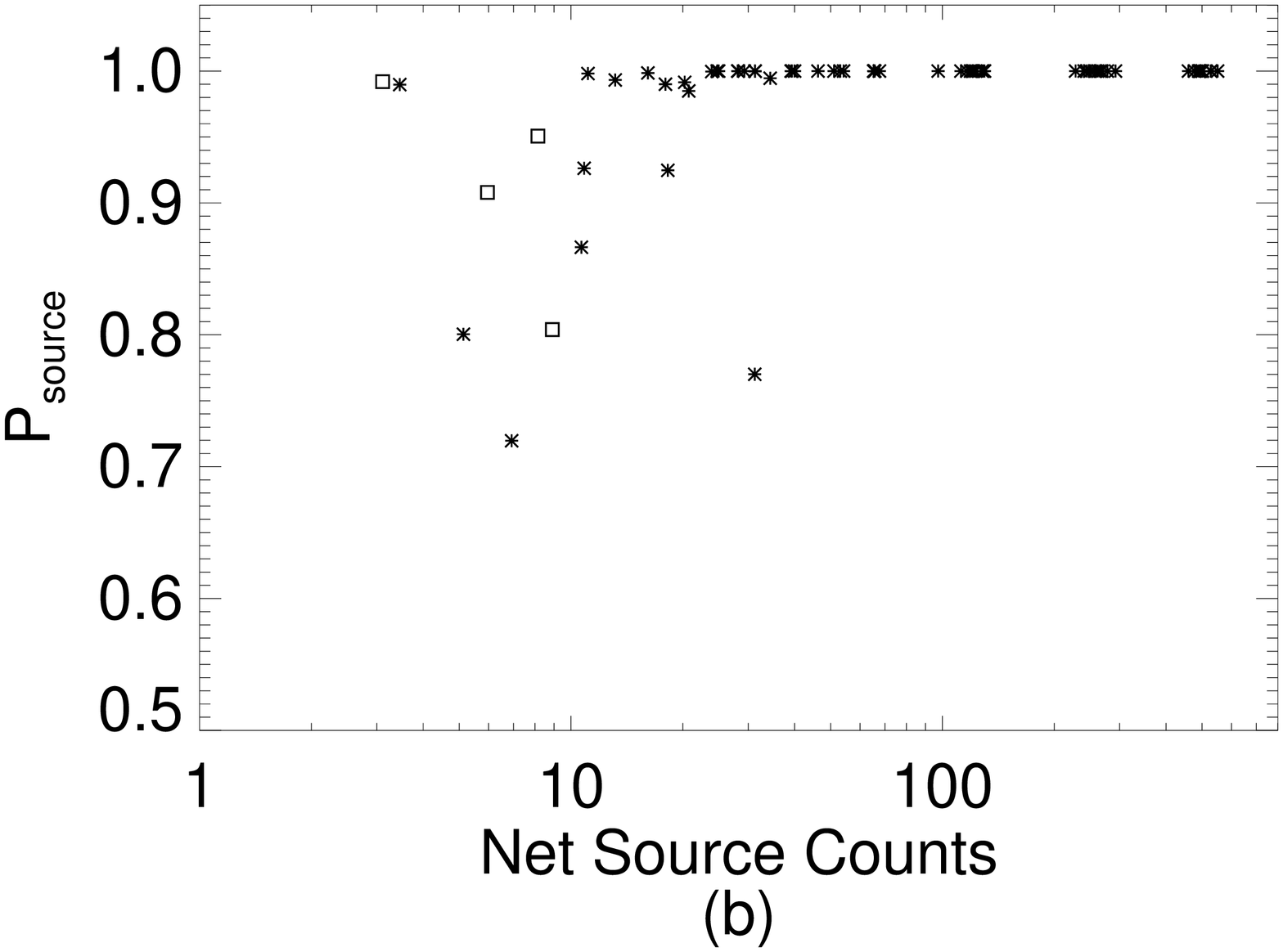} 
\includegraphics[scale=0.28,angle=90]{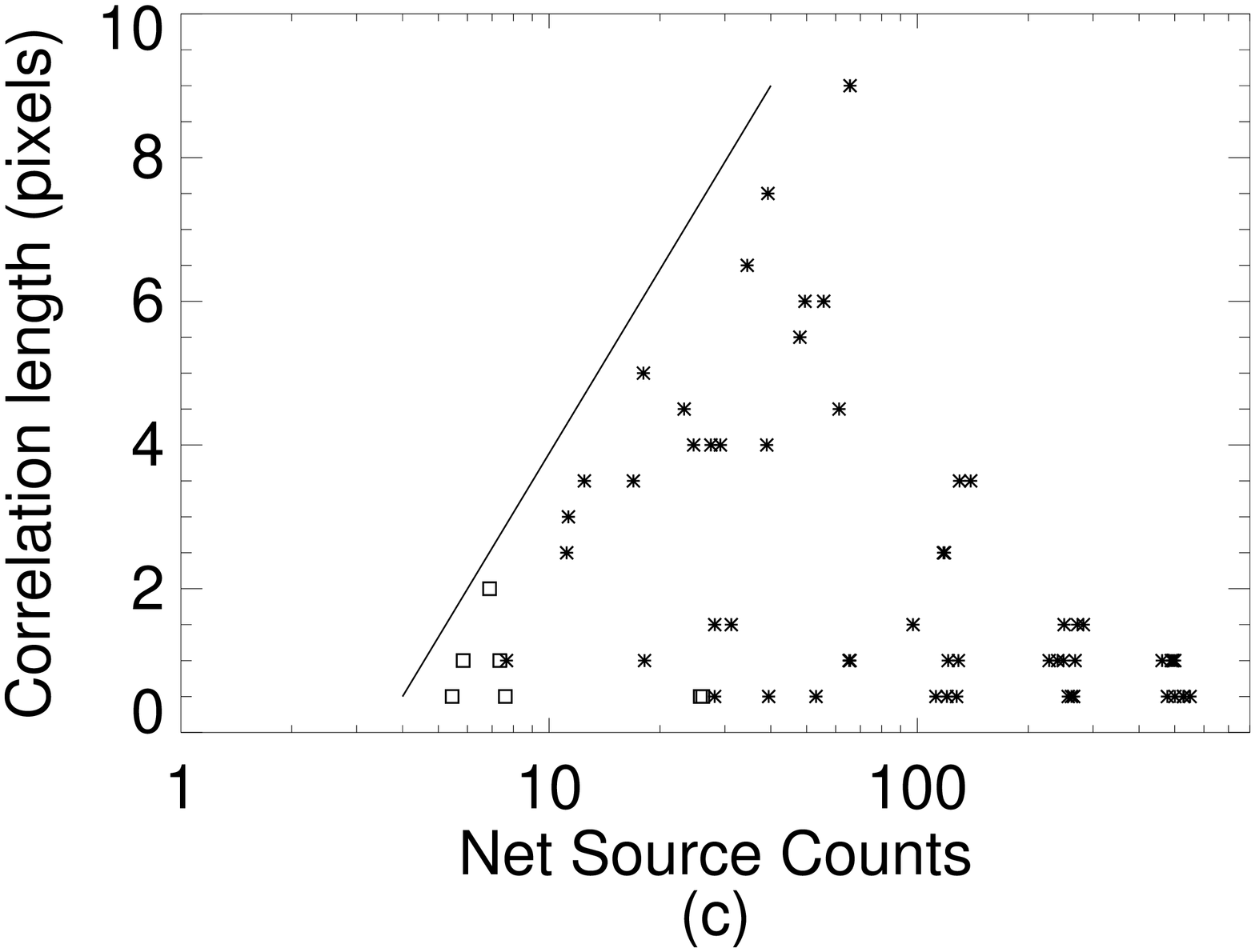}\includegraphics[scale=0.28,angle=90]{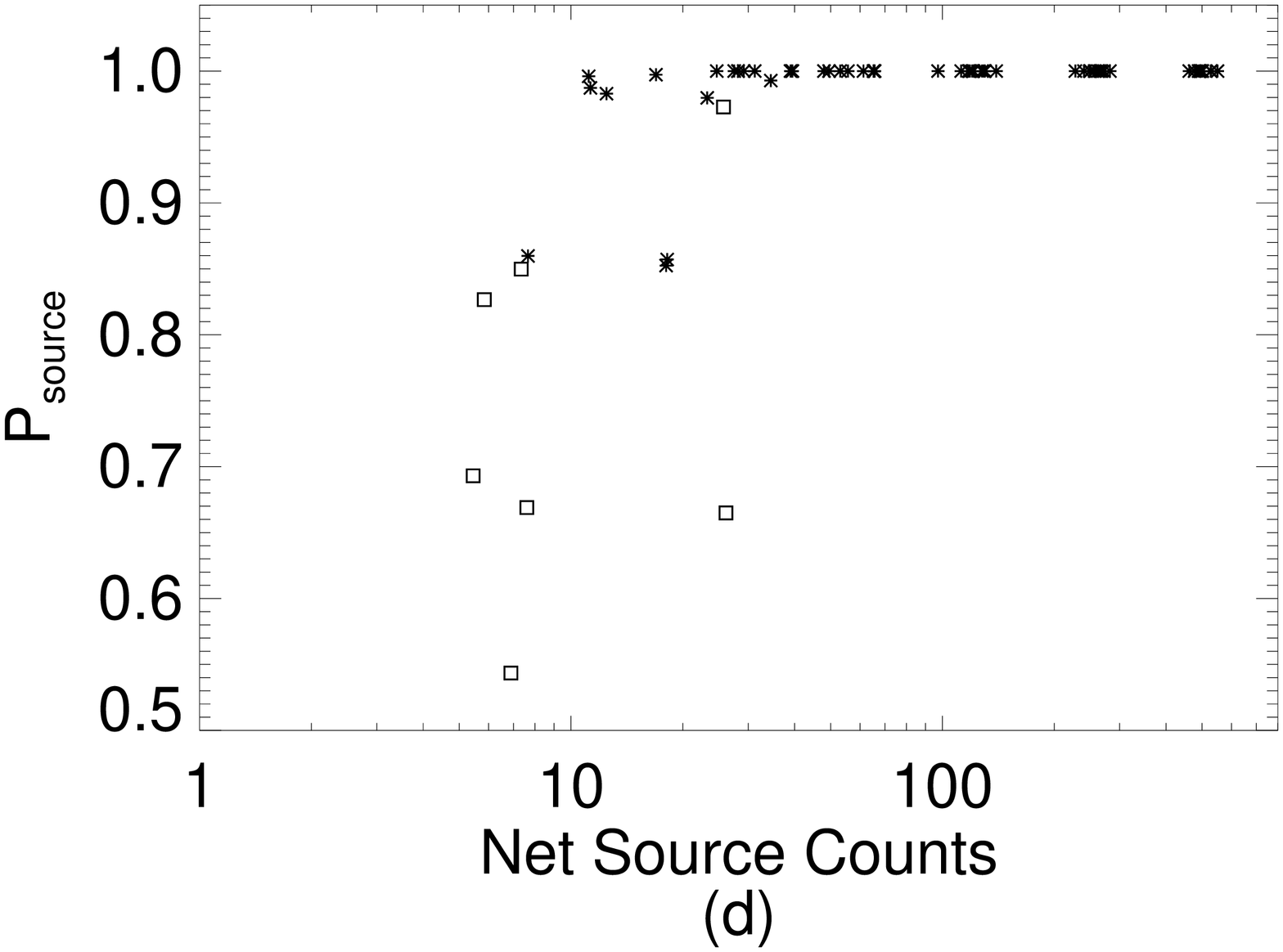}
\caption{Results on simulated
  data with small background employing the {\em exponential prior} pdf (panels $(a)$ and
  $(b)$) and the {\em inverse--Gamma function prior} pdf (panels $(c)$ and $(d)$). 
Panels $(a)$ and $(c)$: correlation length in pixel units versus the net source
  counts. Panels $(b)$ and $(d)$: source probability versus
  net source counts. 
Sources matched with the simulated input catalogue are indicated
  with an asterisk. A square indicates false positives in source detection.
}
\label{simb01_cl} 
\end{figure*}
\begin{figure*}
\begin{center}
\includegraphics[scale=0.255,angle=90]{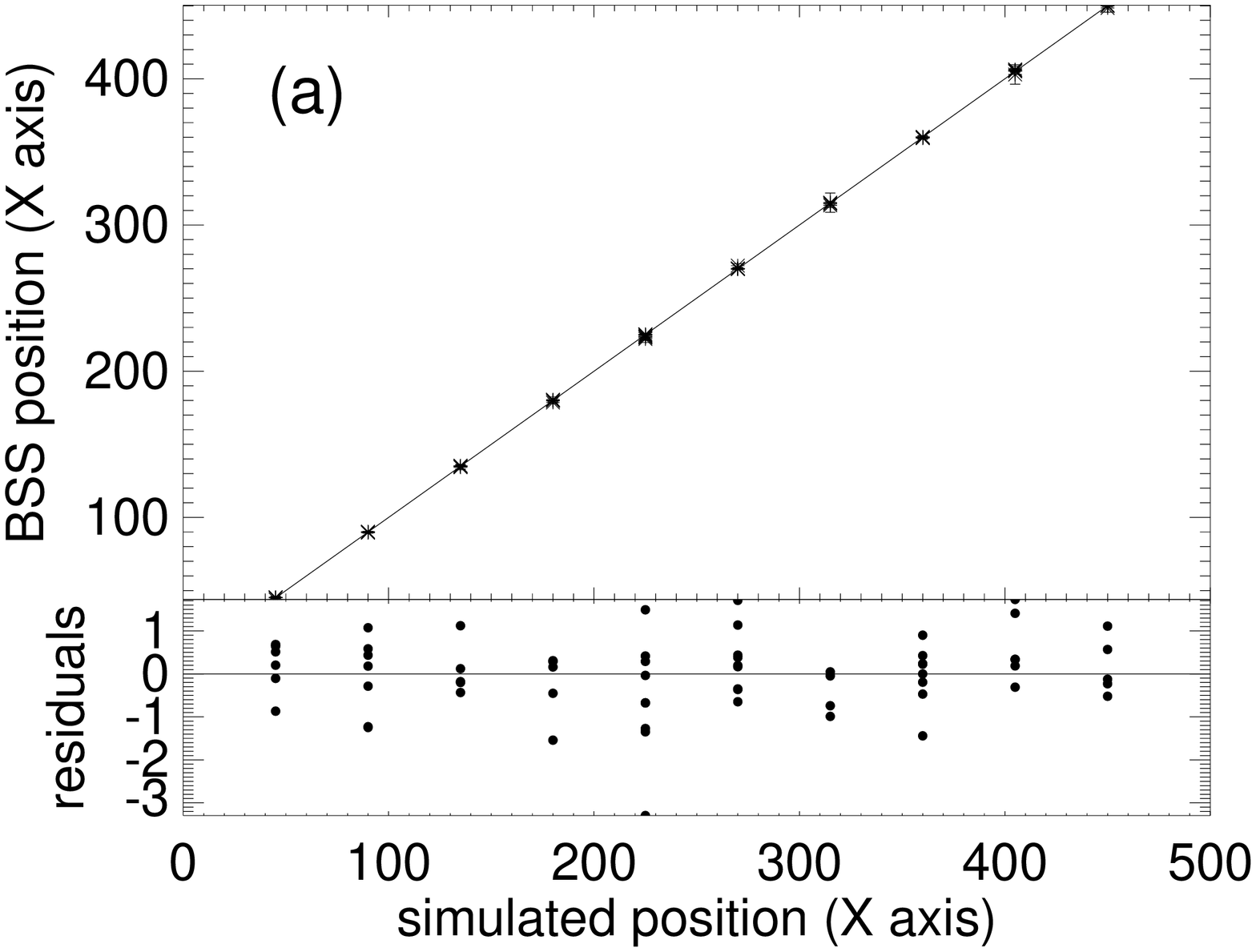}\includegraphics[scale=0.255,angle=90]{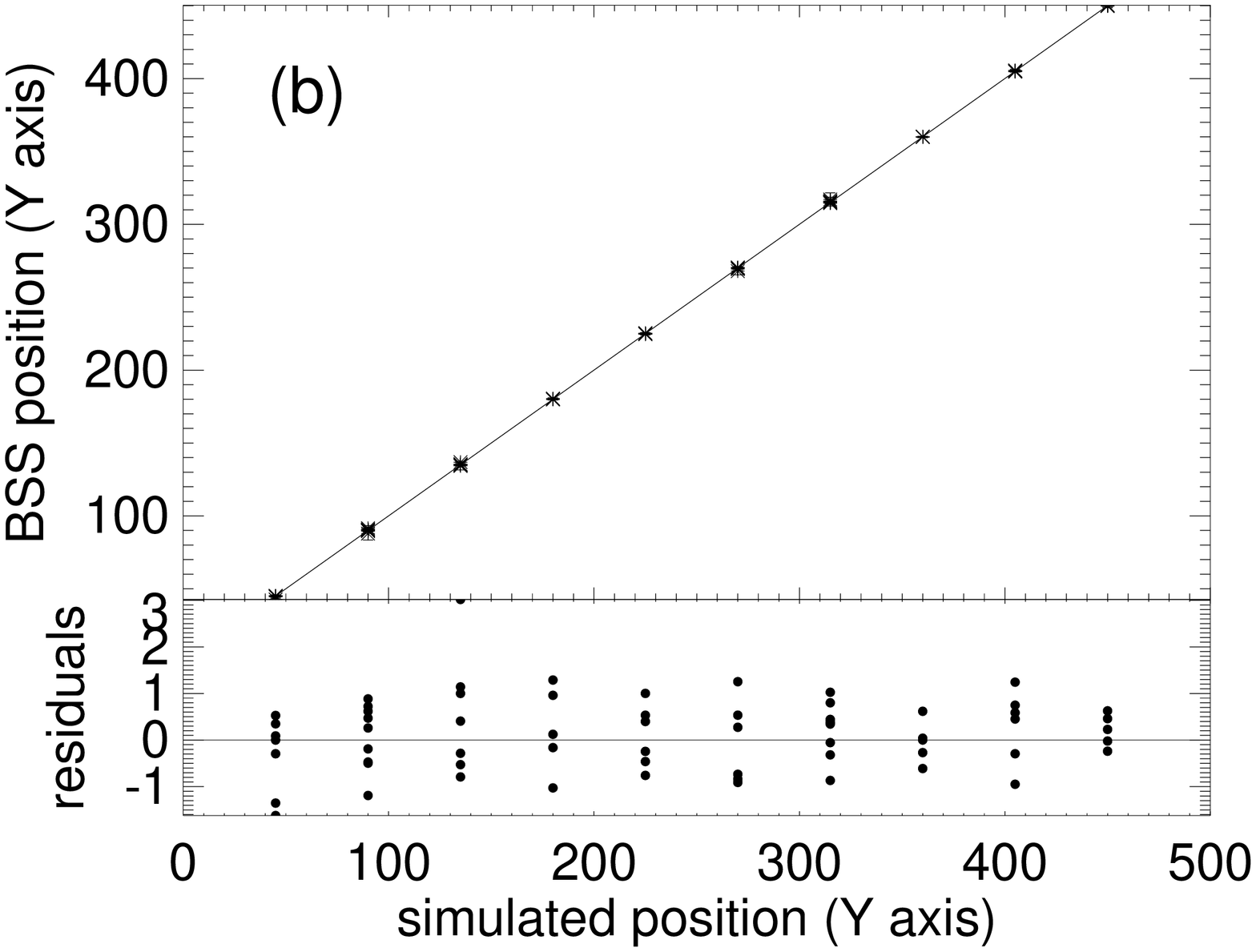}\includegraphics[scale=0.255,angle=90]{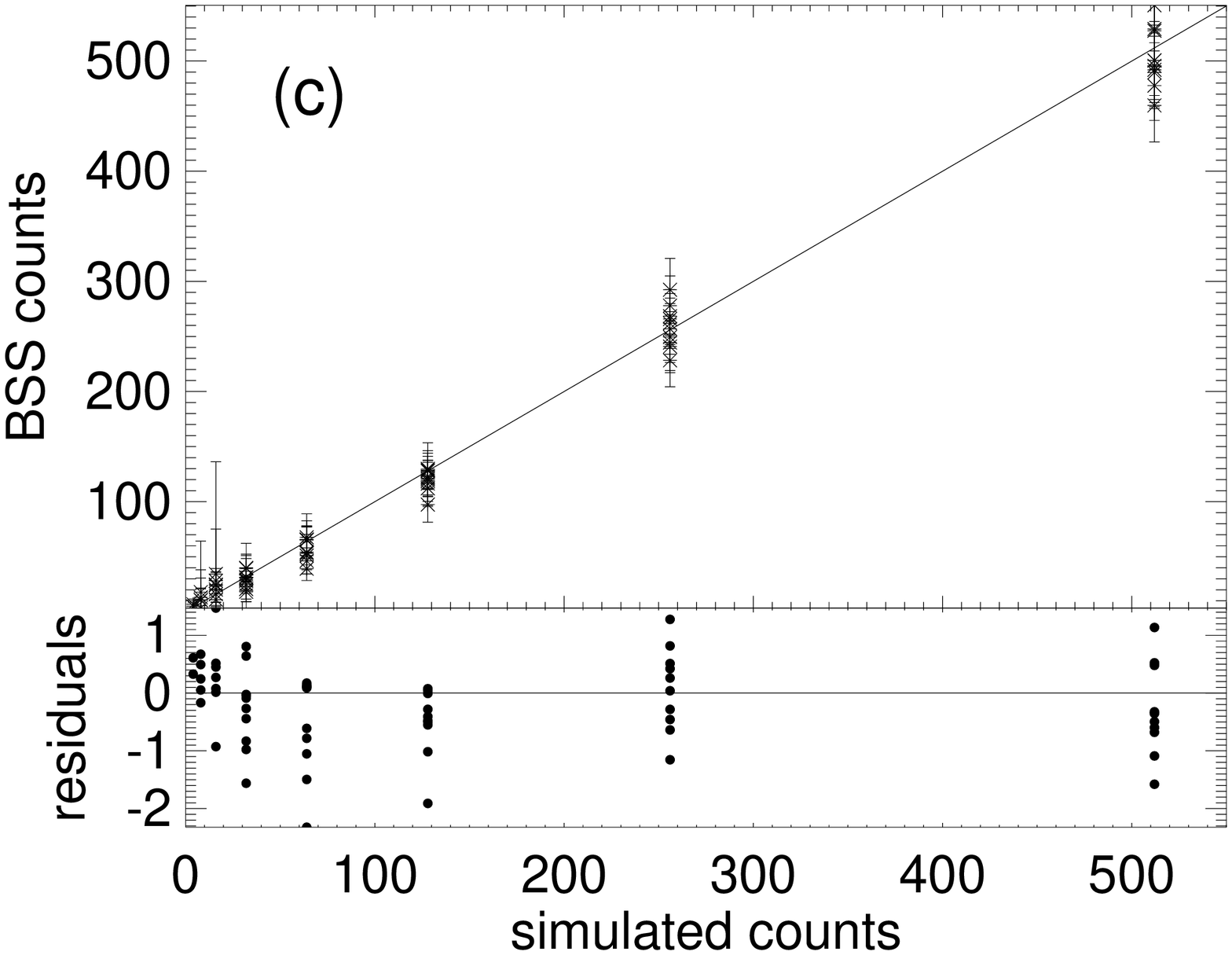}
\includegraphics[scale=0.255,angle=90]{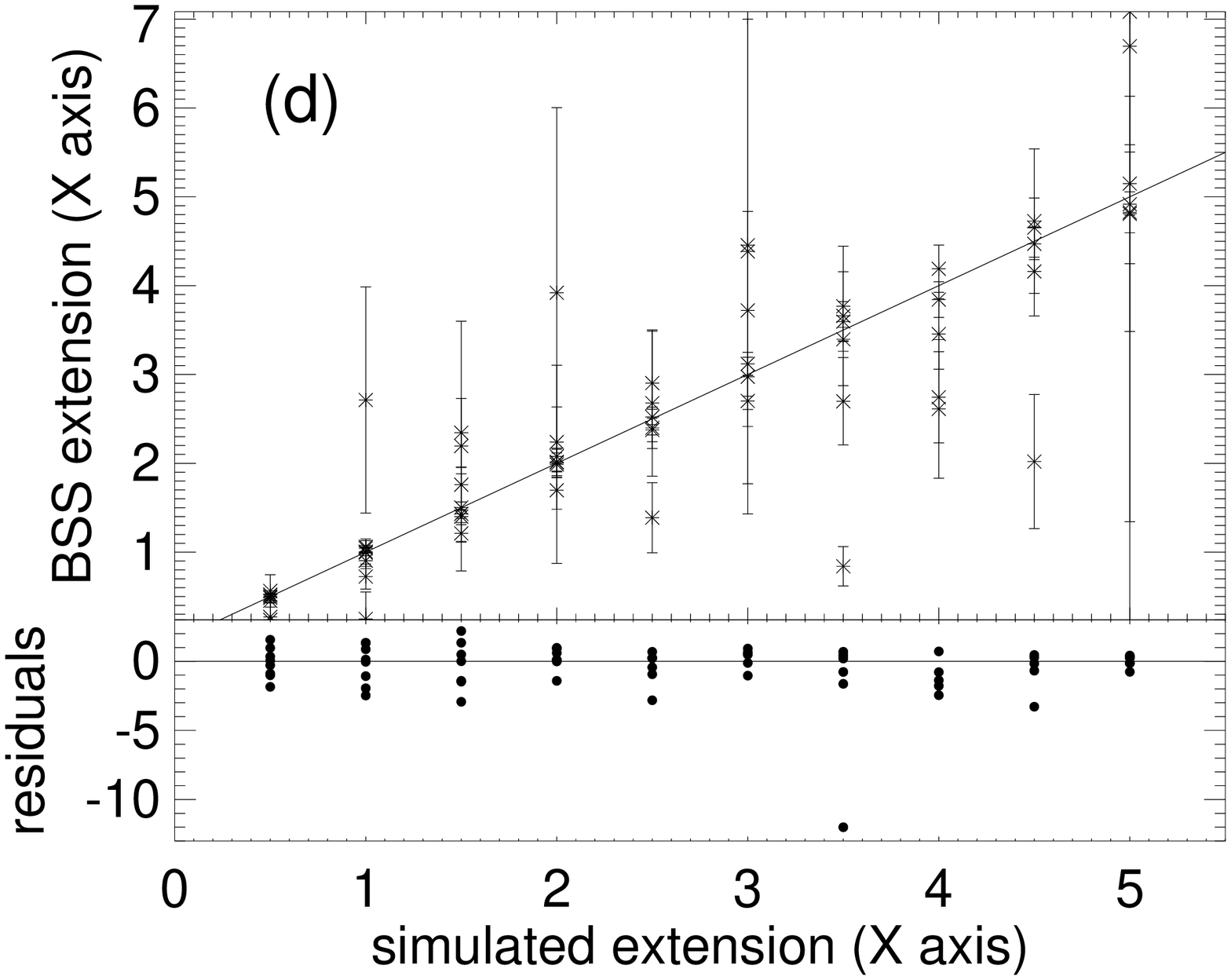}\includegraphics[scale=0.255,angle=90]{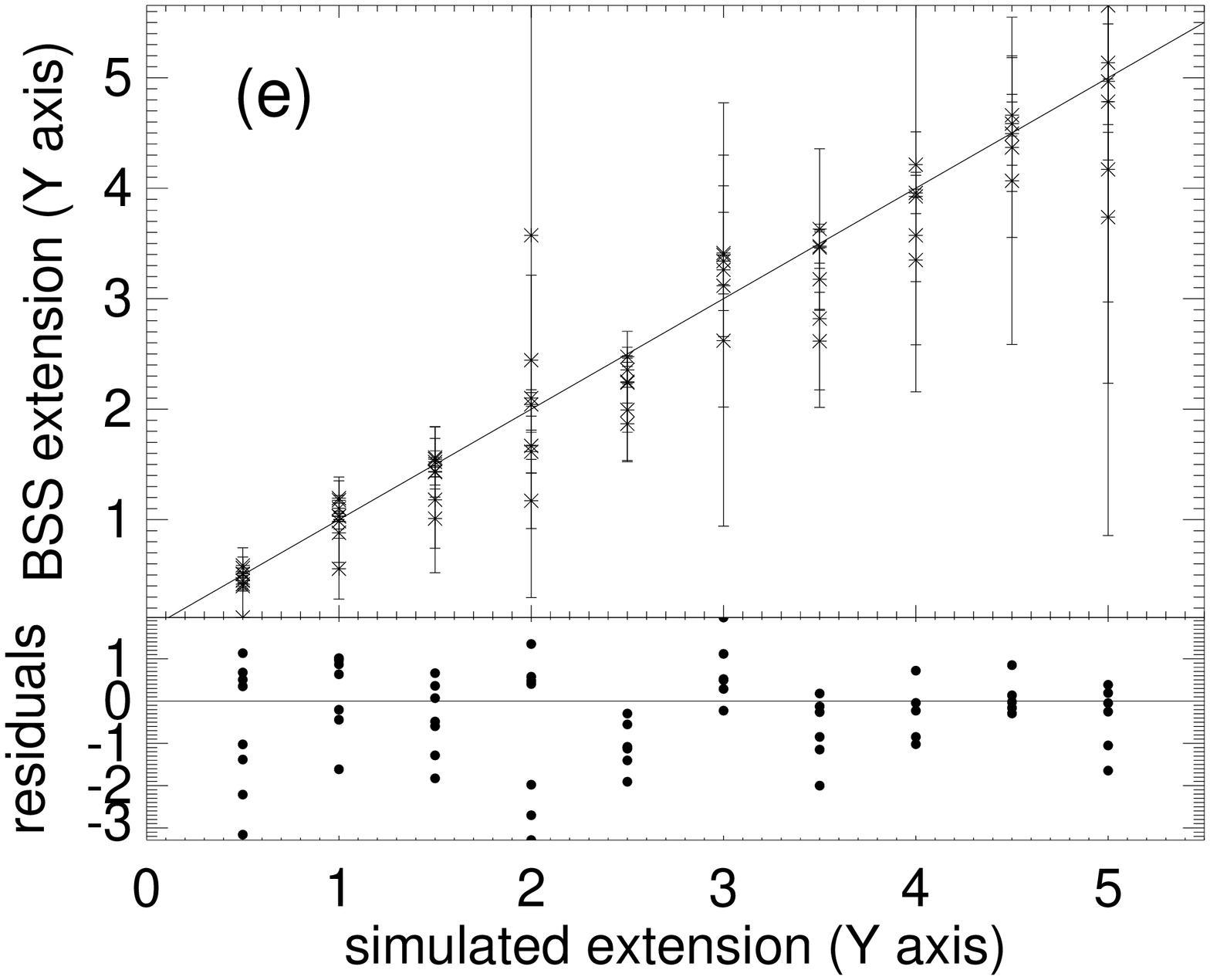}
\end{center}
\caption{Relation between the simulated sources with small background and the 
measured sources as in output
  from the processing with the developed method employing the {\em exponential
  prior} pdf. Plots (a) and (b) show the 
comparison of the measured source positions with the simulated input positions 
on the X axis and on the Y axis. 
Plot (c) displays the relation of the measured source photon
  counts versus the simulated intensities. 
A comparison of the estimated source extensions $\sigma_{\rm x}$ and $\sigma_{\rm y}$ 
with the input source extensions are displayed in the plots (d) and (e). 
The errors estimated for the source parameters are overplotted. 
The error bars on the estimated values denote the 68 per cent confidence 
limit of the corresponding posterior distribution. 
The lower panel in each image shows the fractional residuals.
}
\label{simb01_rel}  
\end{figure*}
In Fig.~\ref{simb01_cl}, 
the estimated source counts are related to their resolution of source
detection (panels $(a)$ and $(c)$) and their source probability (panels $(b)$ and
$(d)$) for the simulated data with small background. These are log--linear plots. 
The results utilizing the exponential and the inverse--Gamma function prior
pdfs are shown in panels $(a)$ and $(b)$ and in panels $(c)$ and $(d)$, respectively. 
The asterisks indicate true sources, while the squares show 
spurious detections.\\
The effect of the cutoff parameter $a$ on source detection is visible in panel
$(c)$. In this example, the value of $a$ is chosen larger than the simulated
background amplitude. Hence, the inverse--Gamma function prior
pdf does not allow to detect sources as faint as the exponential prior. \\  
The plots on panels $(a)$ and $(c)$ denote the source detection technique at several
correlation lengths. In both plots, a line is drawn only with the purpose 
to guide the eye. Left to the line, sources can not be detected. 
Right to the line, sources can be found. Very faint sources with few 
photon counts ($\le 10$) are resolved by small correlation lengths. 
In the range ($10-100$) net source counts, sources are detected at decreasing 
resolutions. Bright sources do not require large correlation lengths for being detected. \\
The plots in panels $(b)$ and $(d)$ of Fig.~\ref{simb01_cl} highlight that
most of the detections occurs at
probabilities larger than $99$ per cent. At this probability value and larger, 
sources matched with the simulated one are strongly separated from false positives.

In Fig.~\ref{simb01_rel} we show the relation between the simulated and the 
estimated source parameters. Good estimates in source parameters are
achieved. The estimated net source counts errors and extension errors can be 
large for faint sources. 

\begin{figure}
\begin{center}
\includegraphics[scale=0.28,angle=90]{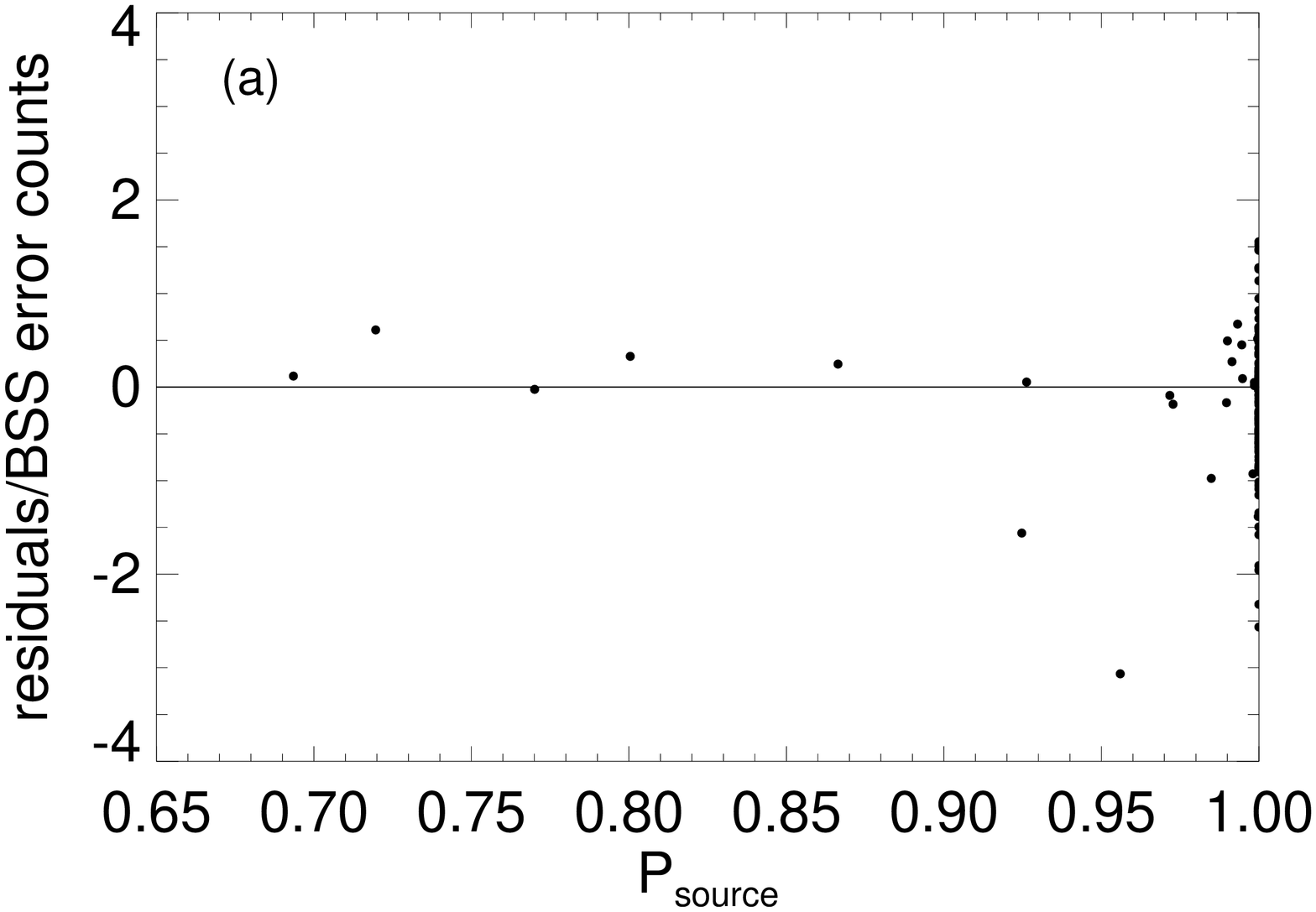}\\ 
\includegraphics[scale=0.28,angle=90]{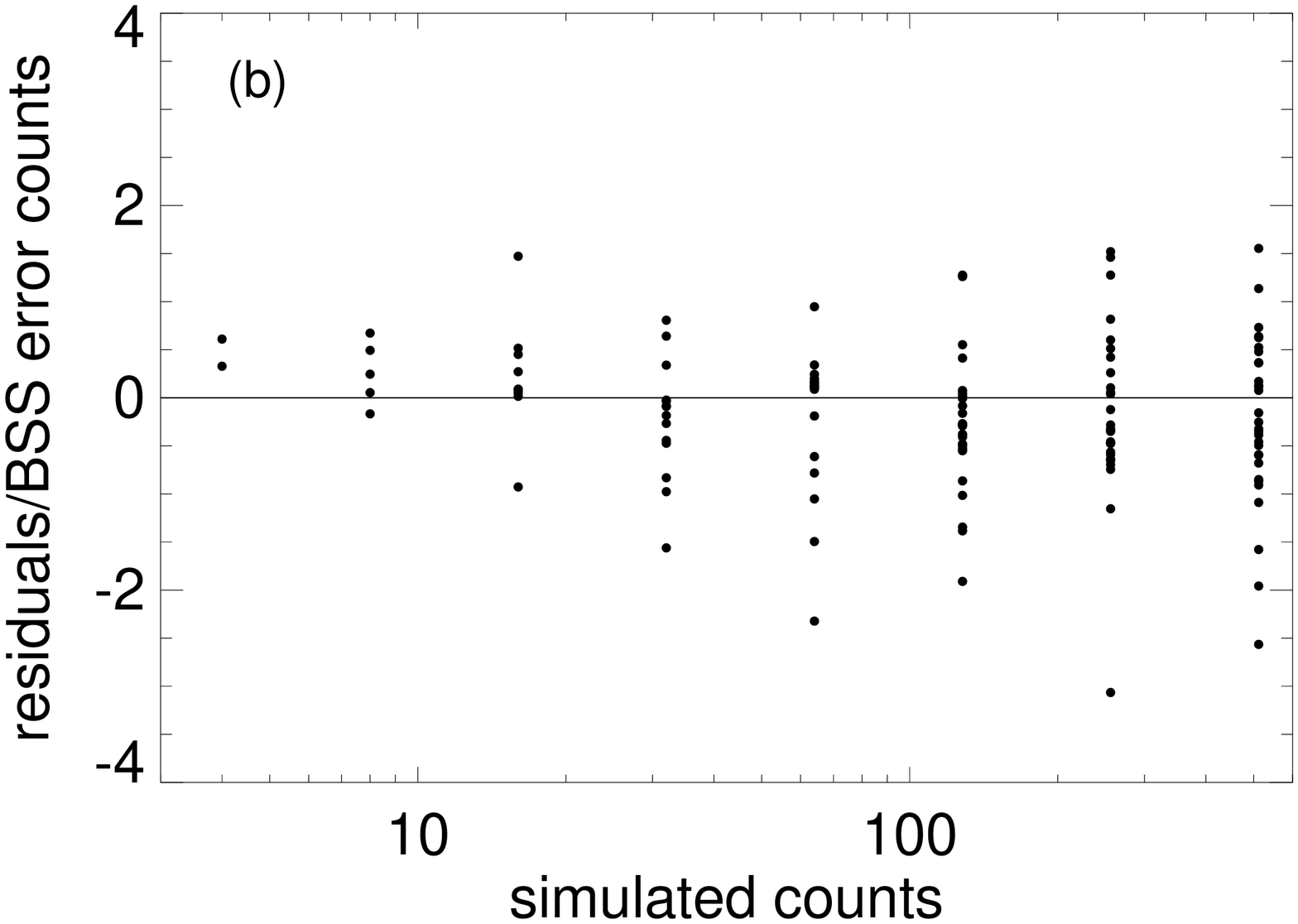}
\end{center}
\caption{Merged information from three simulated datasets of the detected 
sources employing the {\em exponential prior} pdf. 
Panel $(a)$, difference between estimated and simulated net source counts
    normalized by the errors on the estimated net source counts versus source
  probability. 
Panel $(b)$, difference between estimated and simulated
  net source counts normalized by the errors on the estimated net source counts versus
  the simulated counts. 
}
\label{alldata_lambda}  
\end{figure}

In Fig.~\ref{alldata_lambda} a summary on the analysis of all the detected sources
employing the exponential prior on the three simulated datasets is 
presented. 
The plot in panel $(a)$ shows the difference between
estimated and simulated net source counts, normalized with the estimated
errors, versus the source probability of the merged data. 
This plot does not contain the information provided by false positives 
in source detection. $87$ per cent of all detections occurred with
probability larger than $99$ per cent.  
The image in panel $(b)$ is a semi--log plot of the
normalized difference between measured and simulated net source counts versus
the simulated net source counts of $98.5$ per cent true sources detected in the three
simulations. The values of two sources, detected in the simulated data with
large background, are outside the selected $y$ range. These two detections are
included in the analysis of verification with existing algorithms (Section 
\ref{compa}). 
The residuals are normally distributed, as expected. They are 
located symmetrically around zero. At the faint end, the results are
only limited by the small number of simulated faint detectable sources. 
Faint and bright sources are equally well detected. 

\subsubsection{False positives}

Until now we have discussed the detections that have counterparts with the
simulated data. We consider now the detection of false positives. 

In Table \ref{summres} the number of detected false positives are listed for each
simulation. At $50$ per cent probability threshold more false positives are found with the
inverse--Gamma function prior compared to the exponential one.  
At a $90$ per cent source probability threshold, the analyses
with the two prior pdfs provide similar results.  
True detections are strongly separated from statistical fluctuations for 
source probability values larger than $99$ per cent. 

When employing the inverse--Gamma function prior pdf, the number of false
positives is sensitive to the cutoff parameter. Less  
false positives are found when $a$ is set larger than the background, because 
it reduces the number of detectable faint sources. It may be worth noting that
even when the cutoff parameter is set larger than the background, a 
probability threshold $\ge 90$ per cent has to be considered (see
Fig.~\ref{simb01_cl} for more details).

False positives in source detections show large
errors in their estimated parameters. The source probability variation and the
source features analyses in the multi--resolution technique provide hints 
of ambiguous detections. 
However, as all methods, the BSS approach is limited by statistics. 
Spurious detections can not be ruled out.

\subsubsection{Choice of the prior pdf of the source signal}
The big difference between the two prior pdfs of the source signal follows on
from one prior pdf having one parameter and the other pdf having two. 

The parameter $\lambda$, indicating the mean intensity in an astronomical
image, introduced with the exponential prior pdf is estimated from the data.

The parameter $\alpha$, that is the shape parameter of the power--law,
given by the inverse--Gamma function prior pdf is estimated from the data. 
Instead the cutoff parameter $a$ is selected to a small value such that
the inverse--Gamma function prior pdf behaves as a power--law. 
Astronomical images can be characterized by a small background. 
It results that $a$ can be chosen from a number of alternatives, ranging from 
values that are above or below the background amplitude. The choice of $a$ 
implies a selection on the detectable sources:    
sources whose intensity is lower than $a$ are not detected;  
sources close to the background amplitude are detected when $a$ is set 
below the background amplitude. 

On real data much more prior information for the cutoff parameter is needed. 
The inverse--Gamma function prior pdf can be employed if a mean value of the 
background amplitude is already known from previous analyses. 

The exponential prior pdf is preferable over the inverse-Gamma function prior
pdf, since no predefined values are incorporated. 
This is also supported by the results obtained with the simulated data. 
However, the inverse-Gamma function prior is a more suited
model to fit the data and it has potentials for improving the detections of 
faint objects. 

One way to improve the knowledge acquired with the inverse-Gamma
function prior is by the estimation of the cutoff parameter from the
data. 
This change in our algorithm is not straightforward and MCMC algorithms have to be
employed. 

\subsection{Summary}\label{summar}
Simulated data are employed to assess the properties of the BSS
technique.
The estimated background and source probabilities depend on the
prior information chosen.
A successful separation of background and sources must depend on the
criteria which define {\em background}.
Structures beyond the defined properties of the background model are,
therefore, assigned to be sources.
There is no sensible background-source separation without defining a
model for the object {\em background}. 
Additionally, prior information on source intensity distributions helps 
to sort data, which are marginally consistent with the background model,
into background or source. 
Therefore, prior knowledge on the background model as well as on the
source intensity distribution function is crucial for successful
background-source separation.

For the background model a two--dimensional TPS representation was
chosen.
It is flexible enough to reconstruct any spatial structure in the background
rate distribution.
The parameters are the number, position and amplitudes of the spline
supporting points.
Any other background model capable to quantify structures which should
be assigned to background can be used as well. 

For the prior distribution of the source intensities the exponential and
the inverse--Gamma function are used as illustrations. 
For both distributions the source probability can be given analytically.
The hyper--parameters of both distributions can either be chosen in
advance to describe known source intensity properties or can be
estimated from the data.
If they are estimated from the data simultaneously with the background
parameters, properties of the source intensity distribution can be
derived, but at the expense of larger estimation uncertainties.
It is important to note that the performance of the BSS method
increases with the quality of prior information employed for the 
source intensity distribution. The prior distribution of the source intensities   
determines the general behaviour of the sources in the field of view and the
hyper--parameters are useful for fine--tuning.

The aim of detecting faint sources competes with the omnipresent
detection of false-positives.
The suppression of false-positives depends both on  
the expedient choice of prior information and on the level of detection
probability accepted for source identification. 
Compared to, e.g., p-values the BSS technique is rather conservative in
estimating source probabilities.
Therefore, a probability threshold of 99 per cent is mostly effective to
suppress false positives. 

The estimated background rates are consistent with the simulated ones.
Crowded areas with regions of marginally detectable sources might
increase the background rate accordingly. 

The SPMs at different correlation lengths are an important feature 
of the technique. The multi--resolution analysis allows one to detect fine 
structures of the sources. 

The source parameters are well determined.
Their residuals are normally distributed. We will show in Section \ref{compa} 
that the BSS technique performs better than frequently used techniques.
Naturally, the estimation uncertainties of parameters for faint sources
are large due to the propagation of the background uncertainty.

\section{Verification with existing algorithms} \label{compa}
In the {\it X}--ray regime, the sliding window technique and the WT 
techniques are widely used. 
However, the WT has been shown to perform better than the sliding window 
technique for source detection. 
The WT improvement in source detection with respect to the sliding window
technique is inversely proportional to the background amplitude 
(\citealt{freeman:2002}). The WT has also other favourable aspects for being
compared with the BSS method developed in our work. The WT allows for the search of faint 
extended sources. The BSS method with the multi--resolution technique is
  close to the WT method. 

Between all the available software employing WT, we choose {\sc wavdetect} 
(\citealt{freeman:2002}), part of the freely available {\sc CIAO} software package.
We use version $3.4$. 

{\sc wavdetect} is a powerful and flexible software package. 
It has been developed for a generic detector. 
It is applicable to data with small background. 
The algorithm includes satellite's exposure variations. It estimates the local
background count amplitude in each image pixel and 
it provides a background map. 

We utilize {\sc wavdetect} on our simulated data described in Section \ref{simu_data}. 
The threshold setting for the significance (`sigthresh') is chosen equal to 
$4.0\times10^{-6}$, in order to detect on the average $1$ spurious source 
per image. The `scale' sizes are chosen with a logarithmic increment from
$2$ to $64$. 
Tests have been made changing the levels of these parameters,
assuring us that the selected values provide good performance of this WT
technique. In Table \ref{wavres}, we report the number of detected sources per
simulated field, separating the sources matched with the simulated one ($true$
$detect$) to the false positives in source detection found employing the above
mentioned threshold setting. The three simulated
datasets are distinguished by their background values (counts). The simulated background
values are reported in column $simulated$ $data$. 
These results are compared with the
ones obtained with the BSS algorithm when employing the exponential prior pdf 
as shown in Table \ref{summres}. 
\begin{table}
\caption{Number of detected sources employing {\sc wavdetect} on three simulated
  datasets.} 
\label{wavres}
\begin{center}
\begin{tabular}{@{}lcc}  \hline 
simulated & true detect & false \\ 
 data & & positives\\
\hline \hline
 0.1 & 56 & 4 \\ \hline
 1.0 & 37 & 1 \\ \hline
 10.0 & 23 & 1 \\ \hline
\end{tabular}
\end{center}
\medskip
The explanation of these columns can be found in Table \ref{summres}. 
\end{table}

\begin{figure*}
\centering
\includegraphics[scale=0.28,angle=90]{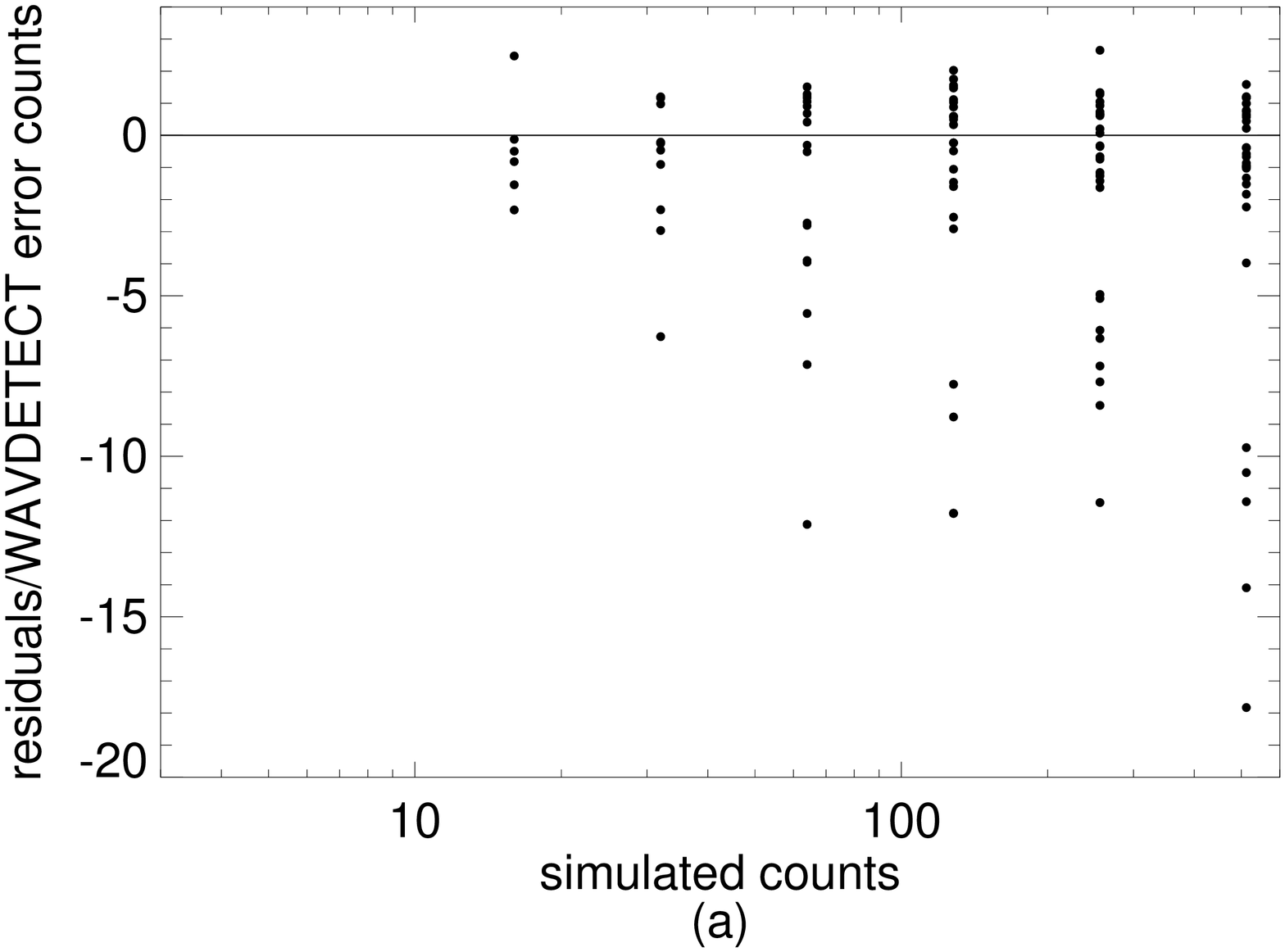}\includegraphics[scale=0.28,angle=90]{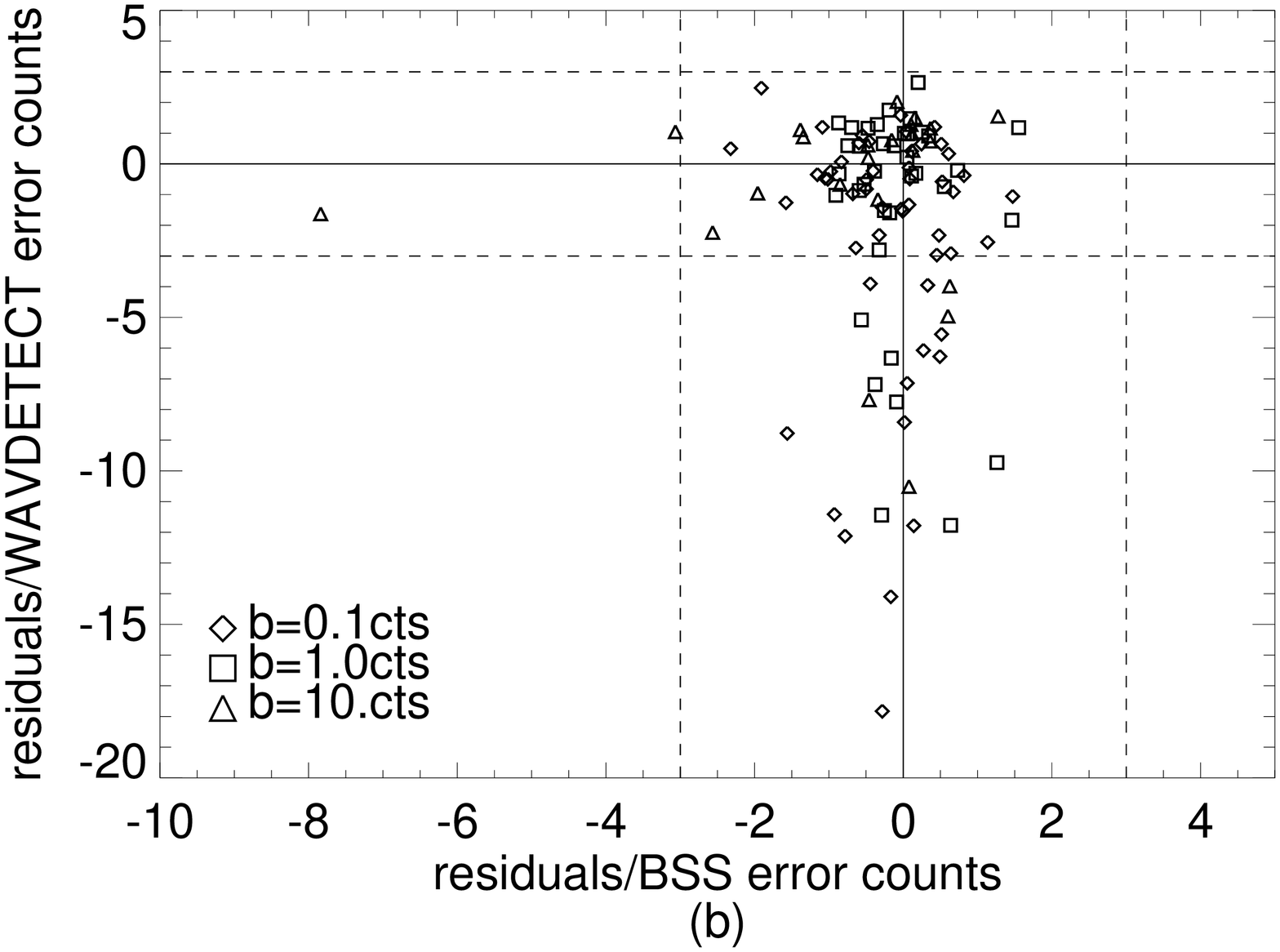} 
\caption{Panel $(a)$: normalized difference between {\sc wavdetect} estimated source
  counts versus simulated counts of all the detected sources in the three
  simulated datasets. Panel $(b)$: comparison between the
  normalized differences of {\sc wavdetect} source counts and simulated counts, on
  the ordinate, and the normalized difference of BSS source counts and
  simulated counts, on the abscissa. Sources detected from the simulated data
  with small background are indicated with a diamond. A square is used to
  highlight sources detected in the dataset with intermediate background. The
  detected sources in the simulated data with large background are indicated
  with a triangle. Dashed lines are drawn as a borderline of the $\pm 3
  \sigma$ detection. The zero line is indicated with a continuous
  linestyle.   
}
\label{compa_res}
\end{figure*}
\begin{figure*}
\centering
\includegraphics[scale=0.28,angle=90]{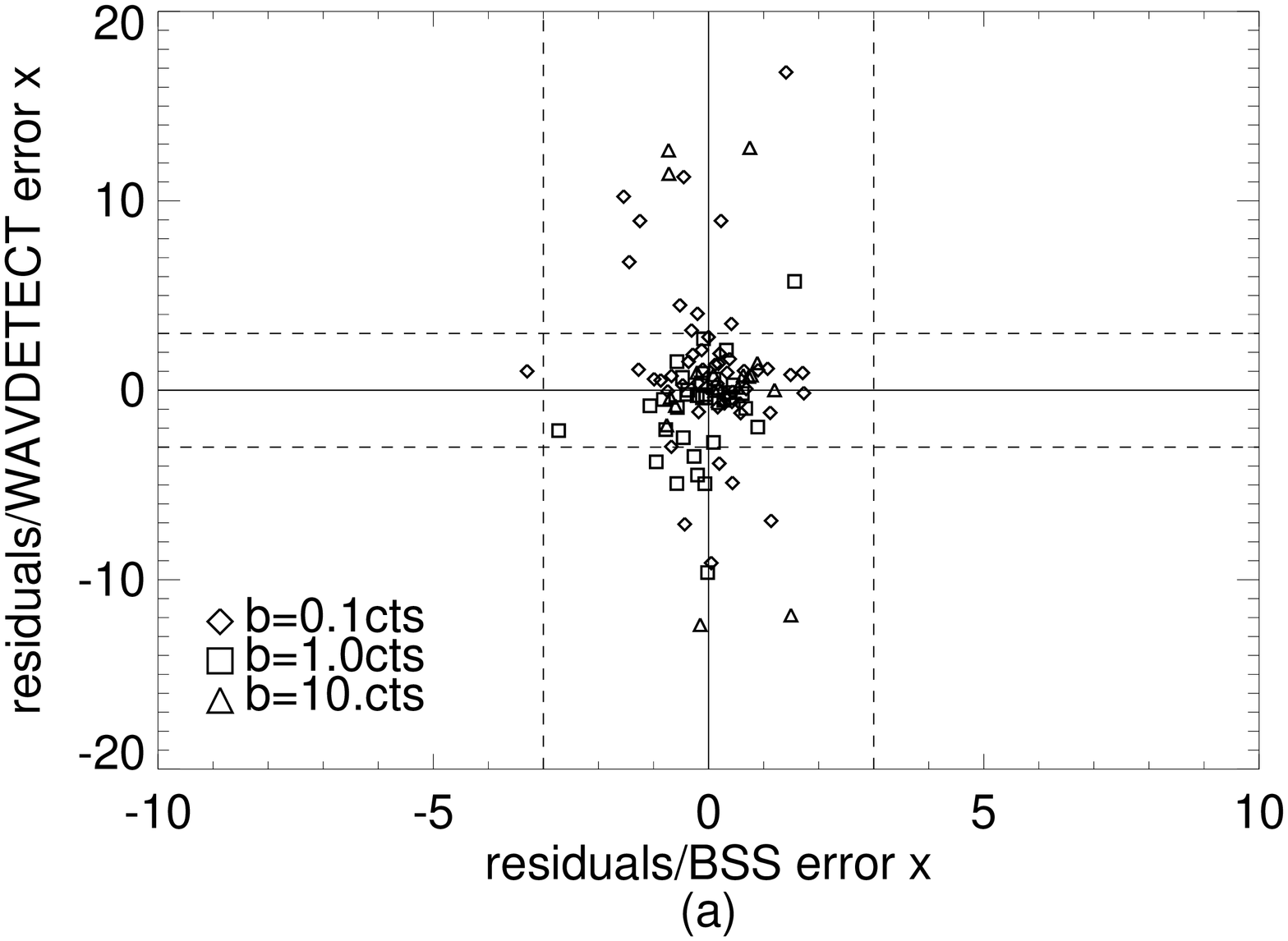}\includegraphics[scale=0.28,angle=90]{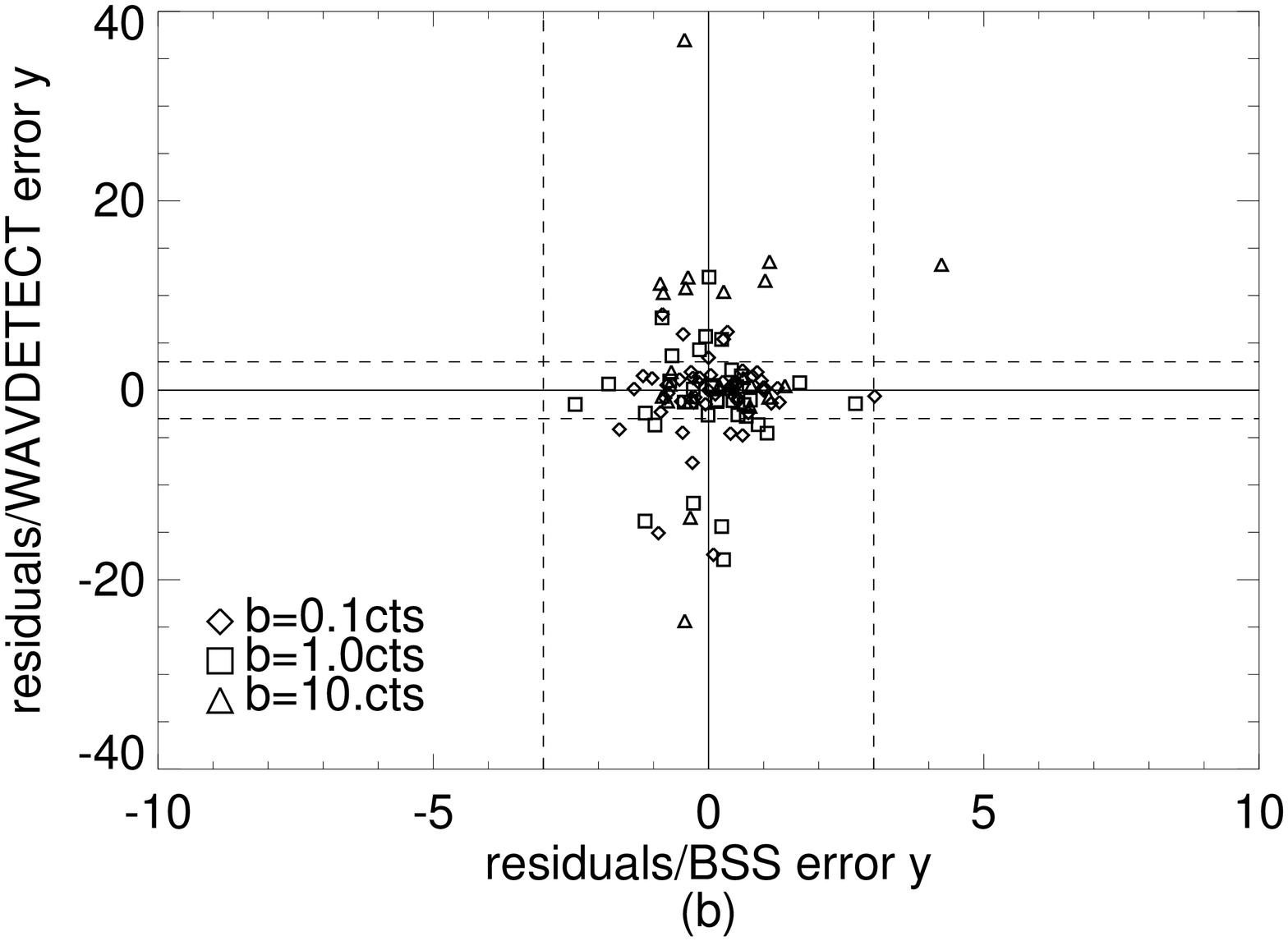} 
\caption{Same explanation as for panel $(b)$ in Fig.~\ref{compa_res}, but
  comparing the estimated source positions: $x$ axis (panel $(a)$), $y$ 
  axis (panel $(b)$).  
}
\label{compa_res2}
\end{figure*}
The BSS technique finds all sources detected by {\sc wavdetect}. In the simulated data 
characterized by small background, the two algorithms find the same number of 
false positives. 
The BSS algorithm reveals $8$ per cent more true detections than {\sc wavdetect}. These sources are
characterized by counts in the range ($4-8$). 
Hence, the BSS technique performs better than {\sc wavdetect} in the low number counts 
regime. 
The performance of the two techniques on the datasets with intermediate and
large background values is similar. 

The explanation for these results is enclosed in the background estimate. 
The {\sc wavdetect} estimates of background values are similar to the results
obtained with the BSS technique in the intermediate and large background datasets. Though, the
backgrounds provided by {\sc wavdetect} present rings due to the Mexican Hat
function employed as a filter on the image data. In the simulated data with small background, 
the {\sc wavdetect} background model is on the average larger than the one estimated
with the BSS method. 
The plots in Fig.~\ref{compa_res} support these conclusions (semi--log plots). 
The image in panel $(a)$ shows that {\sc wavdetect} fluxes are underestimated for
$\sim20$ per cent of all
detected sources. In addition, {\sc wavdetect} sensitivity for source detection is limited to $16$
counts per source within these simulated datasets. In
Fig.~\ref{alldata_lambda}, panel $(b)$ presents the sensitivity achieved
by the BSS method. 
The plot in panel $(b)$ of Fig.~\ref{compa_res} displays a relation between
the sources detected by {\sc wavdetect} (ordinate) and by BSS (abscissa), both
matched with the simulated data. Most of the {\sc wavdetect} underestimated sources
are coming from the simulated data with small background. BSS presents only two
sources underestimated and detected in the simulated data with large
background. By chance, the triangle located at ($-8$,$-2$)
indicates the detection of two sources. Both sources where simulated
with $256$ source counts and a circular extension of $4$ pixels one, $5$
pixels the other. 
The estimated source positions are also improved 
with BSS (Fig.~\ref{compa_res2}, semi--log plots). \\
The residuals provided by the BSS technique are a factor of $10$ smaller than the ones from
{\sc wavdetect}.
Their estimates have many outliers. Our estimates are normally distributed. 

Though the comparison between the two detection methods is not yet
  carried out on real data, these results are encouraging. The BSS method detects at least as many 
sources as {\sc wavdetect}. The simulations prove that the 
developed Bayesian technique ameliorates the detections in the low count
regime. The BSS estimated positions and counts are improved. Finally, we expect that 
the BSS technique will refine {\sc wavdetect} sensitivity on real data, because the BSS technique
is designed for modelling highly and slowly varying backgrounds taking into account
instrumental structures. 

\section{Application to observational data} \label{odata}

In this Section we employ RASS data to show the effectiveness of our method 
at large varying background, exposure values and  
where exposure nonuniformities are consistently taken into account.  
We present results obtained combining SPMs at different energy bands. 
We demonstrate that our algorithm can detect sources independently of 
their shape, sources at the field's edge and sources overlapping a 
diffuse emission. Last, the BSS technique provides evidence for
celestial sources not previously catalogued by any detection technique in the
{\it X}--ray regime. 

\subsection{{\it ROSAT} PSPC in Survey Mode data} \label{ROSATdata}

\begin{figure*}
\centering
\includegraphics[width=0.34\linewidth,angle=90]{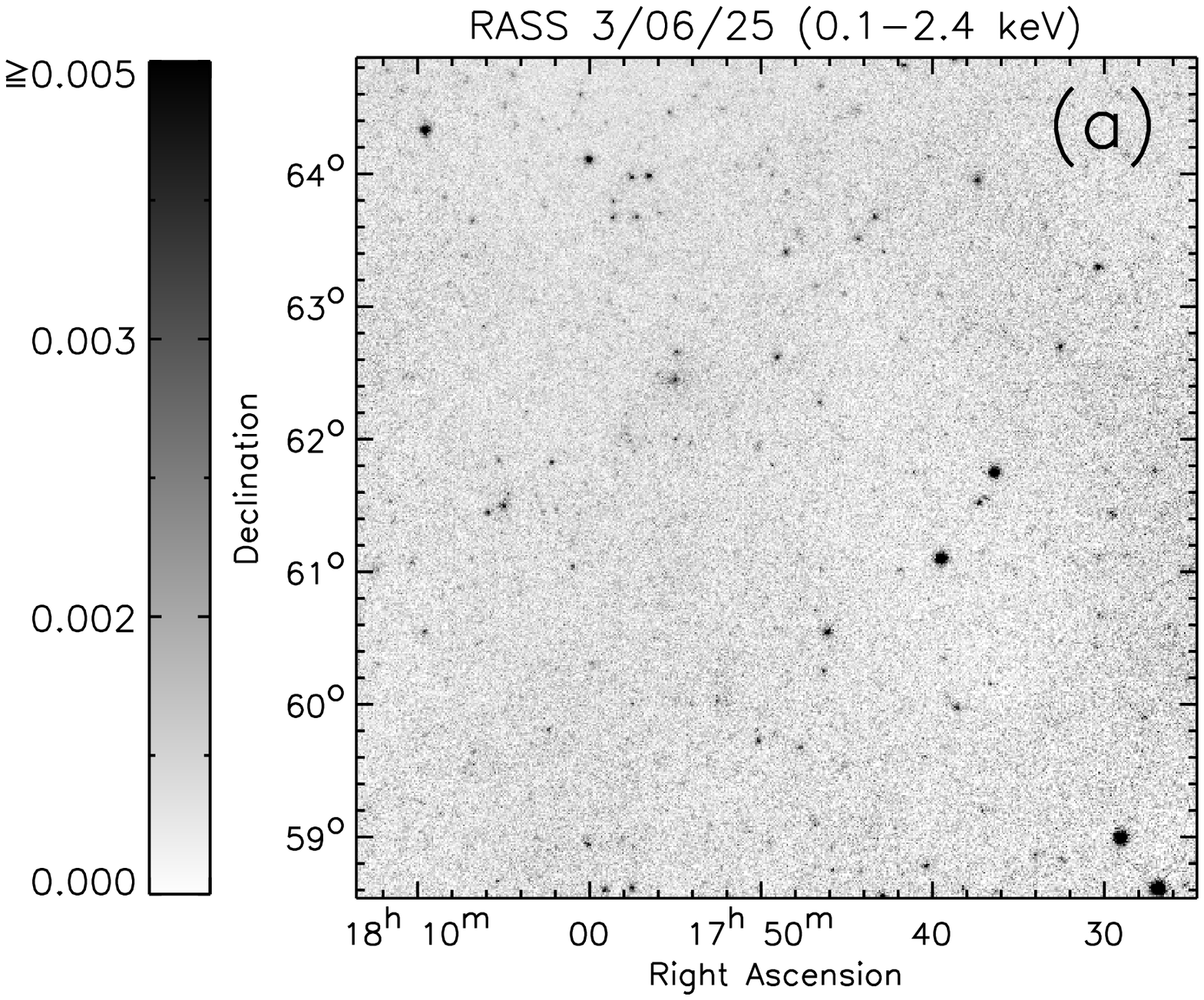}\includegraphics[width=0.34\linewidth,angle=90]{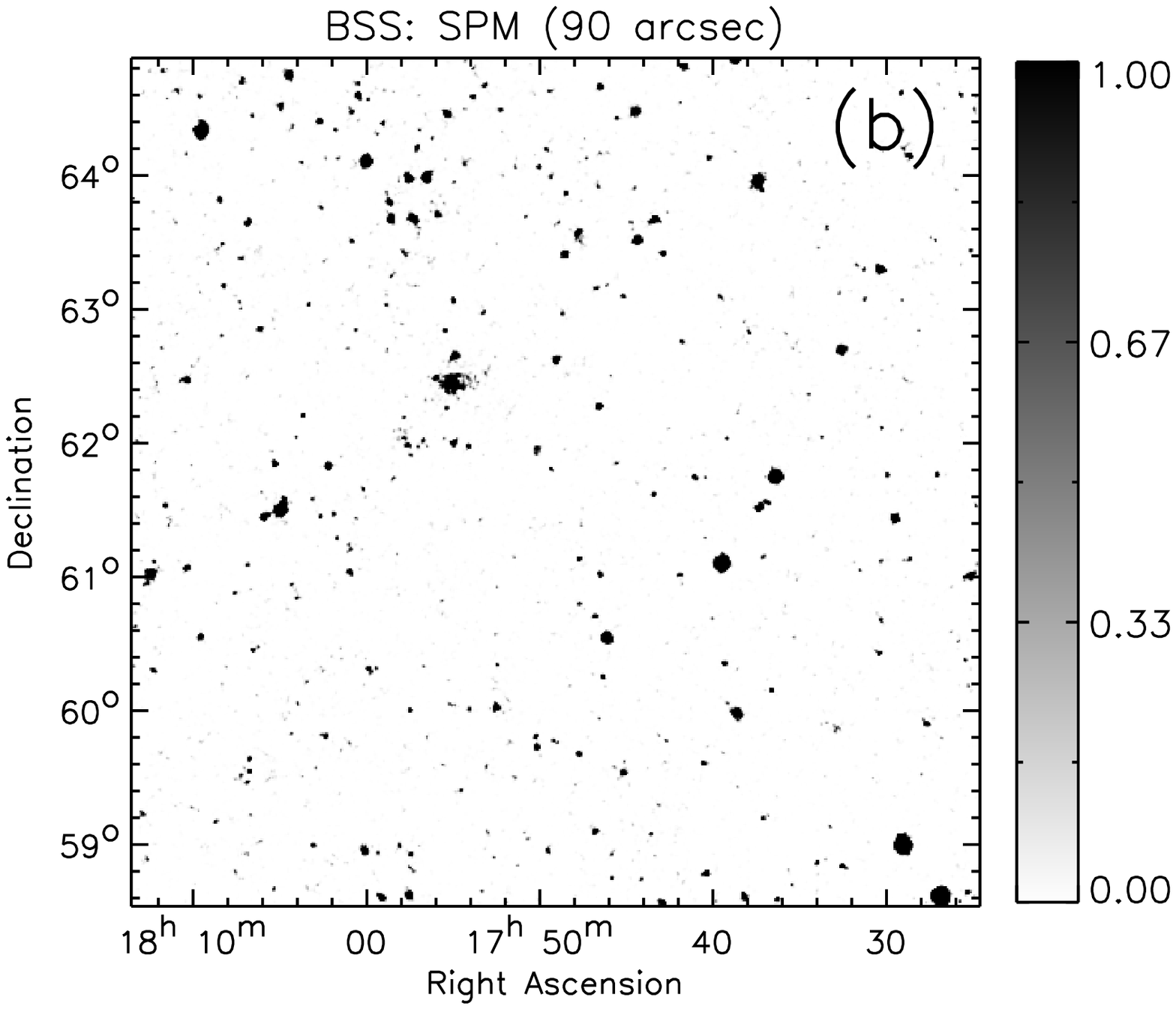}
\includegraphics[width=0.34\linewidth,angle=90]{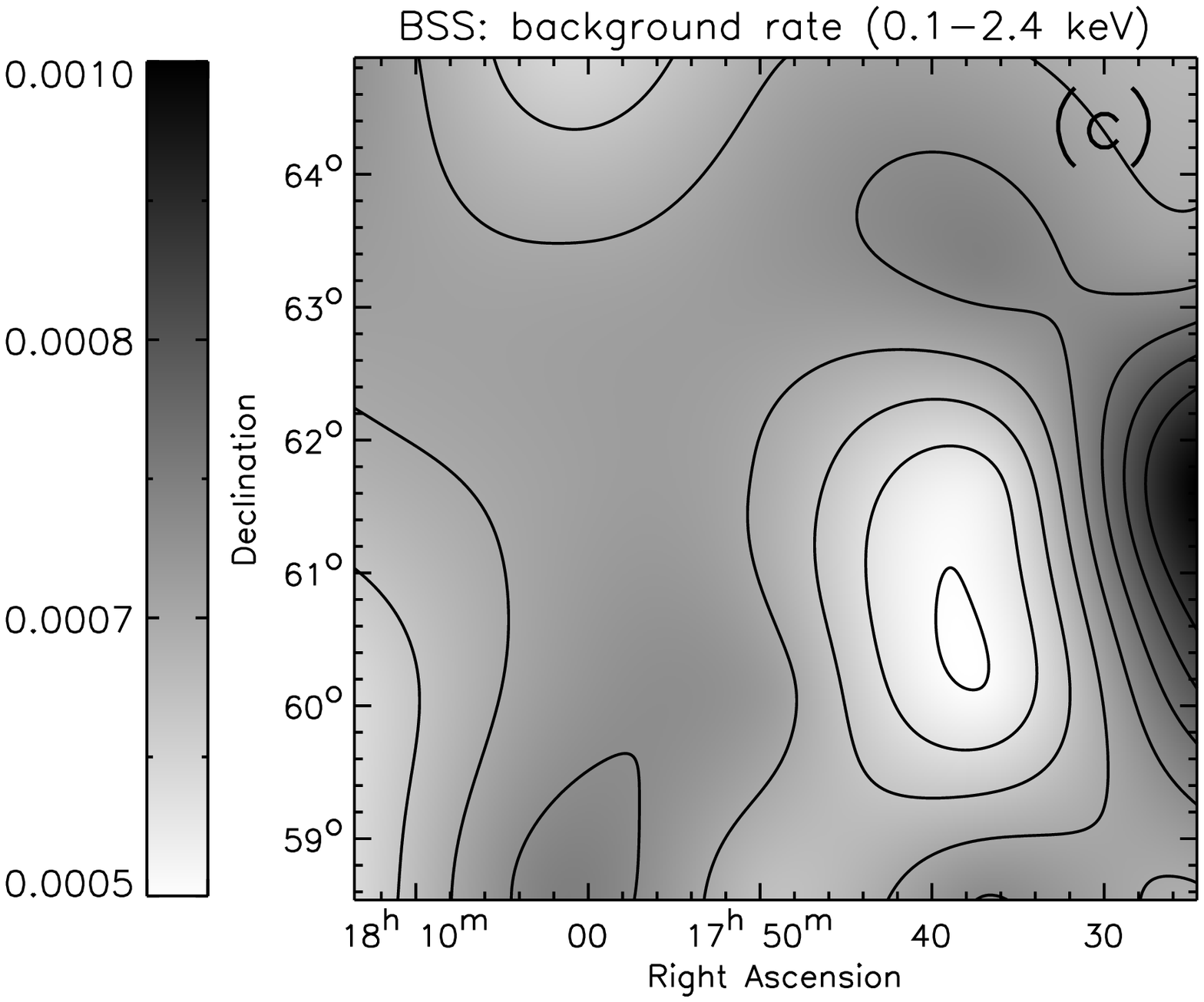}\includegraphics[width=0.34\linewidth,angle=90]{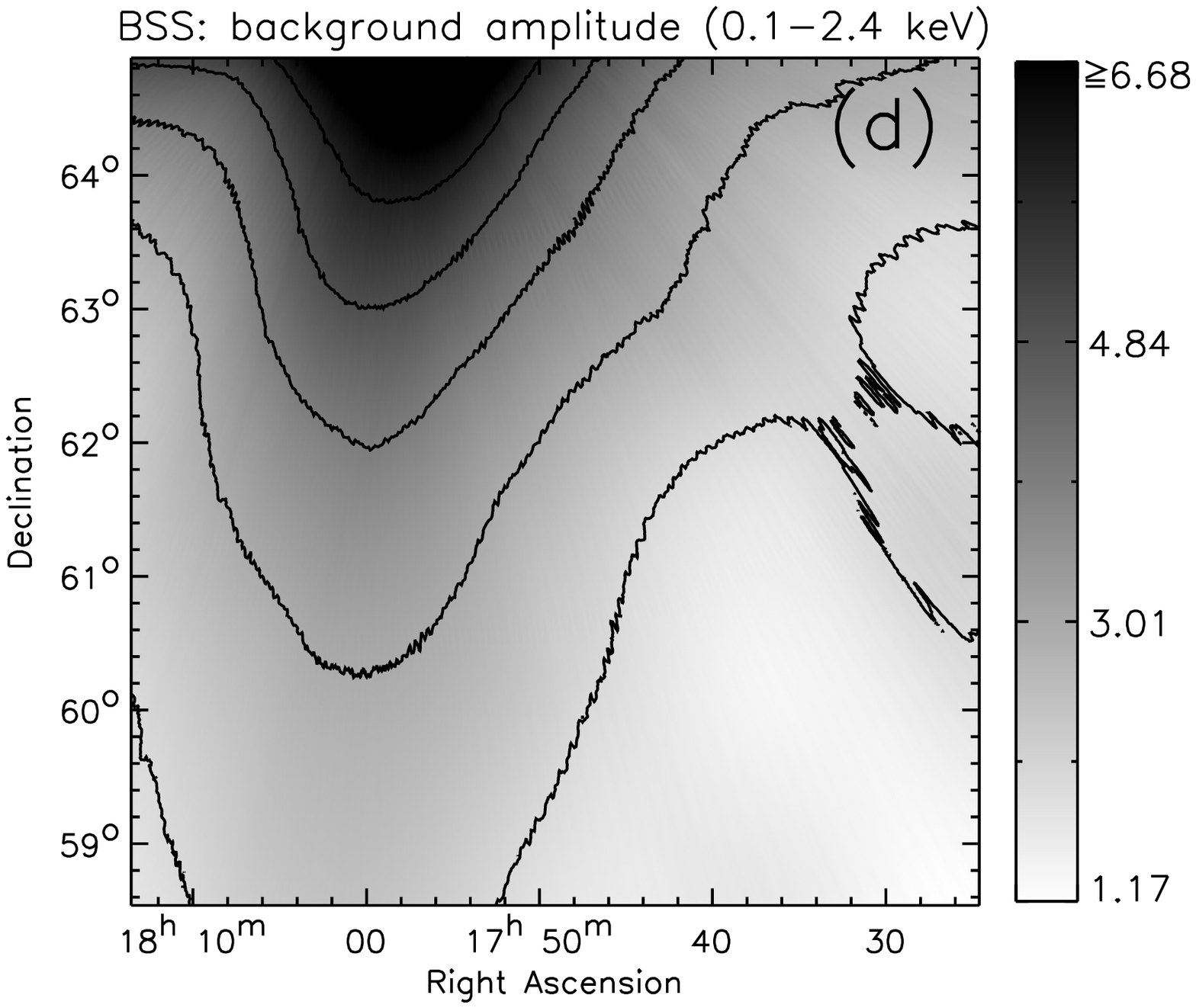}
\includegraphics[width=0.34\linewidth,angle=90]{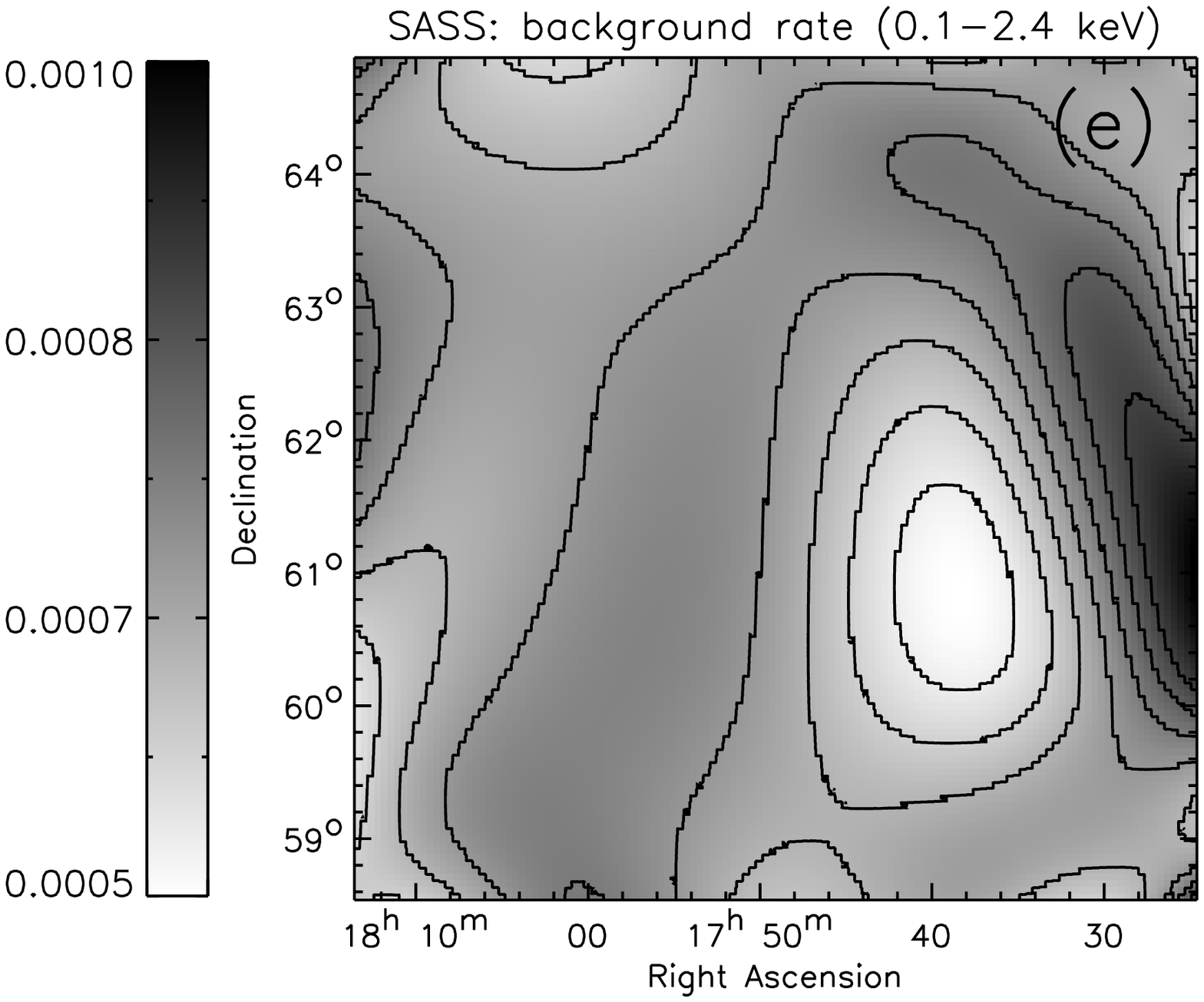}\includegraphics[width=0.34\linewidth,angle=90]{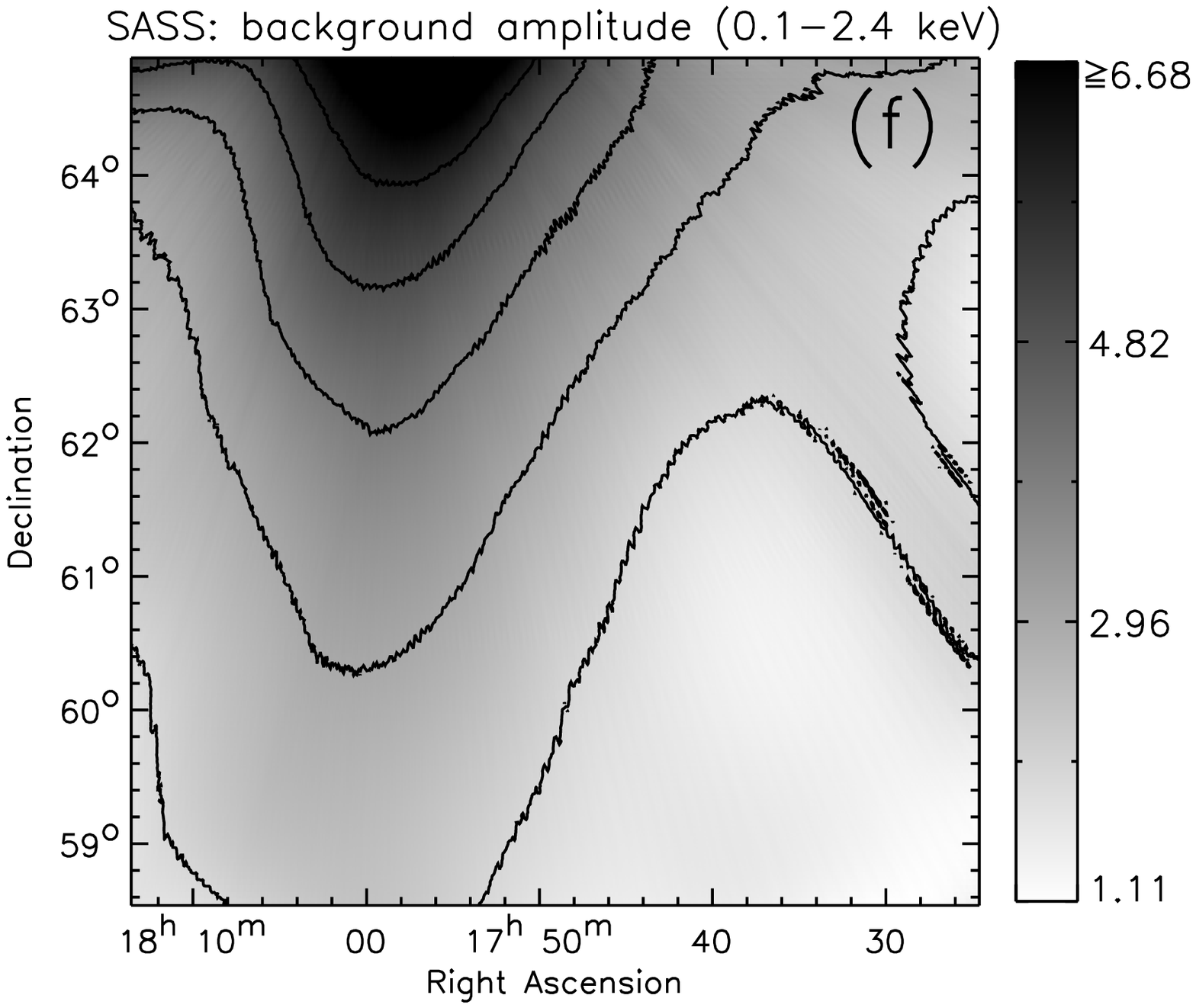}
\caption{Analysis of RS930625n00, {\it ROSAT} PSPC in Survey Mode field. 
} 
\label{rs930625_field}
\end{figure*}
We apply the BSS algorithm on a data sample coming from the 
Position Sensitive Proportional Counter (PSPC) on board of the {\it ROSAT}
satellite in survey mode ($0.1-2.4$ keV). 
{\it ROSAT} satellite provided the only all--sky survey realized using an imaging {\it X}--ray
telescope (Voges et al.~1999, 2000). 

RASS data supply a unique map of the sky in the {\it X}--ray
regime. The map of the sky was divided in $1,378$ fields each of $6.4^{\circ}
\times 6.4^{\circ}$, corresponding to $512 \times 512$ pixels. 
In addition, the satellite's exposure time can vary between about 0.4 to 
40 ks and some parts of the sky are without observations due to the 
satellite's crossing of the auroral zones and
of the South Atlantic Anomaly. Therefore, RASS data provide a wide range of 
possibilities for testing the BSS algorithm.

We present the results obtained analysing three {\it ROSAT} fields, whose IDs are:
RS930625n00,  RS932209n00 and RS932518n00.  

\subsubsection{RS930625n00} \label{twocolor_search}
This {\it ROSAT} field 
is located at $\alpha=17^{h} 49^{m} 5^{s}$, $\delta=+61^{\circ} 52^{\prime}
30''$. It is characterized by large 
variations of the satellite's exposure ranging from 1.7 to 13.5 ks. 
The count rate image of RS930625n00 in the broad energy band ($0.1-2.4$ keV) 
is displayed in panel $(a)$ of Fig.~\ref{rs930625_field}.  
The count rates range over ($0.-0.11$) photon count$\cdot$s$^{-1}\cdot$pixel$^{-1}$. The image is
here scaled in the range ($0.-0.005$) photon count$\cdot$s$^{-1}\cdot$pixel$^{-1}$ in order to
enhance the sources. 

Our results employing the BSS algorithm are shown in Fig.~\ref{rs930625_field}
with a SPM, the TPS map and the background map. 

In panel $(b)$, the SPM is obtained combining statistically the soft ($0.1-0.4$
keV) and the hard ($0.5-2.4$ keV) energy
  bands. The SPM here displayed is obtained employing the exponential prior
  pdf and accounts for the width of the {\it ROSAT} instrumental PSF. 
The information of pixels is combined with the box filter method with
  the cell shape of a circle. This image corresponds to a correlation length of $1.5$
  arcmin. Sources are identified in terms of probabilities. 
The image is in linear scale. 

In panel $(c)$, the TPS map is modelled from RS930625n00 field, broad energy band. 
The TPS models the background rate. Only $25$ support points are used. 
The background rate ranges ($0.0005-0.001$) photon count$\cdot$s$^{-1}$$\cdot$pixel$^{-1}$. 
The contours are superposed for enhancing the features relative to the
modelled background rate. The innermost and the outermost 
contours indicate a level of $0.0005$ and $0.001$ photon 
count$\cdot$s$^{-1}$$\cdot$pixel$^{-1}$, respectively.    

The corresponding background map estimated from the selected {\it ROSAT} field is displayed 
in panel $(d)$. Its values are in the
range ($1.17-8.53$) photon count$\cdot$pixel$^{-1}$. The image is scaled in the range
($1.17-6.68$) photon count$\cdot$pixel$^{-1}$. The contour levels close to black and light gray 
delineate $6.0$ and $2.0$ photon count$\cdot$pixel$^{-1}$, respectively. 
The background map shows the prominent variation due to the heterogeneous 
satellite exposure time. 

The lower row of Fig.~\ref{rs930625_field} shows the 
background rate (panel $(e)$) and the background amplitude (panel $(f)$) as
obtained analysing the broad energy band with the Standard Analysis Software
System ({\sc SASS}). {\sc SASS} is the 
detection method utilized for the realization of RASS. {\sc SASS} combines the
sliding window technique with the maximum likelihood 
PSF fitting method for source detection and characterization, respectively.  
The background rate image is in the range ($0.0005-0.00097$)
photon count$\cdot$s$^{-1}$$\cdot$pixel$^{-1}$. 
The innermost and the outermost 
contour levels delineate $0.0006$ and $0.0009$ photon count$\cdot$s$^{-1}$$\cdot$pixel$^{-1}$, respectively.
The background amplitude image estimated by {\sc SASS} has values in the range
($1.11-8.27$) photon count$\cdot$pixel$^{-1}$. The image is scaled as the one
obtained from the BSS technique. The contour levels indicate from $2.0$
to $6.0$ photon count$\cdot$pixel$^{-1}$.

The BSS technique allows for more flexibility in the background model and the
edges are more stable than the ones obtained with {\sc SASS}. Hence, celestial objects located
on the edges are not lost during source detection. 

\begin{figure}
\centering
\includegraphics[width=0.7\linewidth,angle=90]{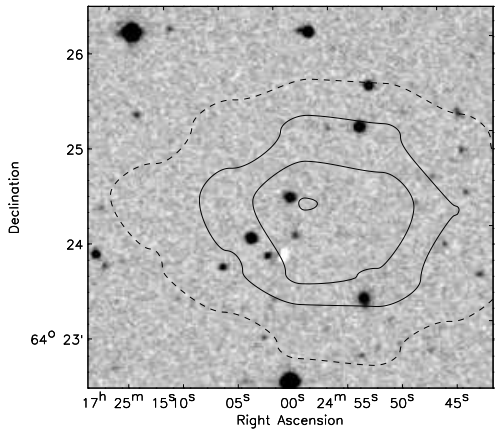} 
\caption{POSS--II red plate with overlaid {\it ROSAT} contours corresponding to
  $3$, $4$, $4.5$ and $5$ $\sigma$ above the local background. In the image
  centre is located SDSS J172459.31+642424.0, a low redshift QSO. 
}
\label{rs930625_797SDSS}
\end{figure}
In Fig.~\ref{rs930625_797SDSS} we show an example of a QSO detection with 
the BSS algorithm analysing the field RS930625n00 in the hard ($0.5-2.4$ keV) band image. 
No counterparts have been found in the {\it ROSAT} Bright and Faint 
source catalogues, Voges et al.~(1999, 2000). 
A counterpart is found with the Sloan Digital Sky Survey. 
This QSO is catalogued as SDSS J172459.31+642424.0 (\citealt{SDSS:2004},
\citealt{hao:2005}, \citealt{berk:2006}). 
It is located at the image centre. 
The image covers a field of view of $\sim5$ arcmin at the side. 
The {\it ROSAT} image, as shown in Fig.~\ref{rs930625_field}, has $45$
arcsec$\cdot$pixel$^{-1}$ resolution. 
The QSO optical position as given by the SDSS is 12 arcsec far from the
BSS position, i.e.~it is within the pixel resolution of the {\it ROSAT} data. 
The estimated source count rate with the BSS algorithm for this source is 
($0.0068\pm0.0023$) photon count$\cdot$s$^{-1}$. Its source detection probability is 
$0.999$. This object is located close to the north--west corner of the field 
RS930625n00 ($\sim45$ arcmin). 
In this region the {\sc SASS} background intensity is $7$ per cent lower than the BSS
background estimate. 

\subsubsection{RS932209n00} 
\begin{figure*}
\centering
\includegraphics[width=0.34\linewidth,angle=90]{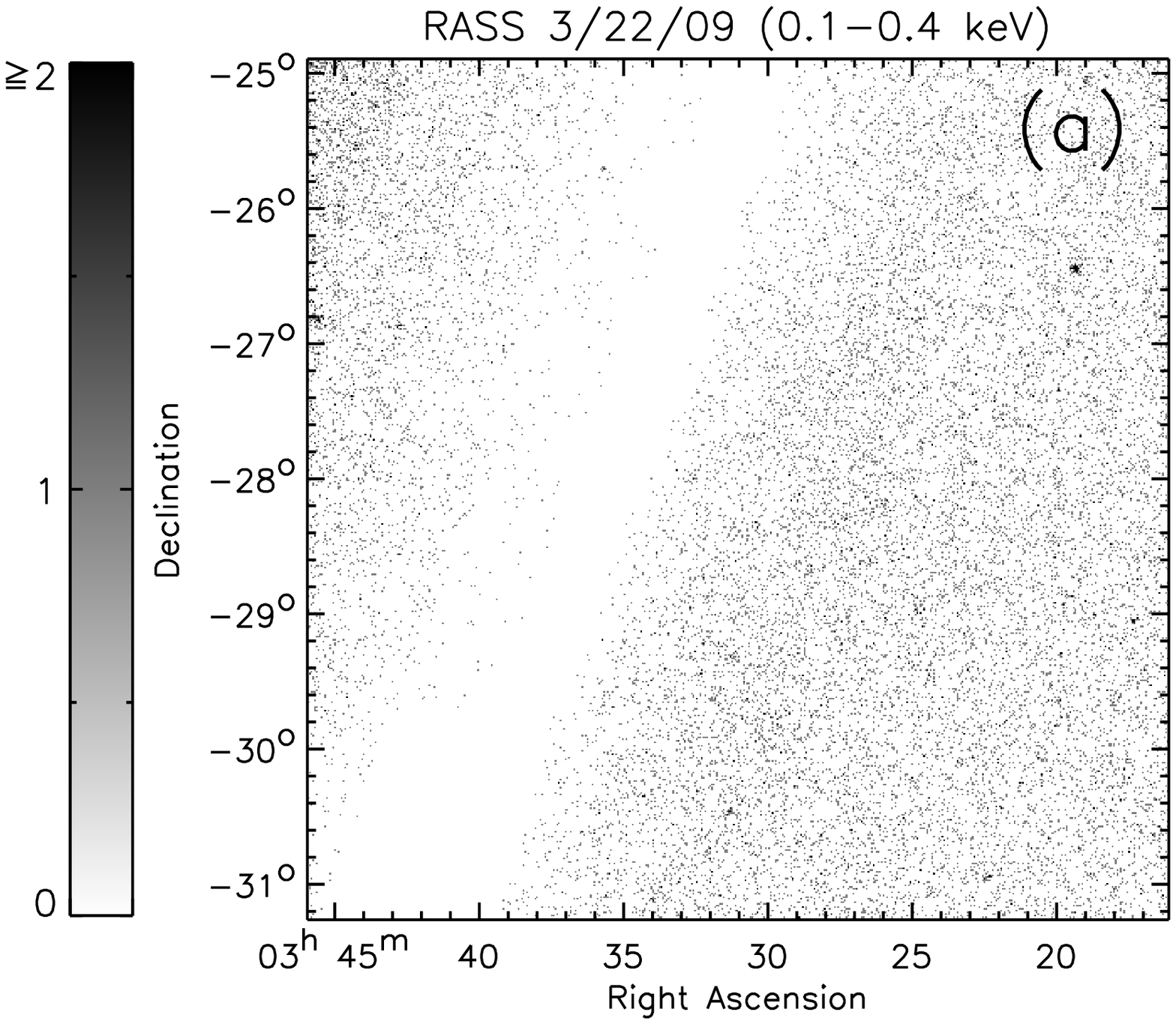}
\includegraphics[width=0.34\linewidth,angle=90]{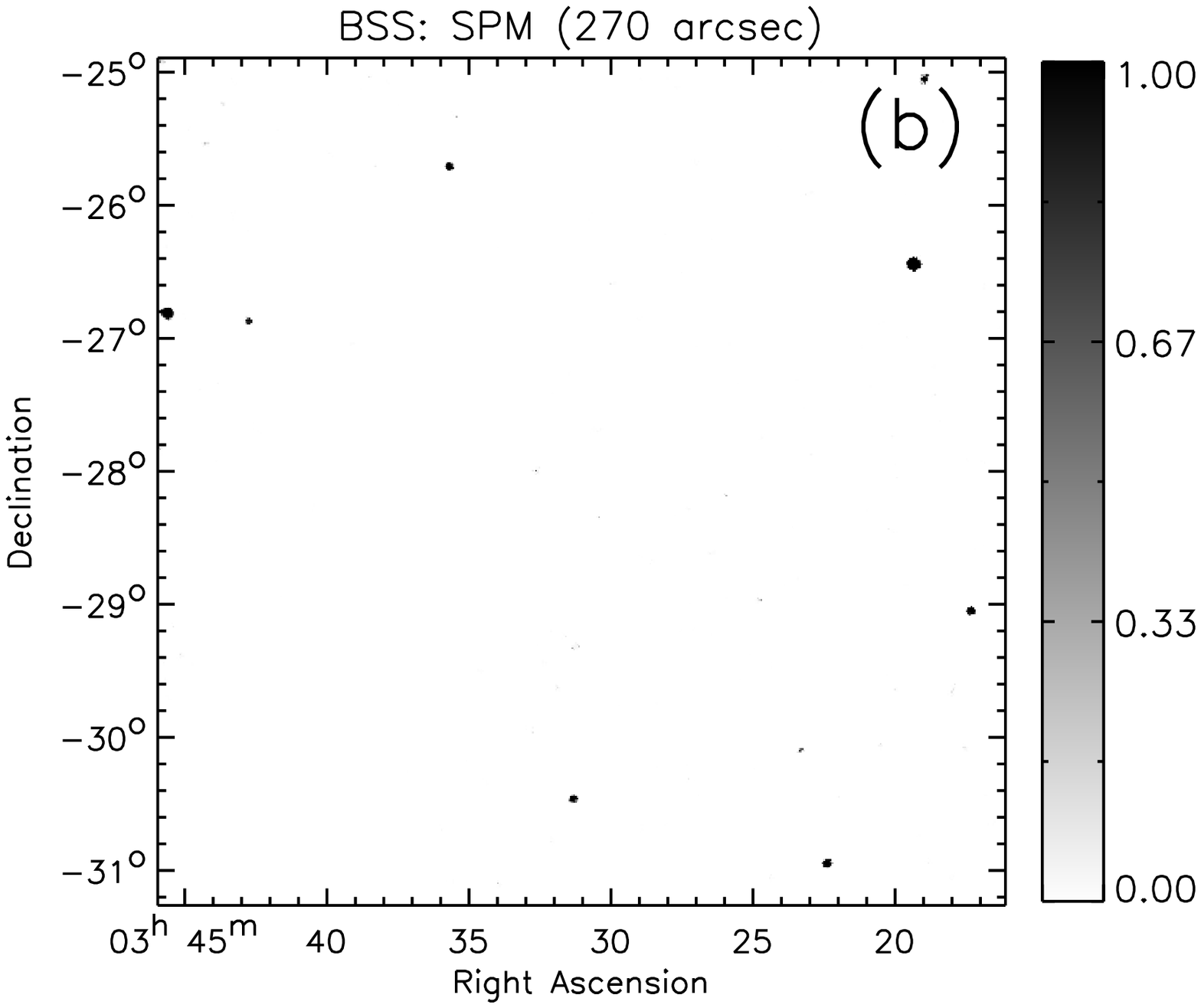}
\caption{Panel $(a)$: {\it ROSAT} all--sky survey {\em photon count image} of the
    field id RS932209n00 in the soft energy band. Panel $(b)$: {\em source
    probability map} as obtained with the developed BSS method.}
\label{rs932209_field_2}
\end{figure*}

\begin{figure*}
\centering
\includegraphics[width=0.41\linewidth,angle=90]{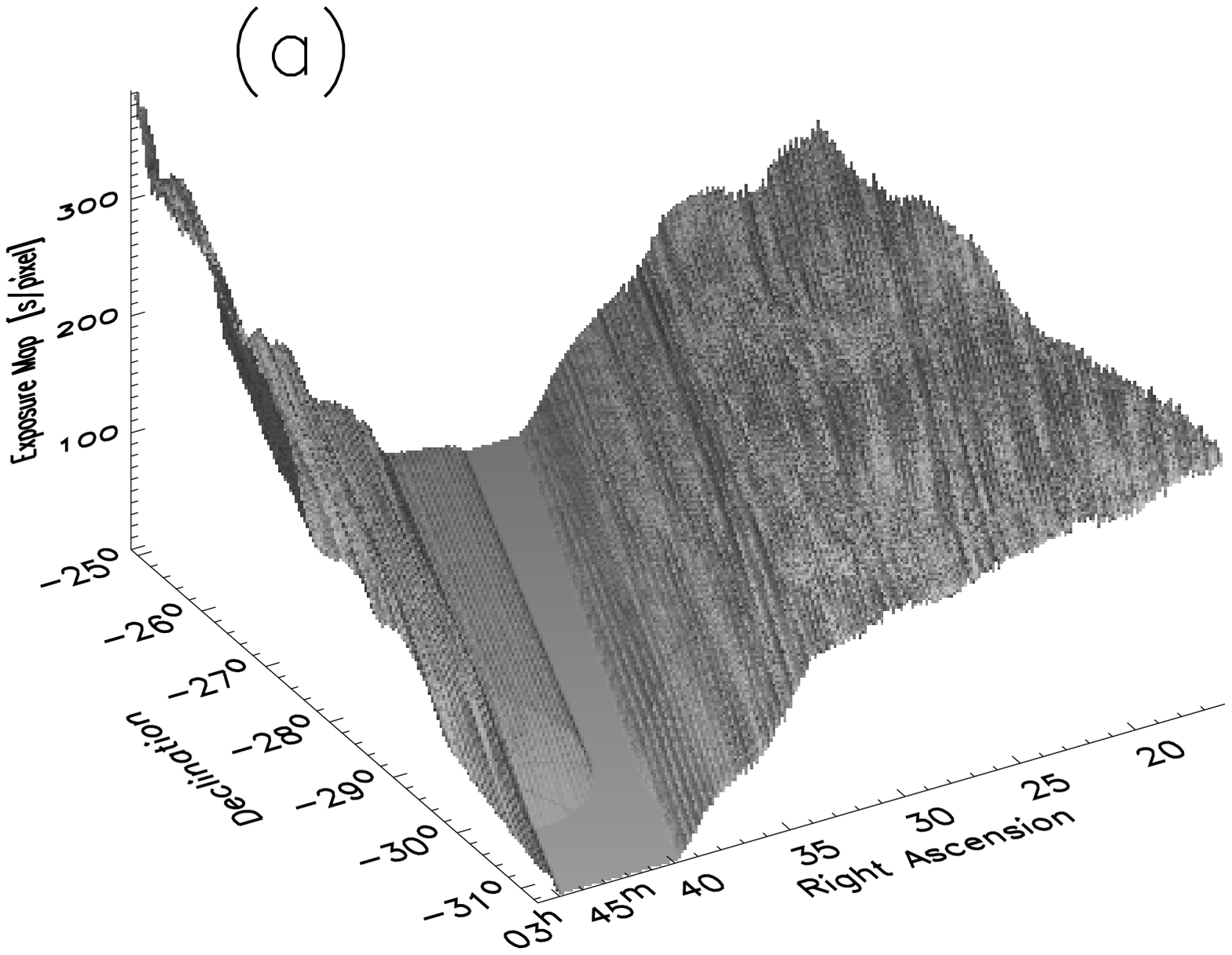}\includegraphics[width=0.41\linewidth,angle=90]{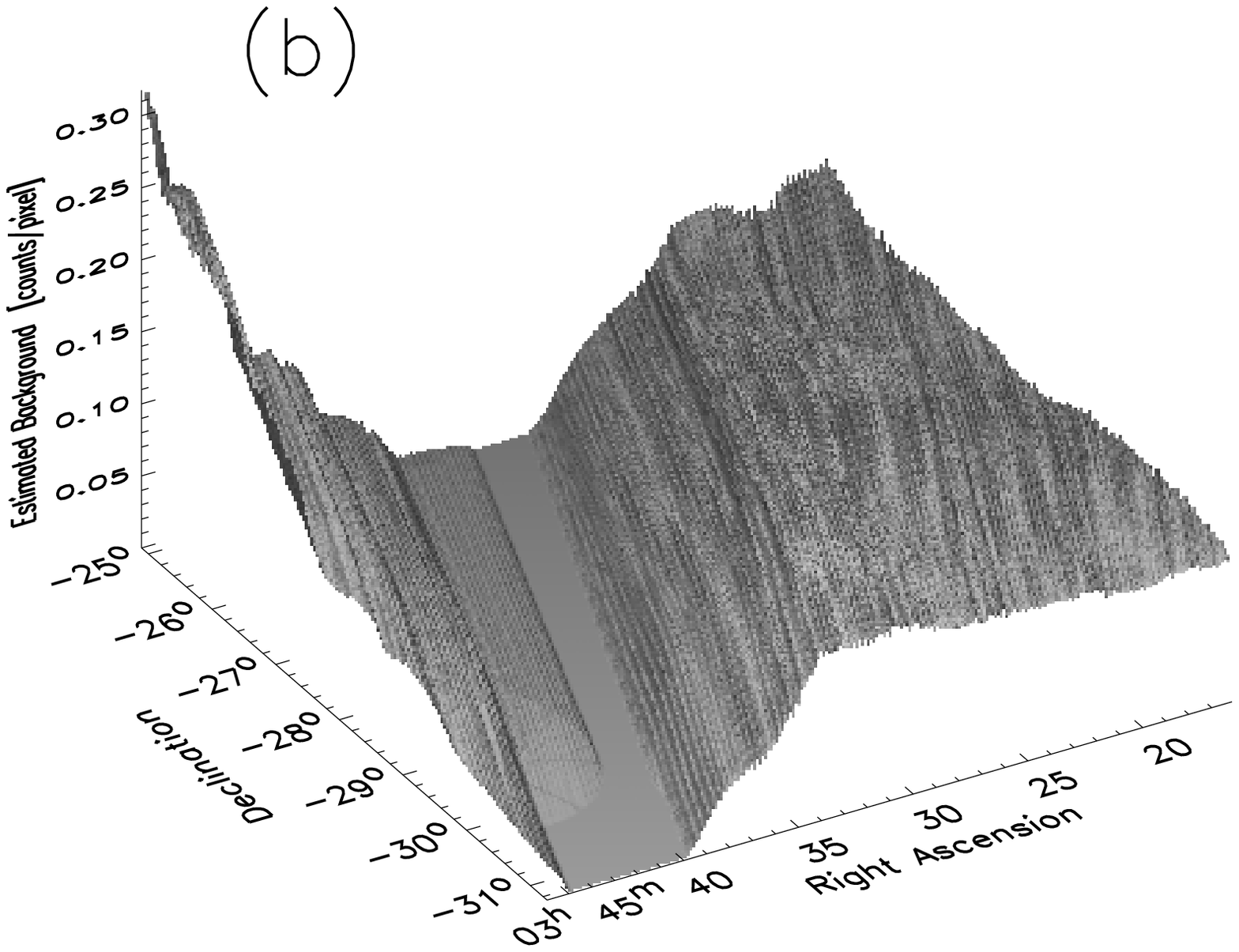}
\caption{Panel $(a)$: {\it exposure map} of the {\it ROSAT} field id RS932209n00
  (Fig.~\ref{rs932209_field_2}). The z axis indicates the exposure time in
  s$\cdot$pixel$^{-1}$. 
Panel $(b)$: {\it estimated background amplitude} with the BSS technique of
  RS932209n00 field in the soft energy band. 
The z axis shows the estimated background count$\cdot$pixel$^{-1}$.
} 
\label{rs932209_field}
\end{figure*}

The detection capabilities of our Bayesian approach on images with exposure 
non--uniformities is presented in Figs.~\ref{rs932209_field_2} and
\ref{rs932209_field}. The analysed {\it ROSAT} field is located at 
$\alpha=3^{h} 31^{m}$, $\delta=-28^{\circ} 07^{\prime} 08''$.
 
In Fig.~\ref{rs932209_field_2}, panel $(a)$, the soft band image ($0.1-0.4$ keV)
is shown. The image accounts for photon count$\cdot$pixel$^{-1}$ in the
range ($0-9$). The image is scaled in the range ($0-2$) photon count$\cdot$pixel$^{-1}$.  
In panel $(b)$, an SPM as in output from the BSS method is displayed. The
inverse--Gamma function prior pdf is used for source detection and background
estimation. This SPM is obtained with a correlation length of $270$ arcsec. 
The box filter method with the cell shape of a circle is employed. 
The SPM shows source detections. 

In Fig.~\ref{rs932209_field}, panel $(a)$, the satellite's exposure is
located. The exposure values are in the range ($0.-0.393$) ks.   
In panel $(b)$ we show the background map estimated with the developed BSS technique. 
Only $16$ pivots equidistantly distributed along the field are employed. 
The background amplitude ranges from $0$ to 
$0.318$ photon count$\cdot$pixel$^{-1}$. The background is estimated with
null value where no satellite's exposure information is provided. 
The estimated background map is similar to the exposure map because the contribution
from the cosmic background is very small. 
 
No artefacts, due to exposure non-uniformities, occur in the background map nor in the
SPMs.

\subsubsection{RS932518n00}
We employ the {\it ROSAT} field id RS932518n00 in the hard energy band ($0.5-2.4$ keV) 
to show the capabilities of our technique on data with large background 
variations.  
This field is characterized by a complex structure of hot gas caused primarily
by the emissions of two galactic {\it X}--ray sources: $(1)$ Vela supernova remnant 
(SNR) located at $\alpha=8^{h} 35^{m} 20.6^{s}$, $\delta=-45^{\circ}
10^{\prime} 35''$; $(2)$ SNR $RX J0852.0-4622$ located at $\alpha=8^{h} 52^{m}$, $\delta=-46^{\circ}
22^{\prime}$, see refs.~\cite{aschenbach:1995} and 
\cite{aschenbach:1998}.  

\begin{figure*}
\centering
\includegraphics[width=0.34\linewidth,angle=90]{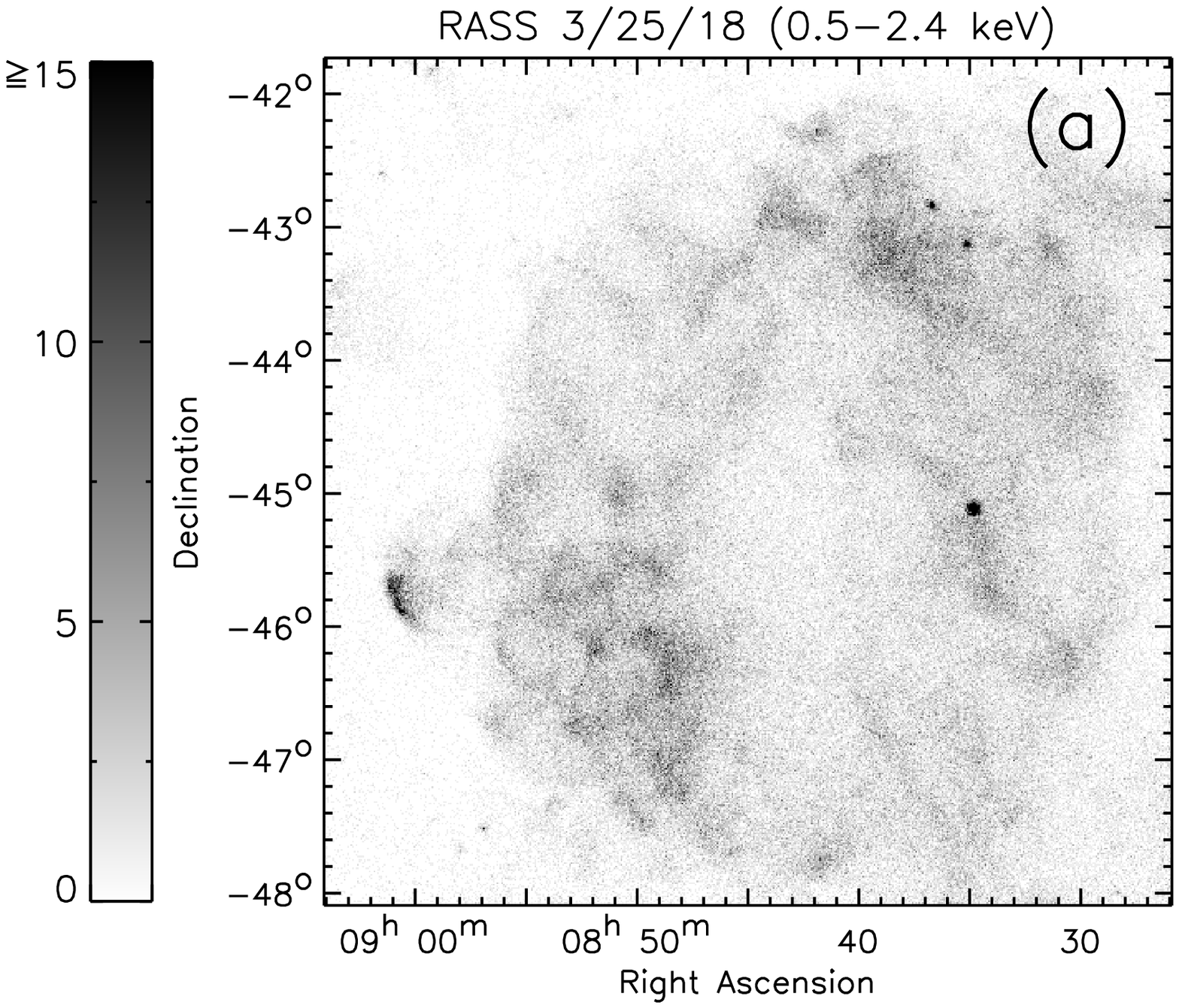}
\includegraphics[width=0.34\linewidth,angle=90]{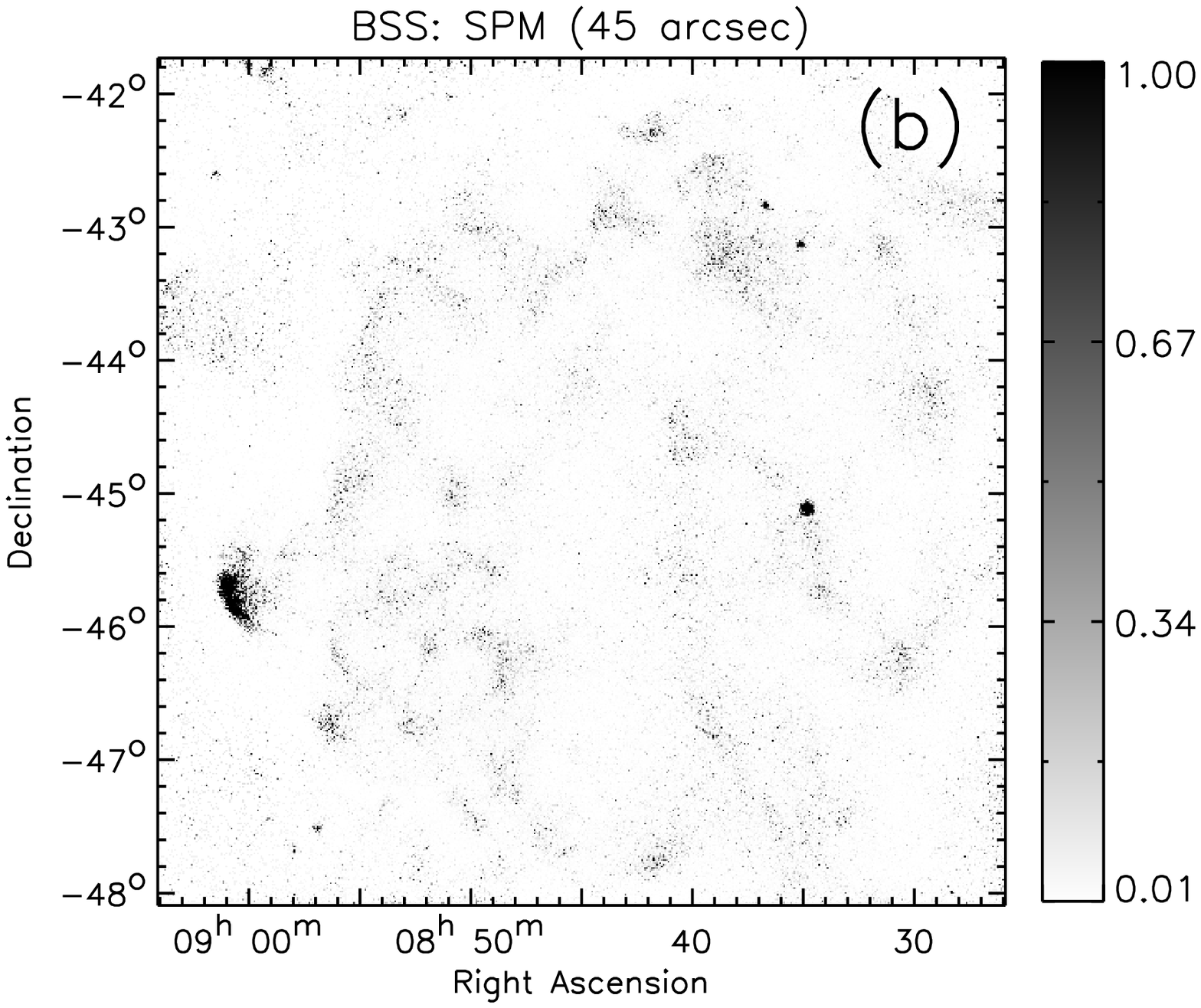}
\includegraphics[width=0.34\linewidth,angle=90]{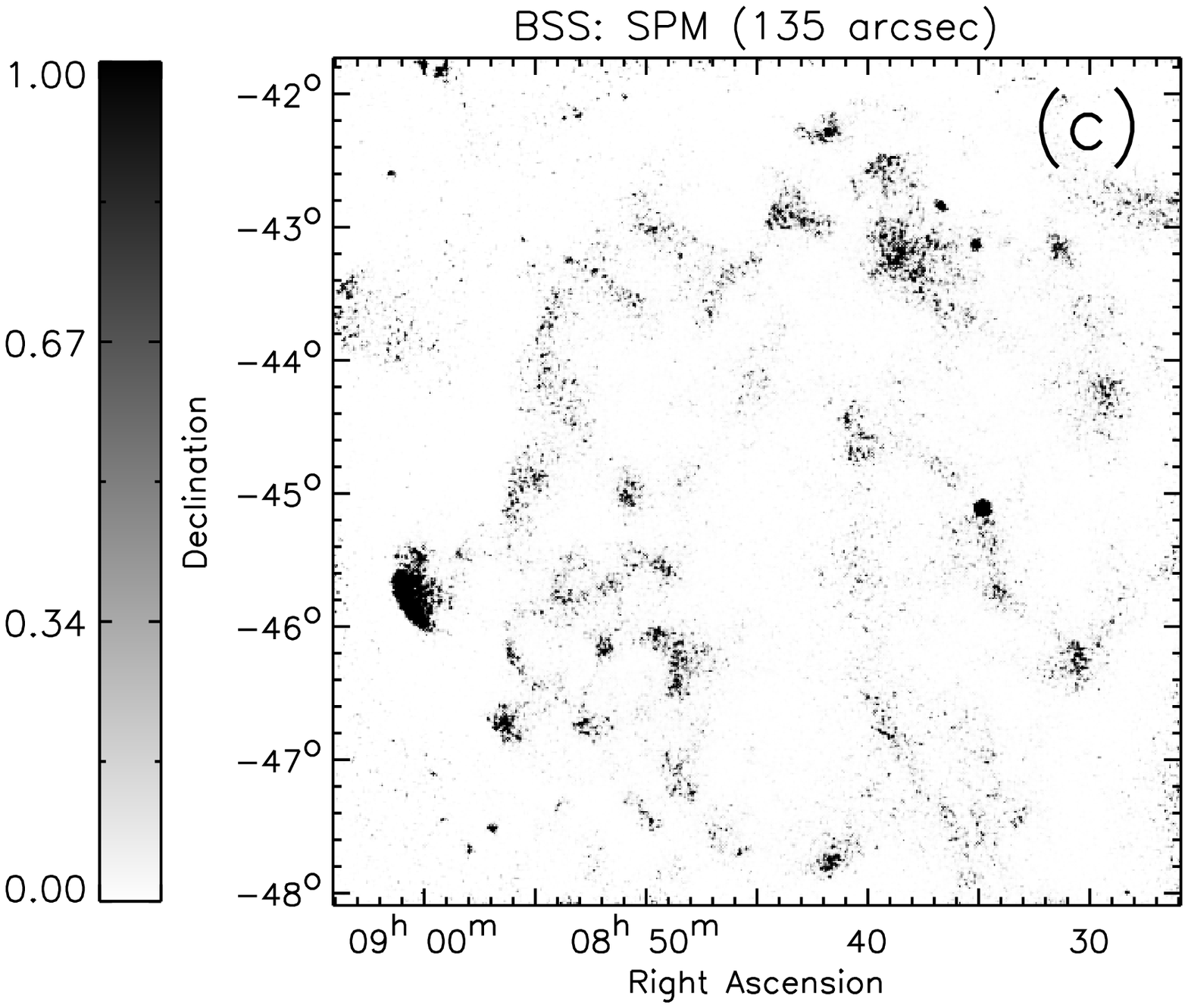}
\includegraphics[width=0.34\linewidth,angle=90]{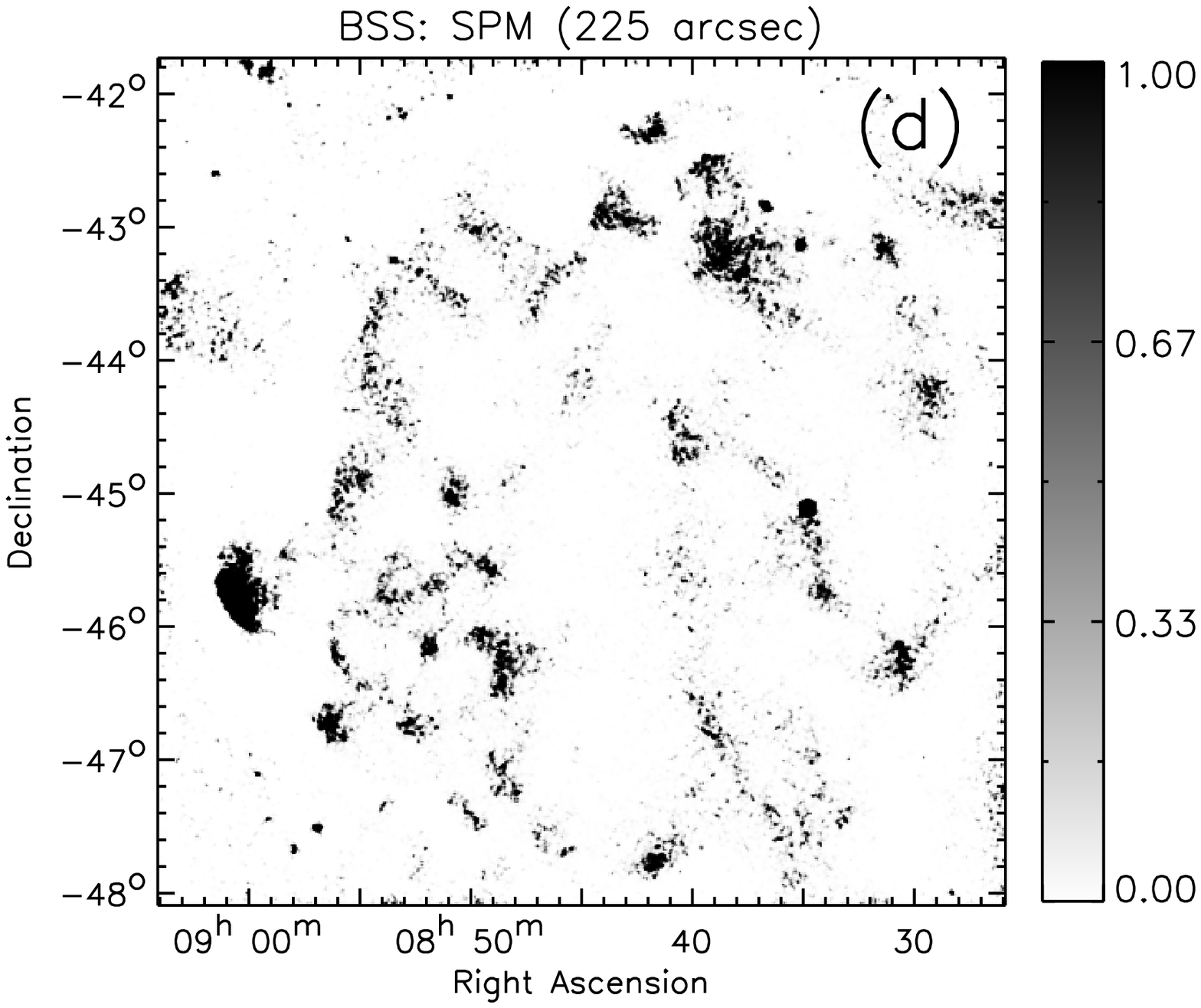}
\caption{Panel $(a)$: {\it ROSAT} all--sky survey {\em photon count image} of the
    field id RS932518n00 in the hard energy band. The analysis of this {\it ROSAT} field with the 
developed BSS method is shown in panels $(b)-(d)$ with {\em SPMs} obtained with
  increasing correlation lengths.} 
\label{rs932518_field_2}
\end{figure*}

\begin{figure*}
\centering
\includegraphics[width=0.41\linewidth,angle=90]{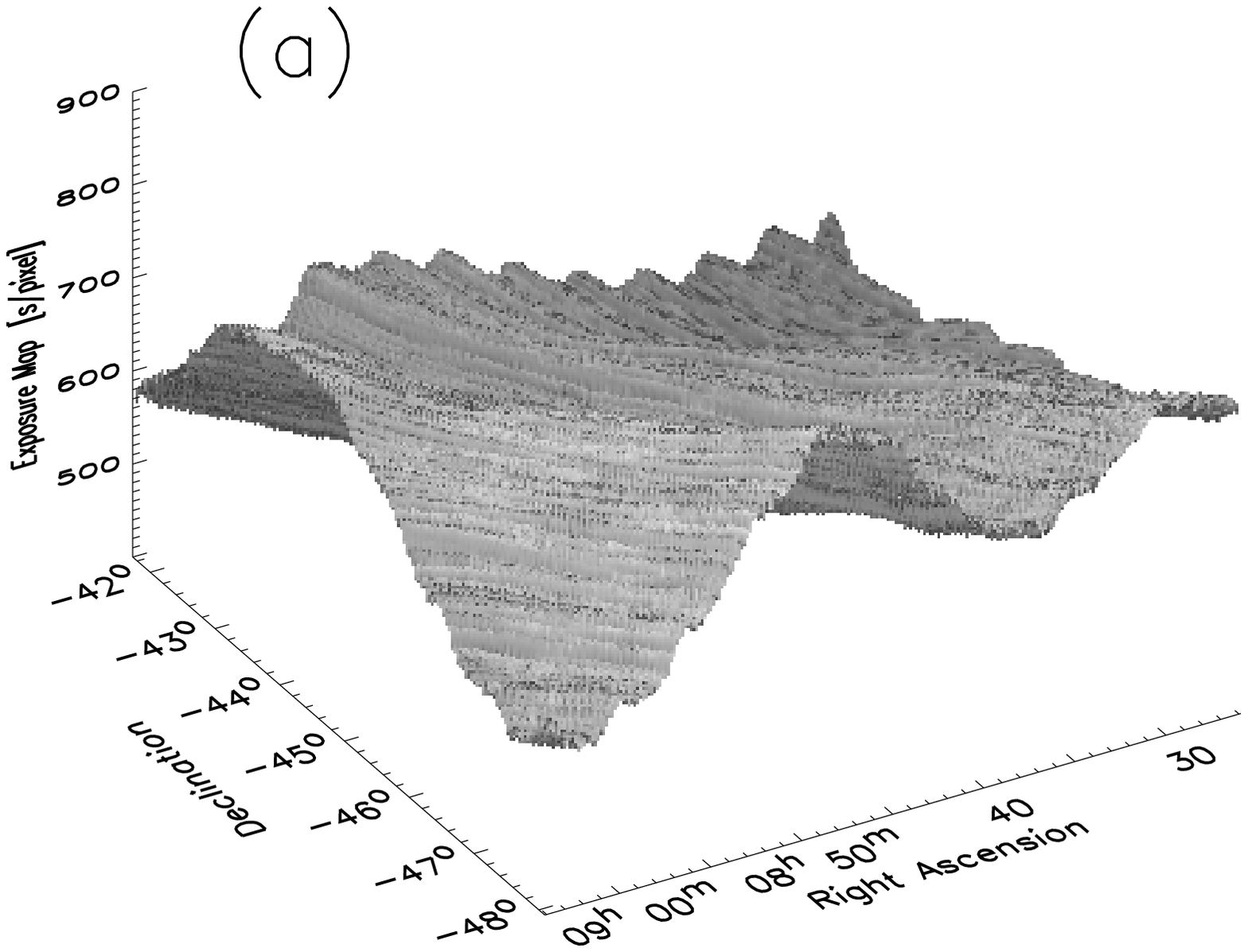}\includegraphics[width=0.41\linewidth,angle=90]{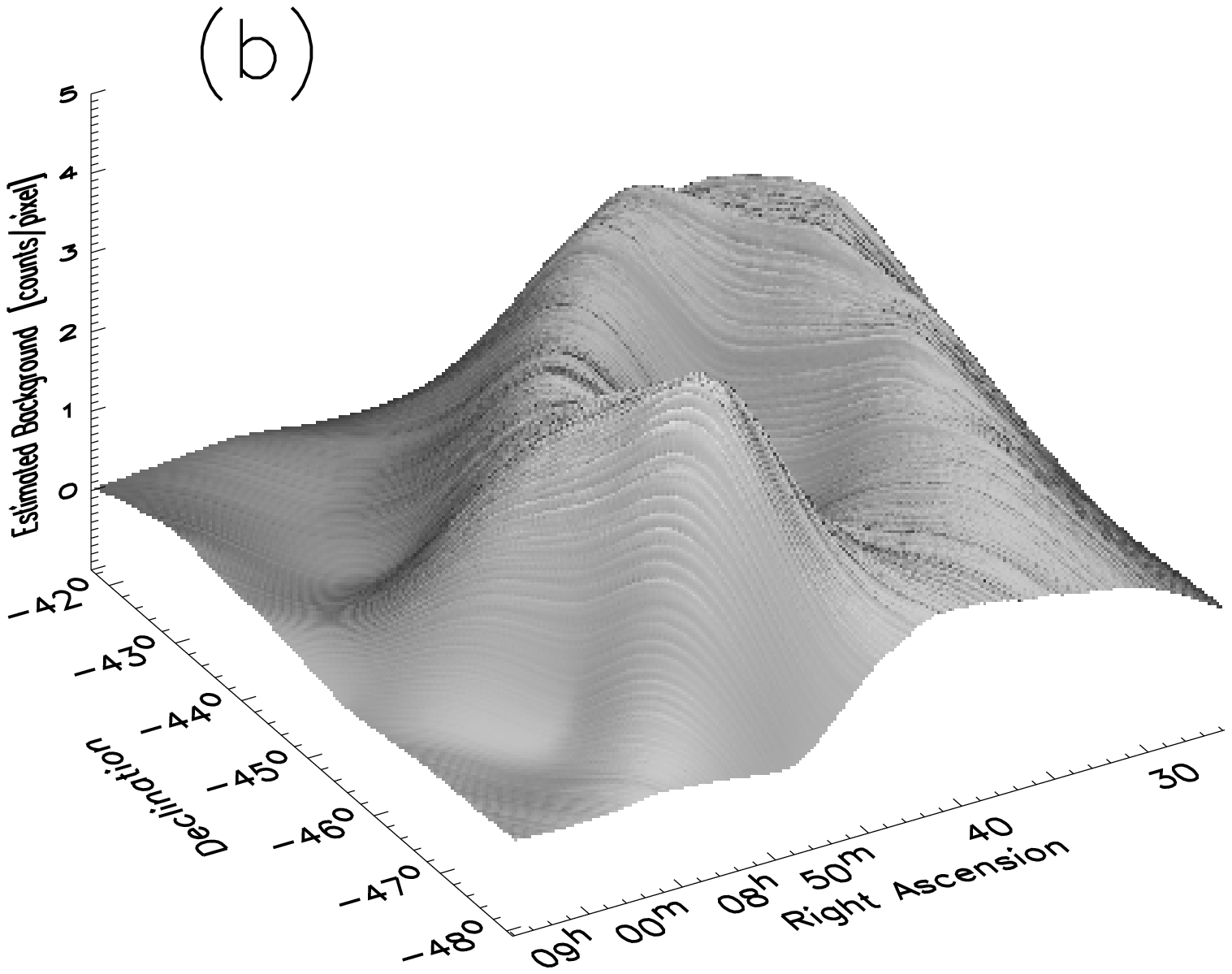}
\caption{Panel $(a)$: {\it exposure map} of the {\it ROSAT} field id RS932518n00
  (Fig.~\ref{rs932518_field_2}). The z axis indicates the exposure time in
  seconds$\cdot$pixel$^{-1}$. 
Panel $(b)$: {\it estimated background amplitude} with the BSS technique of
  RS932518n00 field in the hard energy band. 
The z axis shows the background counts$\cdot$pixel$^{-1}$.
} 
\label{rs932518_field}
\end{figure*}

In Fig.~\ref{rs932518_field_2}, panel $(a)$, we display the photon count image
of RS932518n00 field in the hard energy band. 
The image presents values in the range ($0-136$) photon
count$\cdot$pixel$^{-1}$.  
It is scaled in the range ($0-15$) photon count$\cdot$pixel$^{-1}$ 
in order to enhance the {\it X}--ray emissions. 
The satellite's exposure map is 
displayed in Fig.~\ref{rs932518_field}, panel $(a)$. The exposure map shows
variations in the range ($0.5-0.8$) ks. 

The analysis of RS932518n00 in the hard energy band with the BSS algorithm is
shown in Figs.~\ref{rs932518_field_2}, panels $(b)$-$(d)$, and \ref{rs932518_field},
panel $(b)$. For background estimation and source detection the exponential
prior pdf is employed. 

In Fig.~\ref{rs932518_field_2}, panels $(b)$-$(d)$, three SPMs are
displayed. The correlation lengths used for their realization is written on
each image. The information of neighbouring pixels is put together with the
Gaussian weighting method. These SPMs are in linear scale. 

The background map (Fig.~\ref{rs932518_field_2}, panel $(b)$) is estimated
utilizing $36$ pivots equidistantly spaced along the field. The estimated
background values range from $\sim 0$ to $4.1$ photon
count$\cdot$pixel$^{-1}$. 
The heterogeneous background is recovered.  

The BSS method combined with the multi-resolution analysis allows one to
identify the detection of point-like sources on top of the diffuse emission and
of extended {\it X}--ray features, such as the Pencil Nebula located at
$\alpha=9^{h} 0.2^{m}$, $\delta=-45^{\circ} 57^{\prime}$ and the SNR $RX
J0852.0-4622$. 
The BSS algorithm detects about $50$ objects on this {\it ROSAT} field. 

In Table \ref{RS932518n00_cat} we present part of this catalogue.
Columns $RA$, $Dec$, $sctr$, $\sigma_{\rm x}$, $\sigma_{\rm y}$ and $\rho$ are the
estimated positions, source count rates and source parameters as in output 
from the developed BSS technique. Each listed object is detected with a
probability larger then $0.999$. 
The column indicated with $matched$ $ID$ corresponds to an $ID$ given by
catalogues created analysing {\it ROSAT} data and whose position matches with the
one obtained with the BSS algorithm. During this selection, we give 
priority to IDs provided by RASS catalogues when available. 
Catalogues IDs highlighted with an asterisk are coming from a point source 
catalogue generated from all {\it ROSAT} PSPC pointed observations. 
\begin{table*} 
\caption{BSS sample catalogue from the analysis of RS932518n00 field in the hard energy band ($0.5-2.4$
  keV)}
\label{RS932518n00_cat}
\begin{tabular}{@{}lccccccc}
\hline
  & RA & Dec & sctr & $\sigma_x$ & $\sigma_y$ & $\rho$ & matched ID\\
    &  &  & (count$\cdot$s$^{-1}$) & (pixel) & (pixel) & \\
\hline \hline
(1) & $08^{h} 35^{m} 20.4^{s}$ & $-45^{\circ} 10^{\prime} 36.37''$ &
 2.7259$\pm$0.0745 & 1.91$\pm$0.05 & 1.93$\pm$0.05 &
 -0.05$\pm$0.04 &  1RXS J083520.6-451035$^{(\dagger)}$\\
(2) & $08^{h} 37^{m} 24.62^{s}$ & $-42^{\circ} 53^{\prime} 55.75''$ & 0.6236$\pm$0.0486 &
 1.49$\pm$0.17 & 1.56$\pm$0.12 & -0.32$\pm$0.10 & 1WGA J0837.3-4253$^{(\ast)}$ \\
(3) & $08^{h} 35^{m} 55.85^{s}$ & $-43^{\circ} 11^{\prime} 11.29''$ & 0.7099$\pm$0.0524 &
 1.55$\pm$0.12 & 1.73$\pm$0.15 & 0.06$\pm$0.11 & 1WGA J0835.9-4311$^{(\ast)}$\\
(4) & $08^{h} 57^{m} 44.81^{s}$ & $-41^{\circ} 51^{\prime} 51.52''$ & 0.2478$\pm$0.0253 &
 3.20$\pm$0.33 & 3.60$\pm$0.38 & 0.07$\pm$0.14 & 1RXS
 J085748.6-415101$^{(\dagger)}$\\
\hline
\end{tabular}
\medskip 
References: $(\dagger)$ \citet{voges:1999}; $(\ast)$ \citet{WGA:1994} 
\end{table*}

\section{Conclusions} \label{conclu}

We have presented a new statistical method for source detection, background
estimation and source characterization in the Poisson regime. 
With this technique we have elaborated a very general and powerful Bayesian method 
applicable to images coming from any count detector. It is particularly
suitable for the search of faint and extended sources in images coming from 
instruments of a new generation. We apply our technique to {\it X}--ray data.

The developed technique does not loose information from the data and
preserves the statistics. The BSS algorithm provides a comprehensive error 
analysis where uncertainties
are properly included in the physical model. 

The background and its errors are estimated on the complete field with a two--dimensional
spline and the knowledge of the experimental data and the exposure map. 
The background model is well--defined along the complete field, so that field edges
and instrumental effects are correctly 
handled and sources located at the edges or gaps are not penalized.

The BSS method does not assume anything about the source size or shape for an 
object to be detected. 
Sources are separated from the background employing BPT combined with the 
mixture model technique. 
Point--like and extended sources are detected independent from their
morphology and the kind of background. 
Consequently, the BSS technique can be applied to large data volumes, 
e.g.~surveys from {\it X}--ray missions. 

We developed the BSS technique utilizing two prior pdfs of the source
signal. We proved that the prior pdfs of the source signal allow one to select what has to be
described as source or as background.

The SPMs are robust features of our technique. They allow for the analysis
of faint and extended objects. SPMs obtained at different energy bands can be
combined probabilistically providing conclusive likelihoods for each detected source. 

The BSS photometry is robust. 
The errors on the estimated source parameters are normally distributed. 

We demonstrated that our technique is capable of coping with spatial exposure 
non--uniformities, large background variations and of detecting sources embedded
on a diffuse emission. We expect to handle consistently the heterogeneities
present in astronomical images with CCD patterns, images superposed and 
mosaic of images.

Finally, the verification procedure with existing algorithms 
through simulations demonstrates that the BSS technique improves the detection of sources
with extended low surface brightness. This is supported by the applications on 
real data. 
The BSS method has good potentials for improving, first, the count rate of
detected sources and, second, the sensitivity reached by other 
techniques also on real data. 
\section{Acronyms}\label{acro}
The following is a list of acronyms frequently appearing in the text.
\begin{itemize}
\item[]{BPT \hspace{.8cm} Bayesian Probability Theory} 
\item[]{BSS \hspace{.9cm} Background--Source Separation}
\item[]{CCD \hspace{.77cm} Charge--Coupled Device} 
\item[]{cdf \hspace{1.09cm} cumulative distribution function} 
\item[]{FWHM \hspace{.4cm} Full--Width at Half Maximum}
\item[]{MCMC \hspace{.45cm} Markov--Chain Monte Carlo} 
\item[]{pdf \hspace{1.05cm} probability density function}
\item[]{PSF \hspace{.88cm} Point--Spread Function}
\item[]{PSPC \hspace{.65cm} Position Sensitive Proportional Counter} 
\item[]{RASS \hspace{.67cm} ROSAT All--Sky Survey}
\item[]{SASS \hspace{.75cm} Standard Analysis Software System} 
\item[]{SPM \hspace{.85cm} Source Probability Map}
\item[]{SNR \hspace{.9cm} supernova remnant}
\item[]{TPS \hspace{.92cm} Thin--plate spline}
\item[]{WT \hspace{1.cm} Wavelet Transform} 
\end{itemize}
\section*{Acknowledgments}
The authors are grateful to Piero Rosati for useful discussions regarding
astrophysical contents and to Andrea Bignamini for making available the results on
our simulated data employing {\sc wavdetect}. 
The first author is thankful to Udo von Toussaint and Silvio Gori for their
comments and suggestions regarding coding issue. \\
This work makes use of EURO-VO software tools, the Digitized Sky
Surveys and the SAOImage DS9 developed by Smithsonian Astrophysical 
Observatory. 
The EURO-VO has been funded by the European Commission through contract 
numbers RI031675 (DCA) and 011892 (VO-TECH) under the 6th Framework 
Programme and contract number 212104 (AIDA) under the 7th Framework
Programme. The Digitized Sky Surveys were produced at the Space Telescope
Science Institute under U.S. Government grant NAG W-2166. 
The images of these surveys are based on photographic data obtained using the
Oschin Schmidt Telescope on Palomar Mountain and the UK Schmidt Telescope. The
plates were processed into the present compressed digital form with the
permission of these institutions. 
The Second Palomar Observatory Sky Survey (POSS-II) was made by the California 
Institute of Technology with funds from the National Science Foundation, the National
Geographic Society, the Sloan Foundation, the Samuel Oschin Foundation, and
the Eastman Kodak Corporation.

\bsp

\label{lastpage}

\end{document}